\pdfoutput=1
\documentclass[logo,tresdirectores,doctorado]{tesis}
\usepackage{amsmath}
\usepackage{float} 
\usepackage{hyperref}
\hypersetup{
	pdftitle={Time-dependent Monte Carlo in fissile systems with beta-delayed neutron precursors},    
	pdfauthor={Jaime Romero-Barrientos},     
	pdfcreator={Jaime Romero-Barrientos},   
	colorlinks=true,       
	linkcolor=blue,          
	citecolor=blue,        
}
\usepackage[utf8]{inputenc} 
\usepackage{color} 
\usepackage{booktabs} 
\usepackage{multicol} 
\usepackage{multirow, makecell}
\usepackage{rotating}
\usepackage{longtable} 
\usepackage{pdflscape} 
\usepackage{caption}
\usepackage{fancyvrb} 
\usepackage{enumitem}		
\setlist[description]{leftmargin=\parindent,labelindent=\parindent} 
\usepackage{adjustbox}
\usepackage{xcolor}	
\usepackage[makeroom]{cancel}	
\usepackage{csquotes}
\usepackage{upgreek}

\begin{document}	

\pagestyle{empty}

\renewcommand{\baselinestretch}{1.0}

\titulo{Time-dependent Monte Carlo in fissile systems with beta-delayed neutron precursors}

\autor{Jaime Alfonso Romero Barrientos}

\fecha{15}{Enero}{2021}
\fechapublico{10}{Marzo}{2021}

\director{Dr. Hugo Arellano Sepúlveda} 
\codirector{Dr. Francisco Gabriel Molina Palacios}

\maketapa

\hypersetup{pageanchor=false}
\begin{glosario}
\vspace{2ex}

\begin{description}
	
	\item[$\beta$-delayed neutron emitter] Nuclei that emits a $\beta$-delayed neutron. 
	
	\item[Analog Monte Carlo simulation] Monte Carlo simulation which follows the natural (i.e. physical) probability distribution function for random sampling (See Sec.~\ref{subsection:variance_reduction_methods}).
	
	\item[Batch] In a Monte Carlo simulation, a batch is a single realization of a tally random variable. In the simulation both the number of batches as well as the number of particles per batch must be specified. 
	
	\item[Configuration] In this work, it refers to the state of neutron multiplication of the system. Configuration can be either subcritical, critical or supercritical.
	
	\item[Criticality calculation] Also called eigenvalue calculation. Monte Carlo simulation where the objective is the determination of the state of neutron multiplication in a fissile system. If $k_{\mathit{eff}}=1$ the system is in a critical configuration. While if $k_{\mathit{eff}}<1$ ($k_{\mathit{eff}}>1$) the system is in a subcritical (supercritical) configuration (See Sec.~\ref{subsection:NTE_Keigenvalue}).
	
	\item[Delayed neutron fraction] Denoted as $\beta$. Represents the fraction of total fission neutrons which are delayed (See Sec.~\ref{section:prompt_and_delayed_neutrons}).

	\item[Effective delayed neutron fraction] Denoted as $\beta_{\mathit{eff}}$. Represents the fraction of fissions caused by delayed neutrons. Its defined as the delayed neutron fraction $\beta$ weighted by the neutron importance, which represents how effective the neutron is in causing fission (See Sec.~\ref{subsection:point_kinetics_eqs}).

	\item[Effective multiplication factor] Parameter that shows the state of neutron multiplication in a fissile system. If $k_{\mathit{eff}}=1$ the system is in a critical configuration. While if $k_{\mathit{eff}}<1$ ($k_{\mathit{eff}}>1$) the system is in a subcritical (supercritical) configuration (See Sec.~\ref{subsection:NTE_Keigenvalue}).	
	
	\item[Fixed-source calculation] Monte Carlo simulation where the initial particle source is known and the resulting neutron distribution is determined.
	
	\item[MCNP]  MCNP is a general-purpose Monte Carlo N-Particle code that can be used for neutron, photon, electron, or coupled neutron/photon/electron transport. Developed and mantained by Los Alamos National Laboratory.
	
	\item[$N$-group structure] Scheme that groups all the $\beta$-delayed neutron precursors into $N$-groups. Each precursor group contains a number of different isotopes. In ENDF/B.VIII.$0$ database, $N=6$. In JEFF-$3$.$1$.$1$, $N=8$ (See Sec.~\ref{section:prompt_and_delayed_neutrons}).
	
	\item[Non-analog Monte Carlo simulation] Monte Carlo simulation which follows a modified (i.e. non-physical) probability distribution function for random sampling in order to reduce the variance of the result obtained when using an analog Monte Carlo simulation (See Sec.~\ref{subsection:variance_reduction_methods}).
	
	\item[OpenMC] OpenMC is a community-developed Monte Carlo neutron and photon transport simulation code. It is capable of performing fixed source, k-eigenvalue, and subcritical multiplication calculations on models built using either a constructive solid geometry or CAD representation. It was originally developed by members of the Computational Reactor Physics Group at the Massachusetts Institute of Technology starting in 2011 (See Sec.~\ref{section:monte_carlo_with_openmc}).
	
	\item[OpenMC(TD)] Time-dependent OpenMC. OpenMC code with added capabilities shown in this work, that is: explicit presence of time, scoring of time dependent quantities, non-analog simulation scheme to simulate the $\beta$-delayed neutron emission, population control to keep the number of particles constant and option to use either the $N$-group precursor structure (with $N=1,6, \text{or } 8$) or $M$-individual precursor structure (with $1 \le M \le 269$). See Chapter~\ref{ch:results}.
	
	\item[pcm] \textit{Per-cent mille} one-thousandth of a percent.

	\item[Point kinetics approximation] Theoretical approximation used to study the kinetic behaviour of a fissile system, where the flux is assumed to be a separable function of space and time (See Sec.~\ref{subsection:point_kinetics_eqs}).
	
	\item[Prompt drop] Fast decrease in neutron population or flux caused by a reduction in the system reactivity. The timescale of this process is of the order of the prompt neutron generation time (Sec.~\ref{subsection:energy_u235_vacuum_sub}).
	
	\item[Prompt jump] Fast increase in neutron population or flux caused by an increase in the system reactivity. The timescale of this process is of the order of the prompt neutron generation time (See Sec.~\ref{subsection:mono_reactivity} and Sec.~\ref{subsection:energy_u235_vacuum_super}).

	\item[Prompt neutron generation time] At $k_{\mathit{eff}}=1$, average time between two generations of prompt neutrons.

	\item[Precursor structure] In this work, it refers to the organization of the $\beta$-delayed neutron emitters. Precursor structure can be either a $N$-group precursor structure or a $N$-individual precursor structure. In the latter, $\beta$-delayed neutrons are emitted from individual precursors, not groups (See Sec.~\ref{subsection:individual_precursor} and Sec.~\ref{section:energy_individual_precursors}).  
	
	\item[Precursor] Fission product ($Z,N$) which decays through a $\beta$-delayed process to another nuclei ($Z+1,N-1$), which in turn decays to the ($Z+1,N-2$) nucleus, emitting a $\beta$-delayed neutron (See Sec.~\ref{section:prompt_and_delayed_neutrons}).  
	
	\item[Skipped cycles] Batches discarded before scores begin to accumulate in a Monte Carlo calculation.
	
	\item[System] In this work, it refers to the simulated structure, characterized by its geometry, materials, moderation, and so on. 
	
	\item[Transient source] In this work, it refers to the initial particle source, comprised of neutrons and precursors, needed to start a transient calculation (See Sec.~\ref{section:initial_particle_source}).
	
	\item[Weight, statistical] Number that represents how many real (i.e. physical) particles a Monte Carlo particle represents. If the statistical weight of a neutron is $2$, then that neutron represents $2$ neutrons.  
	
\end{description}
\end{glosario}
\newpage
\hypersetup{pageanchor=false}
\resumenCastellano{En el campo de la f\'isica de reactores nucleares, los fen\'omenos transientes suelen estudiarse usando m\'etodos deterministas o h\'ibridos. Estos m\'etodos requieren de variadas aproximaciones, tales como: discretizaciones de la geometr\'ia, del tiempo y de la energ\'ia; homogeneizaci\'on de materiales; y suposici\'on de condiciones de difusi\'on, por mencionar algunas. En este contexto, las simulaciones Monte Carlo son especialmente adecuadas para estudiar estos problemas. Los retos que se presentan al usar simulaciones Monte Carlo en cin\'etica espacio-temporal de sistemas fisibles son las escalas de tiempo inmensamente distintas involucradas en la emisión de neutrones inmediatos y retardados, lo que implica que los resultados obtenidos tienen asociada una gran varianza si se utiliza una simulaci\'on Monte Carlo an\'aloga. Adem\'as, tanto en simulaciones deterministas como en Monte Carlo, los precursores de neutrones retardados est\'an agrupados en una estructura $6$ u $8$ grupos, pero hoy en d\'ia no hay una raz\'on s\'olida para mantener esta agrupaci\'on.
	
En este trabajo, y por primera vez, se han implementado los datos de precursores individuales en una simulaci\'on Monte Carlo, incluyendo expl\'ictamente la dependencia temporal relacionada con la emisi\'on $\beta$ retardada de neutrones. Esto fue logrado modificando el c\'odigo abierto Monte Carlo OpenMC. En el c\'odigo modificado --\textit{Time Dependent} OpenMC u OpenMC(TD) -- se abord\'o la dependencia temporal relacionada con la emisi\'on retardada de neutrones originada de la desintegraci\'on $\beta$. La varianza del valor esperado de observables, como el flujo neutr\'onico, asociada a las diferentes escalas de tiempo entre los neutrones inmediatos y prompts, fue reducida forzando la desintegraci\'on de una nueva part\'icula Monte Carlo a\~nadida al código, el precursor, dentro de cada intervalo temporal, incrementando intencionalmente el n\'umero de neutrones retardados en la simulaci\'on. Dado que hay una producci\'on continua de neutrones retardados, se tuvo que imponer el control de poblaci\'on. Esto se logr\'o usando el m\'etodo de \textit{combing} al final de cada intervalo temporal.

Los datos de secciones eficaces dependientes de la energ\'ia vienen de la biblioteca JEFF-$3$.$1$.$1$. Los datos de los precursores individuales fueron tomados de las bibliotecas JEFF-$3$.$1$.$1$ (\textit{yields} cumulativos) y ENDF-B/VIII.$0$ (probabilidades de emisi\'on de neutrones retardados y espectros de energ\'ia de neutrones retardados).

OpenMC(TD) fue probado en: i) un sistema monoenerg\'etico; ii) un sistema sin moderaci\'on y dependiente de la energ\'ia donde los precursores se tomaron individualmente o en grupos; y finalmente iii) un sistema moderado por agua liviana dependiente de la energ\'ia, usando $6$-grupos de precursores, $50$ precursores y $40$ precursores individuales.     
}

\clearpage
\resumenIngles{In the field of nuclear reactor physics, transient phenomena are usually studied using deterministic or hybrids methods. These methods require many approximations, such as: geometry, time and energy discretizations, material homogenization and assumption of diffusion conditions, among others. In this context, Monte Carlo simulations are specially adequate to study these problems. Challenges presented when using Monte Carlo simulations in space-time kinetics in fissile systems are the immensely different time-scales involved in prompt and delayed neutron emission, which implies that results obtained have a large variance associated if an analog Monte Carlo simulation is utilized. Furthermore, in both deterministic and Monte Carlo simulations delayed neutron precursors are grouped in a $6$- or $8$- group structure, but nowadays there is not a solid reason to keep this aggregation. 
	
In this work, and for the first time, individual precursor data is implemented in a Monte Carlo simulation, explicitly including the time dependence related to the $\beta$-delayed neutron emission. This was accomplished by modifying the open source Monte Carlo code OpenMC. In the modified code -- Time Dependent OpenMC or OpenMC(TD) -- time dependency related to delayed neutron emission originated from $\beta$-decay was addressed. The variance of the expected values of observables, such as neutron flux, associated to the different time scales between prompt and delayed neutrons was reduced by forcing the decay of a new Monte Carlo particle-like added to the code, the precursor, within each time interval, intentionally increasing the number of delayed neutrons in the simulation. Since there is a continuous production of delayed neutrons, population control had to be enforced. This was accomplished by using the \textit{combing method} at the end of each time interval.

Continuous energy neutron cross-sections data used comes from JEFF-$3$.$1$.$1$ library. Individual precursor data was taken from JEFF-$3$.$1$.$1$ (cumulative yields) and ENDF-B/VIII.$0$ (delayed neutron emission probabilities and delayed neutron energy spectra).

OpenMC(TD) was tested in: i) a monoenergetic system; ii) an energy dependent unmoderated system where the precursors were taken individually or in a group structure; and finally iii) a light-water moderated energy dependent system, using $6$-groups, $50$ and $40$ individual precursors. 
}

\pagestyle{myheadings}
\newpage
\pagenumbering{roman}
\tableofcontents
\setcounter{page}{2}

\listoffigures
\listoftables
\newpage

\pagenumbering{arabic}

\chapter{Introduction}
\label{ch:introduction}

Nuclear fission is a process where the atomic nucleus splits in two or three fission products (lighter weight nuclei) and neutrons. In heavy nuclei this process can happen as a spontaneous desintegration (${}^{252}$Cf), or it can be induced by the reaction with a neutron. If fission is induced in a nucleus by a thermal energy neutron, then the nucleus is said to be fissile (${}^{235}$U or ${}^{239}$Pu). If the nucleus requires neutrons with a certain threshold energy to be fissioned, then it is said to be a fisssionable nucleus (${}^{238}$U or ${}^{232}$Th). 

In fission reactions two types of neutrons are emitted: prompt and delayed neutrons. Prompt neutrons are emitted almost instantaneously ($\sim\!10^{-14}$~s) after fission occurs with energies of the order of a few MeV. Meanwhile, delayed neutrons are emitted from milliseconds to tens of seconds after fission with energies of the order of hundreds of keV. Delayed neutron emission is associated to the decay of isotopes from the decay chain of fission products. These nuclei emitters of $\beta$-delayed neutrons, are called precursors. For example, for ${}^{235}$U there are about $540$ fission products and $270$ precursors~\cite{BROWN20181}. 

If there is enough fissile material, a neutron can induce fission in other nucleus and initiate a chain reaction. This chain reaction can be sustained in time depending on the density of fissile material, neutron energy at the moment of fission and the fission reaction rate. The Neutron Transport Equation models the propagation of neutrons in a fissile system. 

The Neutron Transport Equation is a linear, integro-diferential equation for the neutron flux which depends of seven variables: three for position in space, two for directions, one for energy and one for time~\cite{nla.cat-vn2675466}. Solving this equation is a complex task, for which there are two possible approaches: deterministic and stochastic methods.
Deterministic methods resort to the discretization of the transport equation with respect to its variables and converting the problem into a system of algebraic equations to be solved. One of the main disadvantages of these methods is the accuracy of its results, due to the discretization of the phase space (mesh or grid resolution). On the other hand, stochastic methods simulate the physical transport problem randomly sampling the physical interaction of neutrons in a material according to its reaction cross sections. Observables such as neutron flux, reaction rates, currents, among others are obtained by the expected value of $N$ realizations of the random sampling.
The advantage of this method lies in the fact that does not resort to any approximation or discretization; its disadvantage is that the associated statistical uncertainty converges slowly as $ {1}/{\sqrt{N}}$, with $N$ the number of particles simulated. In this thesis, stochastic Monte Carlo method was used to solve the Neutron Transport Equation in fissile systems, approaching to be used in a complete nuclear reactor model. 

While Monte Carlo methods are widely used in criticality and fixed source calculations, where the system is supposed to be in stationary state, only recently there have been studies to include time dependence in neutron transport, taking advantage of the better computing capabilities available. Some examples of these studies are the work of Snejitzer~\cite{doi:10.13182/NSE12-44}, Mylonakis~\cite{MYLONAKIS2017103} and Faucher~\cite{faucher:tel-02406396}, all of them focused in the inclusion of time dependence together with the coupling of feedback from thermal-hydraulics calculations.         

These investigations have in common the use of the customary group structure for all the precursors. Each precursor group, which contains a number of different isotopes, is characterized by a grouped: i) decay constant; ii) relative yield; and iii) energy spectrum for the delayed neutron emission. This structure was proposed in 1957 by Keepin~\cite{KEEPIN1957IN2}, and it is based in the assumption that the decay of the delayed neutron activity can be represented by a linear superposition of exponential decay periods. 
Although this grouping is routinely used when performing deterministic or Monte Carlo simulations, it limits the possibility of studying the effect of changes in quantities such as the time evolution of the neutron flux, stimulated by new and improved nuclear data on individual precursors. 

There has been a renewed interest in the measurement of nuclear decay properties of the most neutron-rich nuclei, such as decay half-lifes, neutron emission probabilities and production yields~\cite{dimitriou}, along with efforts from the International Atomic Energy Agency Coordinated Research Project on a \textit{Reference Database for $\beta$-delayed Neutron Emission}~\cite{IAEA_CRP}. This scenario brings the opportunity to explore how the new individual precursor data impacts on simulations of fissile systems individually or in different precursor groupings.

The objective of this work is to explicitly include the time dependence related to the $\beta$-delayed neutron emission from individual precursors in a Monte Carlo simulation. This entails two challenges: to simulate the delayed emission from precursors, and the inclusion of individual precursor data in the simulation. To include these modifications, the open source Monte Carlo code OpenMC was chosen~\cite{ROMANO201590}.

This work is divided in $5$ chapters and $5$ appendices. In Chapter~\ref{ch:theoretical_framework}, the theoretical framework behind this work is presented. In particular, the Neutron Transport Equation (NTE) is examined, including the $k$-eigenvalue form and the point kinetics equation approximation. After that, the main features and differences between prompt and $\beta$-delayed neutrons are discussed, along with the important role of $\beta$-delayed emission for nuclear reactor operation. The $N$-group structure for delayed neutron precursors is also examined. Afterwards, the nuclear parameters needed to understand the $\beta$-delayed neutron emission from individual precursors are described, together with the nuclear data libraries used in this work, JEFF-$3$.$1$.$1$~\cite{jeff311} and ENDF/B-VIII.$0$~\cite{BROWN20181}. Finally, two of the approaches used to solve the NTE are discussed: deterministic and Monte Carlo methods. Related to the latter, a description of variance reduction techniques is presented. 

In Chapter~\ref{ch:methodology}, methods used and developed to include the $\beta$-delayed neutron emission from individual precursors in a Monte Carlo simulation are discussed. The first point addressed is about OpenMC, the code chosen to include the modifications needed to achieve the objectives of this work. This modified code will be known as Time-dependent OpenMC or OpenMC(TD). Afterwards, details on the methodology to include time dependence in a Monte Carlo simulation are explained. The next part shows that time delay of $\beta$-delayed neutron emission, in an analog Monte Carlo simulation, entails large variance in the results obtained, due to the different time scales between the emission of prompt and delayed neutrons. To solve this problem, forced decay of precursors is implemented in OpenMC(TD), but this strategy requires population control of the neutron and precursor population. Regarding the inclusion of individual precursors, the steps taken to include them in OpenMC(TD) are: defining a precursor importance, so in the event of delayed neutron emission in the simulation, it can be chosen which precursor will decay. This decay will have its respective precursor decay constant associated and the corresponding delayed neutron energy will be the average energy from the precursor delayed neutron spectrum.
 
In Chapter~\ref{ch:results}, first the OpenMC(TD) code is tested in the context of time dependence and inclusion of individual precursors. With the tests successfully passed, OpenMC(TD) is used to obtain the neutron flux as a function of time in different systems, with different configurations and using different precursor structures. The first system studied was a monoenergetic fissile system with $1$-group precursor structure, in subcritical, critical and reactivity insertion configurations. Afterwards, an energy dependent, unmoderated ${}^{235}$U system was studied. This case was no longer monoenergetic, but energy dependent, using cross sections from JEFF-$3$.$1$.$1$ nuclear database. Two configurations were considered, subcritical and supercritical, and for each the $\beta$-delayed neutron energies simulated were JEFF-$3$.$1$.$1$ and ENDF-B/VIII.$0$ databases, in $1$-group, $6$-group, $8$-group and $50$ individual precursor structures.
The last part of this chapter was related to simulations conducted in a light-water moderated, energy dependent system in a critical configuration with $\beta$-delayed neutron emission from $6$-group, $50$ individual, and $40$ individual precursors. 

Finally, in Chapter~\ref{ch:summary_conclusions} the conclusions and future perspectives of this work are presented.
\chapter{Theoretical Framework}
\label{ch:theoretical_framework}
In this chapter the theoretical framework behind this work is summarized. In Section~\ref{section:NTE} the Neutron Transport Equation (NTE) is examined, including the $k$-eigenvalue form (see Sec.~\ref{subsection:NTE_Keigenvalue}) and the point kinetic equation approximation (see Sec.~\ref{subsection:point_kinetics_eqs}). Then, in Section~\ref{section:prompt_and_delayed_neutrons} characteristics and differences between prompt and $\beta$-delayed neutrons are described, along with their important role for nuclear reactor operation and the $N$-group structure for delayed neutron precursors (see Sec.~\ref{subsection:importance_of_delayed_neutrons}). Afterwards, in Section~\ref{section:nuclear_data} the nuclear parameters needed to describe the $\beta$-delayed neutron emission from individual precursors are shown, including the nuclear data libraries used in this work (See Sec.~\ref{subsection:nuclear_libraries}). Finally, in Section~\ref{section:approaches_solve_NTE} two of the approaches used to solve the NTE are discussed, namely, deterministic (See Sec.~\ref{subsection:deterministic_methods}) and Monte Carlo methods (See Sec.~\ref{subsection:monte_carlo_NTE}), where also a description of variance reduction techniques is presented (See Sec.~\ref{subsection:variance_reduction_methods}). Emphasis is given to time dependent phenomena, which are central to the challenges met throughout this work.

\section{The Neutron Transport Equation (NTE)}
\label{section:NTE}
\subsection{General form}
\label{subsection:NTE_general_form}
The determination of the neutron distribution is the main problem of nuclear reactor theory because it determines the rate at which several nuclear reactions occur within a fissile system. The knowledge about the neutron distribution also gives information about the stability of the fission chain reaction. 
The most general equation that governs the process of neutron transport through a medium, this is, the motion of neutrons as they stream through a system, is the Neutron Transport Equation (NTE) equation~\cite{nla.cat-vn2675466}
\begin{multline}
\left[ \frac{1}{v} \frac{\partial}{\partial t} + \mathbf{\hat{\Omega}} \cdot \mathbf{\nabla} + \Sigma_{tot} (\mathbf{r}, E) \right] \psi(\mathbf{r},E, \mathbf{\hat{\Omega}}, t) \\
=   \int_0^{\infty} dE' \int_0^{4 \pi} d \mathbf{\hat{\Omega'}} \Sigma_S(E' \to E, \mathbf{\hat{\Omega'}} \to \mathbf{\hat{\Omega}}) \psi(\mathbf{r},E', \mathbf{\hat{\Omega'}}, t) + S(\mathbf{r},E, \mathbf{\hat{\Omega}}, t).
\label{eq:transport_general}
\end{multline}
In this equation the quantity to be determined is the neutron flux $\psi$, with $E$ the neutron kinetic energy, $\mathbf{\hat{\Omega}}$ the flux angular direction and $v$ is an average neutron speed. In this equation $\Sigma_{tot}$ and $\Sigma_s$ are the macroscopic total and scattering cross sections, respectively. Any external neutron source, such as fission neutrons, are represented by $S$.

\subsection{NTE with fission neutrons as an external source}
\label{subsection:nte_with_fission_neutrons}
When including fission neutrons explicitly in Eq.~\eqref{eq:transport_general}, two contributions to the source should be accounted for. The first term accounts for the prompt neutrons produced in the nuclear fission process and is given by    
\begin{equation}
\chi_p (E) \sum_i (1-\beta^i) \int_0^{\infty} dE' \int_0^{4 \pi} d \mathbf{\hat{\Omega'}}  \nu(E') \Sigma_f(\mathbf{r}, E') \psi(\mathbf{r},E', \mathbf{\hat{\Omega'}}, t).
\label{eq:prompt_source}
\end{equation}
Here, $\Sigma_f(\mathbf{r},E)$ is the macroscopic fission cross section, $\nu(E)$ is the average number of neutrons produced per fission, $\beta^i$ is the effective delayed neutron fraction per precursor group $i$, and $\chi_p(E)$ the fast fission neutron spectrum. The second contribution to the source term accounts for the delayed neutrons produced after the fission reaction and it reads
\begin{equation}
\sum_l \chi_l(E) \lambda_l C_l (\mathbf{r},t),
\label{eq:delayed_source} 
\end{equation}  
where the precursors are grouped in $l$ groups according to their decay constant $\lambda_l$, $C_l(\mathbf{r}, t)$ represents the $l$-th precursor concentration and $\chi_l(E)$ is the delayed neutron energy spectrum for the $l$ group. The precursor concentration, $C_l(\mathbf{r}, t)$, changes in time as
\begin{equation}
\frac{\partial}{\partial t} C_l(\mathbf{r},t) = \sum_i \beta_l^{i} \int \, dE' \int \, d \mathbf{\hat{\Omega}}\, \nu(E') \Sigma_f(\mathbf{r}, E') \psi(\mathbf{r},E', \mathbf{\hat{\Omega'}},t) - \lambda_l C_l(\mathbf{r},t),  
\label{eq:precursor_conc.}
\end{equation}
where the first term on the right hand of Equation~\eqref{eq:precursor_conc.} stands for the produced precursors while the left hand of the equation stands for decayed precursors.
Taking Eqs.~\eqref{eq:prompt_source},~\eqref{eq:delayed_source} and~\eqref{eq:precursor_conc.} into account, Eq.~\eqref{eq:transport_general} for the neutron flux is reduced to~\cite{osti_5538794}
\begin{align}
\begin{split}
\left[ \frac{1}{v} \frac{\partial}{\partial t} + \mathbf{\hat{\Omega}} \cdot \mathbf{\nabla} + \Sigma_{tot} (\mathbf{r}, E )  \right] & \psi(\mathbf{r},E, \mathbf{\hat{\Omega}}, t) \\
&=  \int_0^{\infty} dE' \int_0^{4 \pi} d \mathbf{\hat{\Omega'}} \Sigma_S(E' \to E, \mathbf{\hat{\Omega'}} \to \mathbf{\hat{\Omega}}) \psi(\mathbf{r},E', \mathbf{\hat{\Omega'}}, t) \\
&+ \chi_p (E) \sum_i (1-\beta^i) \int_0^{\infty} dE' \int_0^{4 \pi} d \mathbf{\hat{\Omega'}}  \nu(E') \Sigma_f(\mathbf{r}, E') \psi(\mathbf{r},E', \mathbf{\hat{\Omega'}}, t)  \\
&+ \sum_l \chi_l(E) \lambda_l C_l (\mathbf{r},t).
\label{eq:transport_full}
\end{split}
\end{align}
It is important to remark that this equation relies on some assumptions: (i) neutrons are point-like; (ii) between two collisions neutrons travel in a straight line; (iii) neutrons do not interact with each other; (iv) collisions are instantaneous; and (v) materials do not change in time.
Since the NTE features derivatives, appropriate initial and boundary conditions must be specified for the neutron flux. 
The initial condition can be the specification of the initial value for the neutron flux for all positions, energies and directions:
\begin{equation}
\psi(\mathbf{r}, E, \mathbf{\hat{\Omega}}, 0) = \psi_0(\mathbf{r}, E, \mathbf{\hat{\Omega}})
\label{eq:initial_condition}
\end{equation}
The boundary condition will depend on the problem being studied, but usually the boundary conditions are: i) vacuum boundary condition; ii) reflective boundary condition; and iii) known surface source.

\subsection{K-eigenvalue form}
\label{subsection:NTE_Keigenvalue}
One of the most useful notations of the NTE is the steady-state form associated with the criticality of the system. In this problem the objective is the determination of the $k$-eigenvalue ($k_{\mathit{eff}}$) that shows the state of neutron multiplication in a fissile system or nuclear reactor. If $k_{\mathit{eff}} = 1$, then the system is said to be in a critical state, if $k_{\mathit{eff}} < 1$ the system in a sub-critical state and if $k_{\mathit{eff}} > 1$ the system is in a super-critical state. In Eq.~\eqref{eq:transport_full} if a stationary solution is required, then it reads:
\begin{align}
\begin{split}
\left[ \mathbf{\hat{\Omega}} \cdot \mathbf{\nabla} + \Sigma_{tot} (\mathbf{r}, E )  \right] & \psi(\mathbf{r},E, \mathbf{\hat{\Omega}}) \\
&=  \int_0^{\infty} dE' \int_0^{4 \pi} d \mathbf{\hat{\Omega'}} \Sigma_S(E' \to E, \mathbf{\hat{\Omega'}} \to \mathbf{\hat{\Omega}}) \psi(\mathbf{r},E', \mathbf{\hat{\Omega'}}) \\
&+ \chi_p (E) \sum_i (1-\beta^i) \int_0^{\infty} dE' \int_0^{4 \pi} d \mathbf{\hat{\Omega'}}  \nu(E') \Sigma_f(\mathbf{r}, E') \psi(\mathbf{r},E', \mathbf{\hat{\Omega'}})  \\
&+ \sum_l \chi_l(E) \sum_i \beta_l^{i} \int \, dE' \int \, d \mathbf{\hat{\Omega}}\, \nu(E') \Sigma_f(\mathbf{r}, E') \psi(\mathbf{r},E', \mathbf{\hat{\Omega'}}).
\label{eq:transport_stationary}
\end{split}
\end{align}
By defining the net disappearance operator $L$ as
\begin{equation}
L f = \mathbf{\hat{\Omega}} \cdot \mathbf{\nabla} f + \Sigma_{tot} (\mathbf{r}, E ) f - \int_0^{\infty} dE' \int_0^{4 \pi} d \mathbf{\hat{\Omega'}} \Sigma_S(E' \to E, \mathbf{\hat{\Omega'}} \to \mathbf{\hat{\Omega}}) f(\mathbf{r},E', \mathbf{\hat{\Omega'}}),
\label{eq:net_disappearance_operator}
\end{equation}
and the total fission operator $F$ as
\begin{align}
\begin{split}
F f &= \chi_p (E) \sum_i (1-\beta^i) \int_0^{\infty} dE' \int_0^{4 \pi} d \mathbf{\hat{\Omega'}}  \nu(E') \Sigma_f(\mathbf{r}, E') f(\mathbf{r},E', \mathbf{\hat{\Omega'}})  \\
&+ \sum_l \chi_l(E) \sum_i \beta_l^{i} \int \, dE' \int \, d \mathbf{\hat{\Omega}}\, \nu(E') \Sigma_f(\mathbf{r}, E') f(\mathbf{r},E', \mathbf{\hat{\Omega'}}),
\label{eq:net_total_fission_operator}
\end{split}
\end{align}
Eq.~\eqref{eq:transport_stationary} can be rewritten as
\begin{equation}
L \psi(\mathbf{r},E', \mathbf{\hat{\Omega'}}) = F \psi(\mathbf{r},E', \mathbf{\hat{\Omega'}}).
\label{eq:eigenfunction_transport}
\end{equation}
By imposing that the system should be critical, the $k$-eigenmodes can be found
\begin{equation}
L \; \psi_k(\mathbf{r},E', \mathbf{\hat{\Omega'}}) = \frac{1}{k_{\mathit{eff}}} \, F \; \psi_k(\mathbf{r},E', \mathbf{\hat{\Omega'}}).
\label{eq:eigenfunction_transport_k}
\end{equation}

\subsection{Point kinetics equations}
\label{subsection:point_kinetics_eqs}
One theoretical approximation used to study the transient behaviour of a nuclear reactor is the point kinetic approximation~\cite{bell_glasstone}. In this case, the flux is assumed to be a separable function of space and time and the equations are obtained by weighting the transport equation by the adjoint flux. 

To obtain the pertinent equations first Eq.~\eqref{eq:transport_full} can be written as
\begin{equation}
\frac{1}{v} \frac{\partial}{\partial t} \psi(\mathbf{r},E, \mathbf{\hat{\Omega}}, t) + L \, \psi(\mathbf{r},E, \mathbf{\hat{\Omega}}, t) =  F_p \, \psi(\mathbf{r},E, \mathbf{\hat{\Omega}}, t) + \sum_l \chi_l(E) \lambda_l C_l (\mathbf{r},t),
\label{eq:time_dep_operators}
\end{equation}
where the prompt fission operator, $F_p$, is defined as
\begin{equation}
F_p \, f = \chi_p (E) \sum_i (1-\beta^i) \int_0^{\infty} dE' \int_0^{4 \pi} d \mathbf{\hat{\Omega'}}  \nu(E') \Sigma_f(\mathbf{r}, E') \psi(\mathbf{r},E', \mathbf{\hat{\Omega'}}, t) \, f.
\label{eq:prompt_fission_operator}
\end{equation} 
The adjoint equation to the $k$-eigenmodes Equation~\eqref{eq:eigenfunction_transport_k} is
\begin{equation}
L^{\dagger} \; \psi_k^{\dagger}(\mathbf{r},E, \mathbf{\hat{\Omega}}) = \frac{1}{k_{\mathit{eff}}} \, F^{\dagger} \; \psi_k^{\dagger}(\mathbf{r},E, \mathbf{\hat{\Omega}}),
\label{eq:eigenfunction_transport_k_adjoint}
\end{equation}
where $\psi_k^{\dagger}$ is the adjoint eigenmode for the neutron flux and the $L^{\dagger}$ is the adjoint of the operator.

To derive the point kinetic equations, the transport Eq.~\eqref{eq:time_dep_operators} is multiplied by $\psi_k^{\dagger}$ and Eq.~\eqref{eq:eigenfunction_transport_k_adjoint} is multiplied by the neutron flux $\psi(\mathbf{r},E, \mathbf{\hat{\Omega}}, t)$. The resulting equations are integrated over space, energy and angle and then are substracted from each other, obtaining
\begin{equation}
\frac{\partial}{\partial t} \langle \ \psi_k^{\dagger}, \frac{1}{v} \psi \rangle = \frac{k-1}{k} \langle \psi_k^{\dagger}, F \, \psi \rangle - \langle \psi_k^{\dagger}, F_d \, \psi \rangle + \sum_l \lambda_l \langle \psi_k^{\dagger}, \chi_d^l \, C_l \rangle.  
\label{eq:dot_product_adjoint}
\end{equation}
It is assumed that the neutron flux can be factorized as an amplitude factor that only depends on time and a time-independent flux shape factor: 
\begin{equation}
\psi(\mathbf{r},E, \mathbf{\hat{\Omega}},t) = s(t) \, \psi_k(\mathbf{r},E ,\mathbf{\hat{\Omega}}),
\label{eq:function_separable}
\end{equation}
where $\psi_k(\mathbf{r},E ,\mathbf{\hat{\Omega}})$ is the fundamental $k$ eigenmode and $s(t)$ is an amplitude factor that only depends on time. Thus, Eq.~\eqref{eq:dot_product_adjoint} the amplitude of the neutron population satisfies 
\begin{equation}
\frac{\partial}{\partial t} s(t) = \frac{\rho - \beta_{\mathit{eff}}}{\Lambda_{\mathit{eff}}} \,s(t) + \sum_l \lambda_l c_l(t),
\label{eq:neutrons_point_kinetic}
\end{equation} 
where $\rho$ is the reactivity given by
\begin{equation}
\rho = \frac{k-1}{k}
\label{eq:static_reactivity},
\end{equation} 
the effective delayed neutron fraction is
\begin{equation}
\beta_{\mathit{eff}} = \frac{\langle \psi_k^{\dagger},F_d \, \psi_k \rangle}{ \langle \psi_k^{\dagger},F \, \psi_k \rangle},
\label{eq:beta_eff_definition}
\end{equation}
the effective mean generation time is
\begin{equation}
\Lambda_{\mathit{eff}} = \frac{\langle \psi_k^{\dagger},\tfrac{1}{v} \, \psi_k \rangle}{ \langle \psi_k^{\dagger},F \, \psi_k \rangle},
\label{eq:Lambda_eff_definition}
\end{equation}
and the effective precursor concentration
\begin{equation}
c_l(t) = \frac{\langle \psi_k^{\dagger},\chi_d^l \, C_l \rangle}{ \langle \psi_k^{\dagger},\tfrac{1}{v} \, \psi_k \rangle}.
\label{eq:precursor_eff_definition}
\end{equation}
By proceeding in a similar way, an equation for the precursor concentration can be derived, which is coupled to Eq.~\eqref{eq:neutrons_point_kinetic},
\begin{equation}
\frac{\partial}{\partial t} c_l(t) = \frac{\beta_l}{\Lambda_{\mathit{eff}}}s(t) - \lambda_l c_l(t),
\label{eq:precursors_point_kinetic}
\end{equation}
where $\beta_l$ is the effective delayed neutron fraction for the precursor family $l$. 

Parameters $\beta_{\mathit{eff}}$ and $\Lambda_{\mathit{eff}}$ are called \textit{effective} because they have been weighted by the adjoint flux $\psi_k^{\dagger}$, which can be interpreted as the neutron importance. Physically, the neutron importance at a given point in phase space is proportional to the asymptotic neutron population of an hypotetical neutron introduced into a critical reactor at the same point in phase space.        

\section{Prompt and delayed neutrons}
\label{section:prompt_and_delayed_neutrons}
In section~\ref{section:NTE} the transport equation was presented. For nuclear fission present in a fissile system, the source term is comprised by two terms, the \textit{prompt} and the \textit{delayed fission term}. In fission events, two types of neutrons are released: prompt and delayed neutrons.

\begin{figure}[h]
	\centering
	\includegraphics[width=12cm]{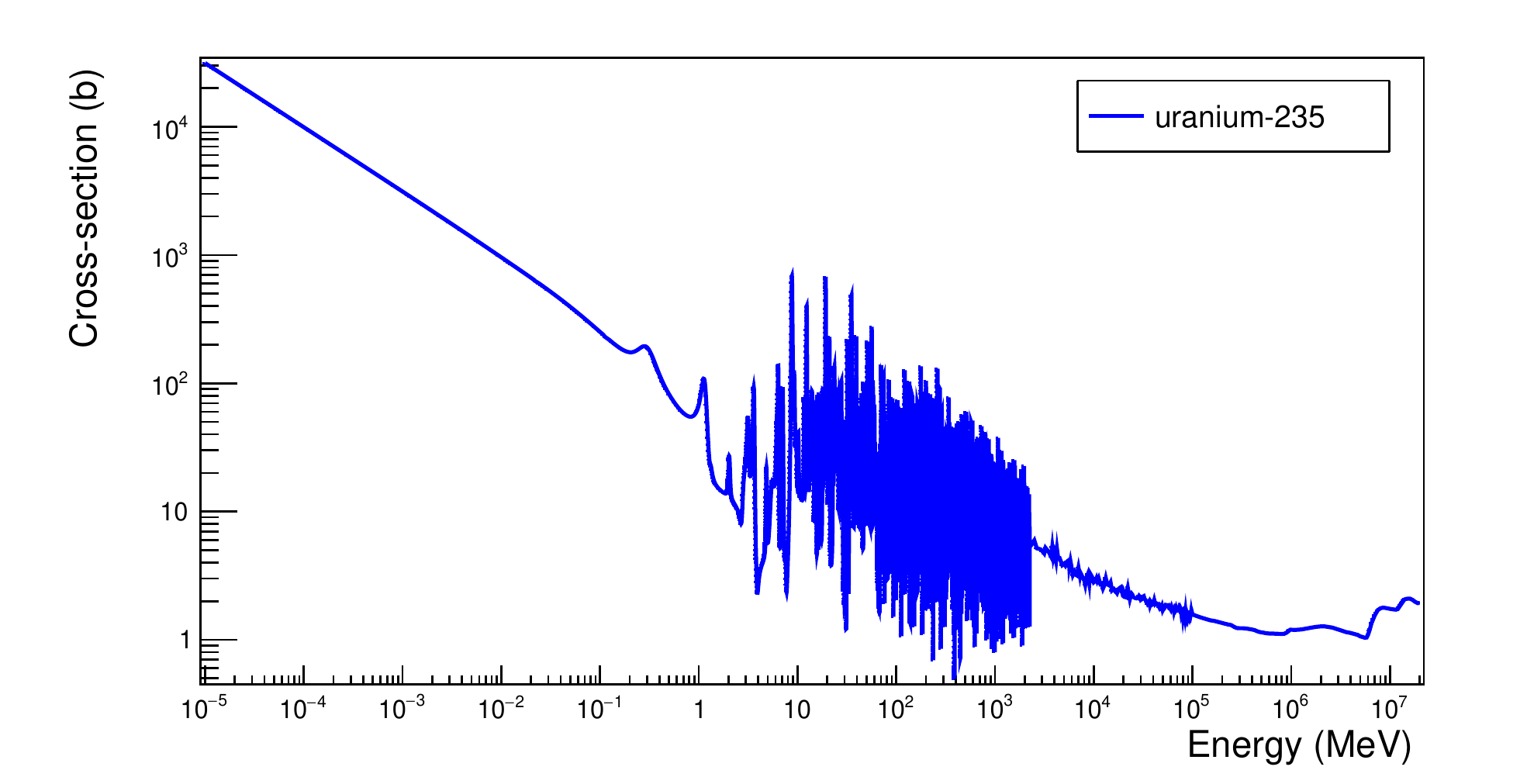}
	\caption{Fission Cross sections for ${}^{235}$~U. Data was retrieved from the JEFF-$3.1.1$ nuclear database.	\label{fig:u235_fission_spectrum}}
\end{figure}

Prompt neutrons are released almost instantaneously ($ \sim 10^{-14}$~s) after fission and are emitted with an average energy of $2$~MeV~\cite{doi:10.13182/NSE73-A23234}. Since the fission cross section for reactor fuel is higher for thermal energies, as it can be seen in Fig.~\ref{fig:u235_fission_spectrum}, prompt neutrons must be slowed down before they can induce fission. The average number of prompt neutrons produced per fission is denoted by $\bar{\nu_p}$. 

\begin{figure}[h!]
	\centering
	\includegraphics[scale=0.35]{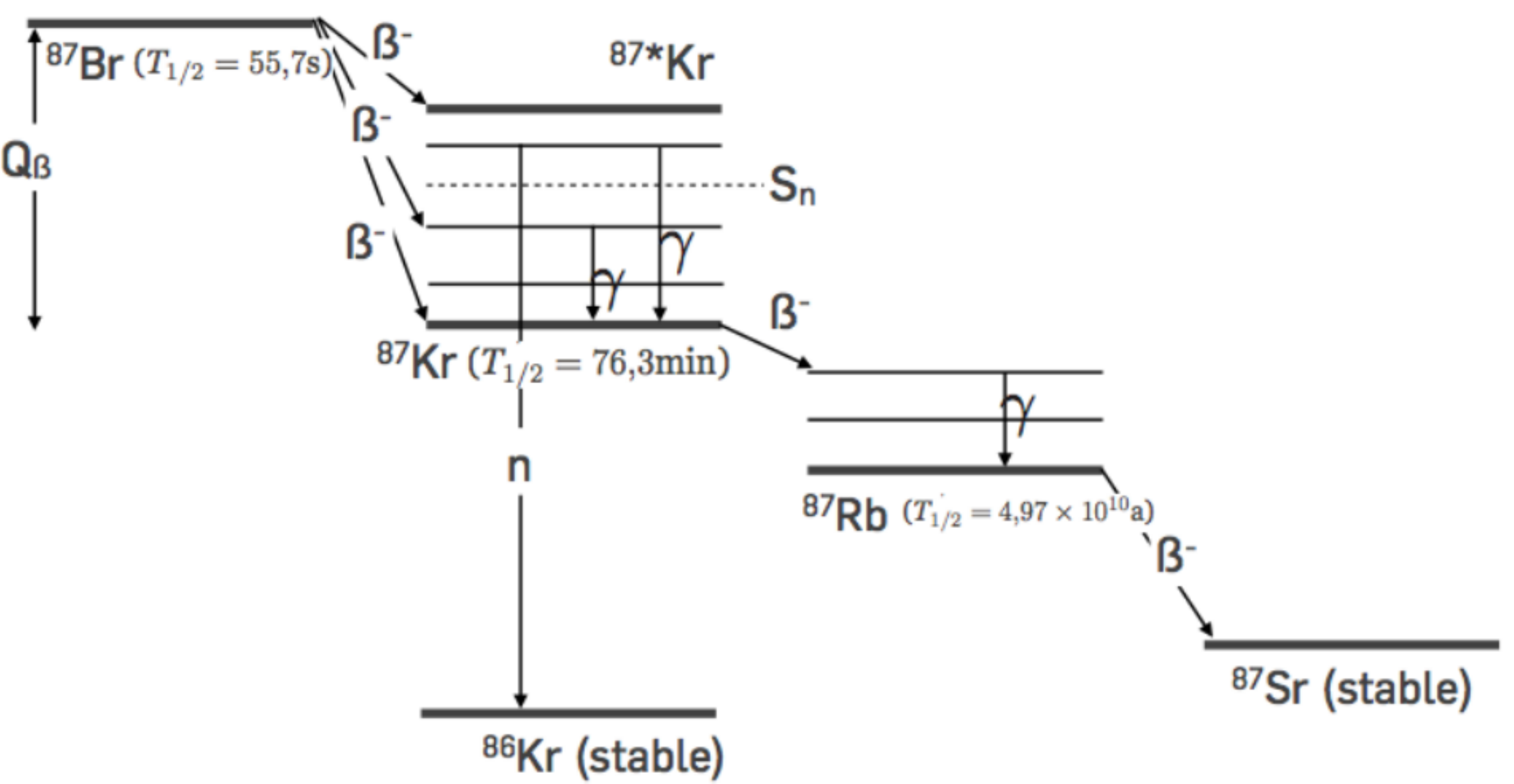}
	\caption{Decay of the ${}^{87}$Br delayed neutron precursor \label{fig:decayBr87}.}
\end{figure}

On the other hand, delayed neutrons are emitted between $10^{-3}$~s and $10^{2}$~s after a nuclear fission event and made up about $1$~\% of the total neutrons released during fission. This delayed emission is produced when a fission product $(Z,N)$ decays through a $\beta^{-}$ process with a father-daughter mass difference $Q_{\beta}$. If $Q_{\beta}$ is greater than the neutron separation energy $S_n$, then excited states in the daughter nucleus $(Z+1,N-1)$ can be populated. This nucleus can in turn decay to the nucleus $(Z+1,N-2)$. Although the neutron emission is instantaneous, the time scale of the emission is related to the half-life of the $\beta^{-}$ decay corresponding to the $(Z,N)$ nucleus. This parent $\beta$-decay nucleus $(Z,N)$ is known as \textit{delayed neutron precursor} or \textit{precursor}. To illustrate this process, in Fig.~\ref{fig:decayBr87} the decay scheme of the precursor ${}^{87}$Br is shown. In this scheme it can be seen that ${}^{87}$Br can decay through $\beta^{-}$ to a state in ${}^{87}$Kr${}^{*}$, followed by the subsequent decay of ${}^{87}$Kr${}^{*}$ to a state in ${}^{86}$Kr via neutron emission. The delayed time of this process is given by the parent half life, which is $55.7$~s. The average number of delayed neutrons emitted per fission is denoted by $\bar{\nu_d}$. The fraction of total fission delayed neutrons is denoted by $\beta$ and is defined as
\begin{equation}
\beta = \frac{\bar{\nu_d}}{\bar{\nu}}.
\label{eq:beta_fraction}
\end{equation} 

\begin{figure}[h]
	\centering
	\includegraphics[width=12cm]{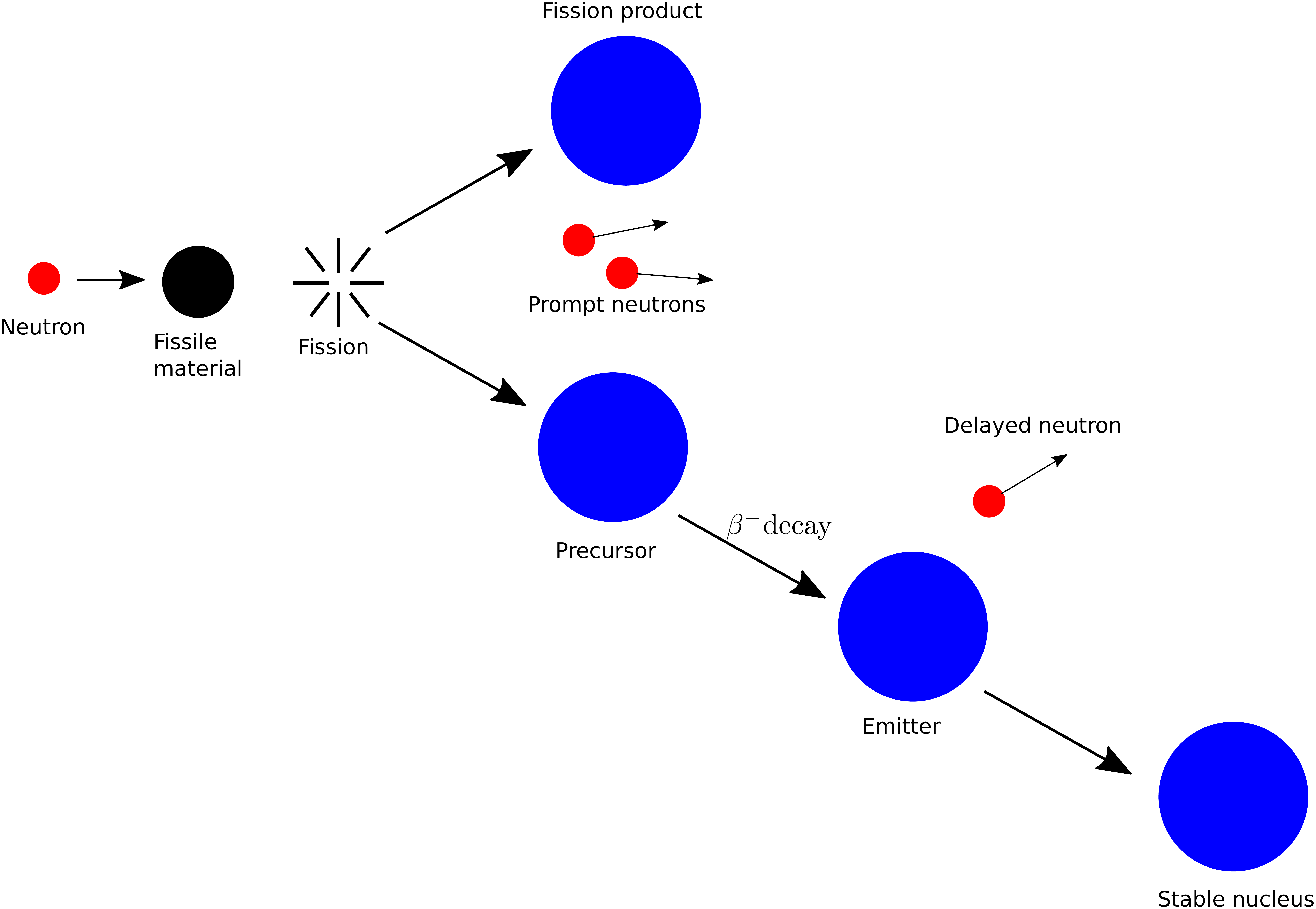}
	\caption{Schematic representation of the prompt and $\beta$-delayed neutron emission.}
		\label{fig:delayed_emission_figure}
\end{figure}

In a fission chain reaction a large number (about $270$ for ${}^{235}$U) of delayed neutron precursor isotopes can be produced~\cite{osti_6187550}. It has been customary to group these precursors in $6$ groups, characterized by its half-lives. Each precursor group contains a number of different isotopes. These groups are described in Table~\ref{table:6groups_precursors} for ${}^{235}$U, where for each group is shown: i) the group mean energy, $\bar{E}$, ii) its half-life, ${{T}_{1/2}}$ and, iii) its relative yield, given by $\beta_i/\beta$, where $\beta_i$ is the delayed fraction considering only the $i$-th group and the total delayed delayed fraction is $\sum_i \beta_i= \beta$.
 This group structure was proposed by Keepin~\cite{KEEPIN1957IN2}, who assumed that the decay of delayed neutron activity with time could be represented by a linear superposition of exponential decay periods. 

\begin{table*}[h!]
	\centering
	\footnotesize
	\begin{tabular}{@{}cccccc@{}}\toprule
		{Group} & {Precursors} &  {$\bar{E}$ (MeV)} & {${{T}_{1/2}}$ (s)} & {$\beta_i / \beta$} \\
		\midrule
		$1$ & ${}^{87}$Br, ${}^{142}$Cs & $0.40$ & $54.51$ & $0.038$\\
		$2$ & ${}^{137}$I, ${}^{88}$Br & $0.47$ & $21.84$ & $0.213$\\
		$3$ & ${}^{138}$I, ${}^{89}$Br, ${}^{93}$Rb, ${}^{94}$Rb  & $0.44$ & $6.00$ & $0.188$\\
		$4$ & ${}^{139}$I, ${}^{143}$Xe, ${}^{93}$Kr, ${}^{94}$Kr, ${}^{90}$Br, ${}^{92}$Br  & $0.55$ & $2.23$ & $0.407$\\
		$5$ & ${}^{140}$I, ${}^{145}$Cs & $0.51$ & $0.496$ & $0.128$\\
		$6$ & Br, As, Rb.  & $0.54$ & $0.179$ & $0.026$\\
		\bottomrule
	\end{tabular}
	\caption{$6$ delayed neutron precursor groups for ${}^{235}$U fission. In $6$-th group several isotopes of Br, As, Rb are included. \label{table:6groups_precursors}}
\end{table*}     

\subsection{Importance of delayed neutrons}
\label{subsection:importance_of_delayed_neutrons}
Delayed neutrons are important in the operation of a nuclear reactor due to the time delay that they introduce to the system, needed to control the state of the reactor through mechanical means, such as control rods. To illustrate this point, the variation of the neutron density without taking into account delayed neutrons is
\begin{equation}
\frac{dn}{dt} = \frac{k-1}{\ell}n(t) = \frac{\Delta k}{\ell} \, n(t),
\label{eq:kinetic_only_prompts} 
\end{equation} 
where $n(t)$ is the neutron density, $k$ is the multiplication factor and $\ell$ is the prompt neutron lifetime, taken as the average time a neutron stays in the system before leaking or being absorbed. The solution to this Eq. is
\begin{equation}
	n(t) = n_0 \exp \left (\frac{\Delta k}{\ell}\, t \right ) \equiv n_0 \exp \left ( \frac{t}{\tau}\right )
	\label{eq:kinetic_only_prompts_sol},
\end{equation} 
with $n_0$ the initial neutron density. The rate at which the reactor power increases is given by $\Delta k / \ell$. The reciprocal of this quantity is the reactor period, $\tau=\ell/\Delta k$, namely the time needed for the reactor power to grow by a factor of $e$. 

If there are no delayed neutrons, then the mean neutron lifetime is the mean prompt neutron lifetime (i.e. $\ell = \ell_p$) which in a light-water reactor is about $10^{-5}$~s~\cite{nla.cat-vn2675466}. Then, if there is a positive reactivity insertion of $10$~pcm\footnote{Per-cent mille: one-thousandth of a percent.}, the multiplication factor would change from $k=1.0000$ to $k=1.0001$ ($\Delta k=0.0001$ or a $0.01\%$ of $k$) and the reactor period is $\tau = 10^{-5}/0.0001 = 0.1$~s. With this period, in one second the reactor power would rise by a factor of $\approx 20000$, making impossible to control the reactor using mechanical control systems. 
When taking into account the delayed neutrons, the neutron lifetime changes because now a fraction $(1 - \beta_{\mathit{eff}})$ of the total neutrons have the prompt neutron lifetime $\ell_p$, while the delayed neutrons, a fraction of $\beta_{\mathit{eff}}$ the neutrons, live longer with a lifetime $(T_{\mathit{avg}} + \ell_p)$. This implies that the neutron lifetime in this case is
\begin{equation}
	\ell = \ell_p (1 - \beta_{\mathit{eff}}) + (T_{\mathit{avg}} + \ell_p) \beta_{\mathit{eff}} \approx \beta_{\mathit{eff}} T_{\mathit{avg}},
	\label{eq:weighted_mean_lifetime}
\end{equation}  
and if Eq.~\eqref{eq:weighted_mean_lifetime} is used in Eq.~\eqref{eq:kinetic_only_prompts_sol} the reactor period in this case would be $\tau = 0.08/0.0001 \approx 100$~s. With this reactor period, in a second the power would rise by a factor of $\approx 0.1$, making easier to control the behavior of the reactor in the face of reactivity insertions.  

Then, due to the effect of delayed neutrons, the period of the reactor increases and the rise in reactor power slows down, making possible to control the reactor by mechanical means.  

\section{Nuclear data}
\label{section:nuclear_data}
In this section the quantities of interest needed to characterize individual precursor nuclides are presented. Then, the two nuclear databases considered in this work, ENDF/B-VIII.$0$ and JEFF-$3.1.1$, are briefly described along with some of the differences found in the course of this work. 
\subsection{Quantities of interest}
\label{subsection:quantities_of_interest}
Although the $6$- or $8$- group structure is widely used in reactor calculations~\cite{rudstam}, nowadays there is not a solid reason to keep this aggregation. More and better nuclear data for the individual precursor nuclei has been available using high luminosity accelerator such as RIKEN~\cite{riken}, exploring neutron rich part of the nuclide chart. These data, combined with high efficiency neutron detector systems would allow to increase the knowledge of the individual parameters relevant for each precursors, such as fission yields or emission probabilities. To this end the quantities that characterize each precursor must be known: the fission yield ($FY$), the precursor decay constant ($\lambda$) and the precursor delayed neutron emission probability ($P_n$). 
To study the effect of individual precursors in a Monte Carlo simulation, information related to the energy distribution of the emitted neutrons is also needed. Now each of this quantities will be reviewed~\cite{foligno_thesis}:

\noindent \textbf{Fission Yield ($\boldsymbol{FY}$)}: Two yields can be defined, one is the \textit{{Independent Fission Yield (IY)}}, which is the average number of atoms of a specified nucleus produced by one fission, after the emission of prompt neutrons and excluding radioactive decay. For example, ${}^{87}$Br is one of the most prominent delayed neutron precursor and its \textit{IY} is $0.0127$, according to the ENDF/B-VIII.0 library. This means that for $10000$ fissions, $127$ atoms of ${}^{87}$Br are produced.  The other is the \textit{{Cumulative Fission Yield (CY)}}, which is the number of atoms of a specific nuclide produced directly and via decay of precursors per one fission reaction.

\noindent \textbf{Precursor Decay Constant ($\boldsymbol{\lambda}$)}: The decay constant represents the probability for a nucleus to decay per time unit. The decay probability of a precursor is proportional to the number of nuclei. 
\begin{equation}
\lambda = -\frac{dN/dt}{N}
\end{equation}

\noindent \textbf{Precursor delayed neutron emission probability ($\boldsymbol{P_n}$)}: For $\beta$-delayed neutron emission to occur, the $\beta^{-}$ decay energy ($Q_{\beta}$) must be larger than the neutron separation energy ($S_n$) of the decay daughter. The precursor delayed neutron emission probability represents the probability of one or more neutron emission.

\noindent \textbf{Precursor delayed neutron spectrum}: the energy distribution of neutrons emitted by each precursor is characterized by its spectrum. At this moment, the ENDF/B-VIII.0 database has only $34$ evaluated experimental spectra while the others come from QRPA calculations~\cite{osti_6187550}. In this work the ``mean energy'' of the delayed neutrons emitted was used. 

\noindent \textbf{Average delayed neutron yield ($\boldsymbol{{\nu_d}}$)}: Also known as the average number of delayed neutrons produced per fission, it can be either measured~\cite{KEEPIN1957IN2}, or calculated using the cumulative yield and the precursor delayed neutron emission probability,
\begin{equation}
	{\nu_d} = \sum_i^N CY_i \, P_{n,i}
	\label{eq:nud_calculation}
\end{equation} 
where, $N$ is the number of precursors. In $1990$ the Nuclear Energy Agency established a Working Party on International Nuclear Data Evaluation Co-operation (WPEC) to ``promote the exchange of information on nuclear data evaluations, validation and related topics. Its aim is also to provide a framework for co-operative activities between members of the  major nuclear data evaluation projects''. The Subgroup 6 (WPEC-$6$) in particular had the objective of reducing the uncertainties on delayed nuclear data and in this context, the recommended value for the average delayed neutron yield for ${}^{235}$U${}_{thermal}$ is $1.62 \times 10^{-2}$~\cite{rudstam}. 

\subsection{Nuclear data libraries}
\label{subsection:nuclear_libraries}

A nuclear data library is a dataset of stored nuclear data in a certain format~\cite{Wu2019}. Nuclear data derived from the combination of experimental data and nuclear physics models are known as evaluated nuclear data libraries. The standard format for the storage of nuclear data is the ENDF-$6$ format (\textit{Evaluated Nuclear Data File})~\cite{osti_981813}. 
In the course of this work nuclear data from two libraries was studied, the \textbf{JEFF} and \textbf{ENDF/B} libraries. 
The JEFF (\textit{Joint Evaluated Fission and Fusion File}) library is created by the OECD/NEA and the version used in this thesis was JEFF-$3.1.1$~\cite{jeff311}. The ENDF/B (\textit{Evaluated Nuclear Data File / B}) library is created by the \textit{Cross Section Evaluation Working Group} and the version used was ENDF/B-VIII.$0$~\cite{BROWN20181}. Both libraries contain radioactive decay data sub-libraries, where the yields, branching ratios and delayed neutron spectra can be found.

It was found that the data from both libraries do not agree with each other and that there are important differences between quantities of interest. To show this, the independent yield, cumulative yield and branching ratio is shown for some of the main precursors from the $6$-group structure~\cite{doi:10.13182/NSE75-A26620}, as it is shown in Table~\ref{tab:yields_and_probs_example}. 

\begin{table}[h!]
\begin{center}
\resizebox{\columnwidth}{!}{%
\begin{tabular}{|*{8}{c|}}  
	\hline
	\multicolumn{2}{|c}{ } & \multicolumn{2}{|c}{\textbf{IY} (\% per fission)} & \multicolumn{2}{|c}{\textbf{CY} (\% per fission)} & \multicolumn{2}{|c|}{$\boldsymbol{\beta_n}$ (\% per fission)}\\ \hline
	\multicolumn{1}{|c}{\textbf{Group}} & \multicolumn{1}{|c|}{\textbf{Nucleus}} & ENDF/B-VIII.$0$ & JEFF-$3.1$ & ENDF/B-VIII.$0$ & JEFF-$3.1$ & ENDF/B-VIII.$0$ & JEFF-$3.1$ \\ \hline
	    $1$ & ${}^{87}$Br & $1.270(36)$ & $1.41(21)$ & $2.030(41)$ & $2.140(49)$ & $2.60(4)$ & $2.51(8)$ \\ 
	\hline
	\multirow{2}{*}{$2$} 
	       & ${}^{88}$Br & $1.390(28)$ & $1.48(30)$  & $1.780(50)$ & $1.82(16)$ & $6.58(18)$ & $6.7(2)$    \\ 
	       & ${}^{137}$I & $2.62(10)$  &  $2.95(54)$ & $3.070(86)$ & $3.57(25)$ & $7.14(23)$ & $6.5(4)$  \\  
	 \hline
	 \multirow{3}{*}{$3$} 
	       & ${}^{89}$Br & $1.040(42)$ & $1.29(33)$ & $1.090(30)$ & $1.36(24)$ & $13.8(4)$ & $14.1(4)$     \\ 
	       & ${}^{92}$Rb & $3.130(63)$ & $2.87(51)$ & $4.820(67)$ & $4.83(14)$ & $8.1017(5)$ & $0$ \\  
	 	   & ${}^{138}$I & $1.420(40)$ & $1.38(42)$ & $1.490(42)$ & $1.47(33)$ & $5.56(22)$ & $5.3(3)$ \\  
	 \hline 
	 \multirow{2}{*}{$4$} 
	       & ${}^{85}$As & $0.121(78)$ & $0.141(47)$ & $0.22(14)$ & $0.143(42)$ & $59.40(24)$ & $22(3)$    \\ 
	       & ${}^{90}$Br & $0.553(33)$ & $0.48(17)$ & $0.564(23)$ & $0.49(15)$ & $25.2(9)$ & $24.6(7)$ \\  
	       & ${}^{94}$Rb & $1.570(44)$ & $1.40(41)$ & $1.650(46)$ & $1.50(32)$ & $10.5(4)$ & $10.1(2)$ \\ 
	       & ${}^{139}$I & $0.771(62)$ & $0.59(20)$ & $0.778(62)$ & $0.60(19)$ & $10.0(3)$ & $9.8(4)$ \\ 
	 \hline 
	 	 \multirow{3}{*}{$5$--$6$} 
	 	   & ${}^{91}$Br & $0.224(25)$ & $0.151(53)$ & $0.224(25)$ & $0.152(52)$ & $0$ & $20(2)$    \\ 
	       & ${}^{95}$Rb & $0.764(31)$ & $0.65(22)$ & $0.770(31)$ & $0.66(20)$ & $8.7(3)$ & $8.6(2)$ \\  
	       & ${}^{96}$Rb & $0.168(13)$ & $0.067(24)$ & $0.206(33)$ & $0.101(26)$ & $13.3(7)$ & $13.4$ \\ 
	 \hline

\end{tabular}%
} 
\end{center}
\caption{Independent yields, cumulative yields and branching ratio values found in JEFF~$3.1$ and ENDF-B/VIII.0 nuclear libraries for some selected precursors.}
\label{tab:yields_and_probs_example}
\end{table}

A summary of the differences between both libraries is shown in Table~\ref{tab:yields_and_probs_deltas}. It can be seen that even for important precursors such as ${}^{87}$Br the difference in the values for the independent yield is the order of $10\%$ and in extreme cases this difference can take values up to $170\%$, as in the case of ${}^{85}$As. 

\begin{table}[h!]
\begin{center}
\footnotesize
\begin{tabular}{|*{5}{c|}}  
\hline
\textbf{Group} & \textbf{Nucleus} & $\boldsymbol{\Delta IY} (\%)$ & $\boldsymbol{\Delta CY} (\%)$ & $\boldsymbol{\Delta \beta_n} (\%)$ \\ \hline
$1$ & ${}^{87}$Br & $10$ & $5$ & $-3.6$  \\ 
\hline
\multirow{2}{*}{$2$} 
& ${}^{88}$Br & $6$ & $2$  & $1.8$     \\ 
& ${}^{137}$I & $11$  &  $14$ & $-9.8$   \\  
\hline
\multirow{3}{*}{$3$} 
& ${}^{89}$Br & $20$ & $20$ & $2.1$      \\ 
& ${}^{92}$Rb & $8$ & $1$ & $-100$  \\  
& ${}^{138}$I & $3$ & $1$ & $-4.9$  \\  
\hline 
\multirow{2}{*}{$4$} 
& ${}^{85}$As & $14$ & $35$ & $-170$     \\ 
& ${}^{90}$Br & $14$ & $14$ & $-2.4$  \\  
& ${}^{94}$Rb & $11$ & $9$ & $-4$  \\ 
& ${}^{139}$I & $24$ & $23$ & $-2$  \\ 
\hline 
\multirow{3}{*}{$5$--$6$} 
& ${}^{91}$Br & $32$ & $32$ & $100$     \\ 
& ${}^{95}$Rb & $15$ & $15$ & $-1.2$  \\  
& ${}^{96}$Rb & $60$ & $50$ & $0.7$  \\ 
\hline
\end{tabular}
\end{center}
\caption{Independent yields, cumulative yields and branching ratio values found in JEFF~$3.1.1$ and ENDF-B/VIII.$0$ nuclear libraries for some selected precursors.}
\label{tab:yields_and_probs_deltas}
\end{table}
Given these differences between quantities of interest, in this work the $CY$'s used were taken from JEFF~$3.1.1$, while the $P_n$ were taken from ENDF-B/VIII.$0$. This pairing is the recommended when comparing the $\nu_d$ calculated using the summation method given by Eq.~\eqref{eq:nud_calculation} with the experimental value~\cite{foligno_thesis}.  

\section{Approaches to solve the Neutron Transport Equation}
\label{section:approaches_solve_NTE}
Basically there exists two different approaches to solve the Neutron Transport Equation: \emph{deterministic} and \emph{stochastic} (Monte Carlo) techniques. Although they aim to study the same physical problem, they are different in their approach and techniques used to solve the problem. They have complementary advantages and disadvantages.
  
\subsection{Deterministic Methods}
\label{subsection:deterministic_methods}
Deterministic methods model the physical problem ignoring the random aspect of individual particle histories and then they solve the NTE by discretizing these equations with respect to each of its variables and converting the problem into a system of algebraic equations that has to be solved~\cite{osti_5538794}. One of the strategies used to calculate the neutron flux use the quasi-static method, developed in the 1950s~\cite{doi:10.13182/NSE58-1}. This method resorts to an approximation where the flux is factored as a product between a shape function and an amplitude function. The shape function can be obtained through stationary state calculations, using discrete ordinates~\cite{doi:10.13182/NSE01-A2235, osti_776452} or Monte Carlo~\cite{Bentley1992DevelopmentOA,SHAYESTEH2009901}. To obtain the time-dependent amplitude function, diffusion~\cite{doi:10.13182/NT10-A9485} or $S_n$ methods~\cite{alcouffe2005} can be used. One feature of all of these methods is that they discretize the phase space: difussion theory assumes that neutrons diffuses through the medium following Fick's law and ignores the angular dependence of the flux. More advanced methods such as the $S_n$ method does take into account the angular dependence of the flux, but this dependence is discretized and neutrons are transported though discrete angles. With the use of these techniques it can be possible to refine the modelling of the angular dependence of the flux, but it can be complex to use the necessary number of angles to obtain a good solution for the flux. One of the main disadvantages of these methods is the constraints in the resolution of the discretization grid, since memory is required to store the unknown variables. With a coarser grid, higher discretization errors are obtained, so limitations in memory limit the accuracy of deterministic methods~\cite{graziani2008}.   

\subsection{Monte Carlo Method for solving the Transport Equation}
\label{subsection:monte_carlo_NTE}
Unlike the methods described in the preceding section, the Monte Carlo method (which is an stochastic method) do not solve the Neutron Transport Equation explicitly, but simulate the physical problem by transporting the neutrons through the medium. The physical processes involved in the evolution of the neutron population is governed by probability distributions. In the application of the Monte Carlo method to neutron transport, a stochastic model is simulated, and then the expected value of some random variable is equivalent to the value of a physical quantity that is to be determined. This quantity is estimated using the average of independent samples that represent the random variable. 

To illustrate this point, the procedure to carry out a Monte Carlo simulation will be outlined~\cite{graziani2008}. For simplicity a time-independent fixed source problem in a homogeneous medium will be considered. In this problem a source and a detector in the phase space must be simulated, and the detector response will be the quantity to be estimated, this is, the contributions of neutrons reaching the detector will be collected. The idea is to simulate $N$ neutrons, sampling the source distribution to find initial energy, position and direction for each neutron. The emitted neutrons are then transported. The distance $d$ that each neutron of energy $E$ travels between two interactions is exponentially distributed and given by
\begin{equation}
d = - \frac{\ln(\xi)}{\Sigma_t(E)},
\label{eq:distance_sampled}
\end{equation}       
where $\xi$ is a random number, sampled from a uniform distribution between $[0,1]$, and $\Sigma_t$ is the total macroscopic cross section. If $d$ is larger than the distance to the boundary of the next volume, then the particle is stopped at that boundary and a new path is sampled using Eq.~\eqref{eq:distance_sampled}. 
At the new position the interacting nucleus $i$ needs to be sampled, which will be chosen with probability
\begin{equation}
	p_i = \frac{\Sigma_{t,i}(E) }{\Sigma_t(E)},
	\label{eq:probability_nucleus}
\end{equation}
where $\Sigma_{t,i}(E)$ is the total cross section for nucleus $i$. Once the interacting nucleus is sampled, the specific interaction occurs with a probability
\begin{equation}
p_{i,x} = \frac{\sigma_{i,x}}{\sigma_{i,t}},
\label{eq:probability_interaction}
\end{equation}
where $\sigma_{i,x}$ is the microscopic cross section for the interaction $x$ and nucleus $i$.

After the interaction the neutron can be eliminated if absorbed or if it leaves the simulation world. Otherwise a new path is sampled and the process starts again. Neutron contributions are accumulated when they reach the detector. After the $N$ particles are transported the process is repeated $M$ times with a different random seed each time.

The tallies collected are averaged over the $M$ experiments. The associated uncertainty is calculated using the variance and its inversely proportional to the square root of the number of particles simulated. This means that the uncertainty of the result obtained in a simulation can be improved by simulating a larger number of particles. Although the number of particles required is problem dependent, it is usually quite large and this implies that Monte Carlo simulations are very time consuming. Fortunately, Monte Carlo algorithms are specially suited for parallel computing~\cite{rosenthal2000parallel}, which allows to speed up, in principle, by the order of the processor availables. The idea is that each processor simulates its own number of particles, and when each processor have completed the transport, the final results are collected.      

\subsection{Variance reduction methods}   
\label{subsection:variance_reduction_methods}
A Monte Carlo simulation as described in section~\ref{subsection:monte_carlo_NTE} requires knowledge of the probability distribution that governs the physical process that is used to calculate the expected value. In other words, in this method the computation describes how a particle would behave in an equivalent physical experiment. This method is known as \textit{analog} Monte Carlo simulation~\cite{osti_5538794}. There are some experimental setups where, for example, the detector counting rate could be too low or, for a shielding problem, there are too few initial particles that reach the region of interest. In those cases, longer detection times or several repetitions of the experiment might be necessary to achieve an acceptable uncertainty.
If one of these physical systems would be simulated using Monte Carlo, large number of particles would be required in order to achieve a reliable estimate of the quantity being studied. But the simulation time is governed by the number of particles simulated, which means that the simulation would require very long computation times. One way to overcome this problem is through a \textit{non-analog} Monte Carlo simulation~\cite{spanier2008monte}, which is a modification of the \textit{analog} Monte Carlo simulation where the physical probability distribution is modified in order to promote the occurrence of a given event (for example, to make that more particles can reach the detector). To keep the results unbiased, a compensation has to be applied elsewhere. For this purpose a \textit{statistical weight} is defined and assigned to each particle at the beginning of the simulation. Then, this weight can evolve along the simulation to counterbalance the changes that occur when the probability of a physical process is altered.   
One of the variance reduction techniques used in this work is \textit{survival biasing}, which must be used in conjunction with a population control technique called \textit{Russian roulette}. 
In survival biasing (also known as \textit{implicit absorption}), absorption reactions are prohibited to occur and instead at every collision the statistical weight of the particle, $w_{\mathit{new}}$ is reduced by the probability that the absorption occurs:
\begin{equation}
w_{\mathit{new}} = w \left ( 1 - \frac{\sigma_a(E)}{\sigma_t(E)} \right),
\label{eq:implicit_capture}
\end{equation} 
where $\sigma_a(E)$ and $\sigma_t(E)$ are the absorption and total microscopic cross sections, respectively, and $w$ is the statistical weight of the particle before the collision.
It is important to notice that survival biasing can reduce the weights of the particles to very low values. In that case, particles of low statistical value slow down the calculation, while contributing very little to the statistics. This means that this method must be combined with another method capable of stochastically \textit{killing}\footnote{In the context of Monte Carlo simulations, to kill a particle is to remove it from the simulation.} particles. This method is called \textit{Russian roulette}. If a particle falls below some threshold weight, then a random number is generated. If the random number is below the initial weight, then the particle is \textit{killed}. Otherwise, it \textit{survives} and its weight is set to some predefined value. 
\chapter{Methodology}
\label{ch:methodology}

In this chapter, the methodology used to include the $\beta$-delayed neutron emission from individual precursors in transient Monte Carlo simulation is discussed. Firstly, in Section~\ref{section:monte_carlo_with_openmc} the Monte Carlo OpenMC code is described, along with explanation why it becomes suitable for this work, and a benchmark calculation result is presented (See Sec.~\ref{subection:benchmarks_other_codes}). After that, in Section~\ref{section:details_time_dependence} a discussion on how the time dependence is treated. Following, Section~\ref{section:delayed_neutron_precursors} addresses precursors, including consequences of $\beta$-delayed neutron emission in the context of a Monte Carlo simulation (see Sec.~\ref{subsection:precursors_and_consequences}). This comprises how individual precursors are implemented in the code (see Sec.~\ref{subsection:individual_precursor}), and the strategy to overcome large variances associated with the different time scales between prompt and delayed neutrons (see Sec.~\ref{subsection:forced_decay}). Afterwards, in Section~\ref{section:initial_particle_source} the issue of how to sample a proper initial source to start a Monte Carlo transient simulation is discussed. Finally, in Section~\ref{subsection:pop_control} the method chosen to enforce population control is described.

\section{Monte Carlo simulations with OpenMC}
\label{section:monte_carlo_with_openmc}
The OpenMC~\cite{ROMANO201590} code is relatively new, an open-source code for particle transport developed at the Massachusetts Institute of Technology in $2013$. This code is capable of simulating neutrons in fixed source, $k$-eigenvalue, and subcritical multiplication problems. The geometry is built using a constructive solid geometry. The code supports both continuous-energy and multigroup transport. The continuous-energy nuclear cross section data follows the HDF5 format~\cite{Koranne2011}  and is generated from ACE files produced by NJOY~\cite{osti_1338791}.
Since this code is open source, its use is not subject to licensing, with no restrictions on modifications, developments and addition of new capabilities. 

\subsection{Benchmarks}
\label{subection:benchmarks_other_codes}
As a first step before beginning the development of new capabilities for OpenMC, benchmark calculations were performed to further validate the code. In order to do this, and during the author's first doctoral internship at the Bariloche Atomic Center, the OpenMC code was used to model and calculate the Effective Multiplication Factor of the RA-$6$ research reactor. The result obtained was compared with the experimental values from the ICSBEP \textit{International Handbook of Evaluated Criticality Safety Benchmark Experiments}~\cite{ICSBEP, bazzana}, and with the result obtained when modeling the reactor using the Monte Carlo transport code MCNP~\cite{GOORLEY201205198}.

Another parameter that can be calculated is the Effective Delayed Neutron Fraction, $\beta_{\mathit{eff}}$, which was mentioned in Sec.~\ref{subsection:point_kinetics_eqs}. Formally, the adjoint neutron flux is required to calculate this parameter, but it can be estimated using the prompt method~\cite{doi:10.13182/NSE03-107}. This method assumes that the value of $\beta_{\mathit{eff}}$ is given by
\begin{equation}
\beta_{\mathit{eff}} \sim 1 - \frac{k_p}{k_{\mathit{eff}}},
\label{eq:prompt_method}
\end{equation} 
where $k_p$ is the effective multiplication factor obtained from a criticality calculation, but without taking into account the contribution from $\beta$-delayed neutron emission. The advantage of this method is that the adjoint flux is not needed to calculate the effective delayed neutron fraction. Thus, the capability to run a criticality calculation without delayed neutrons was added to OpenMC, which enabled the estimation of $\beta_{\mathit{eff}}$ in two steps. In MCNP$6$ the same feature can be achieved by using the \verb|TOTNU NO| card to perform a criticality calculation only with prompt neutrons.

\noindent \textbf{Description of the RA-$6$ reactor}

The RA-$6$ (Spanish acronym for \textit{Argentina Reactor, Number 6}) is an open pool research reactor with a nominal power of $3$~MW, located at Bariloche Atomic Center, a nuclear research center in San Carlos de Bariloche, Río Negro, Argentina.
The core of the reactor is made up of an array of flat plates MTR-type fuel elements with $20$ \% enriched uranium located inside a stainless steel tank filled with demineralized water that acts as a coolant, moderator, reflector and shielding in the axial direction. Four Ag-In-Cd absorber elements are the control elements. The model was the one included for the ICSBEP benchmark evaluation, with added graphite reflectors~\cite{bazzana} and --since in Monte Carlo codes it is possible to model the reactor geometry in detail-- fuel elements were modeled explicitly, such as cadmium wires, water gaps, guides and nozzles. The model also included the supporting grid for the core and BNCT filter.

\noindent \textbf{Simulation parameters and results for $\boldsymbol{k_{\mathit{eff}}}$ calculation}

In OpenMC and MCNP the criticality calculation was peformed using $8050$ batches\footnote{the total number of source particle simulated is broken up into a number of batches.}, $50$ skipped\footnote{skipped cycles will be discarded before data accumulation begins.} and $10000$ particles per batch. The neutron cross section database used was ENDF/B-VII.$1$. Results obtained are summarized in Table~\ref{table:keff_kp_beta} 

 \begin{table}[h!]
\centering
\begin{tabular}{@{}|l|r|r|r|@{}}
\hline
Magnitude     & OpenMC      & MCNP        &   Benchmark         \\ 
\hline
$k_{\mathit{eff}}$     & $1.0050(1)$ & $1.0045(1)$ &   $1.0026(25)$ \\
$k_p$         & $0.9975(1)$ & $0.9971(1)$ &      $-$                   \\
$\beta_{\mathit{eff}}$ & $746(15)$   & $737(13)$  & $782(7)$         \\ 
\hline
\end{tabular}
\caption{Results obtained for the effective multiplication factor and the effective delayed neutron fraction for the RA-6 reactor.}
\label{table:keff_kp_beta}
\end{table}

\section{Details on the inclusion of time dependence}
\label{section:details_time_dependence}
As stated in the Introduction, the main objective of this thesis is to study the inclusion of time-dependence in a Monte Carlo simulation, considering the delayed emission from the neutron precursors present in a fissile system. To this end several issued must be addressed, which will be discussed in the remainder of this chapter. 

\subsection{Time evolution of the neutrons}
\label{subsection:neutron_time_evolution}
In a stationary Monte Carlo transport simulation time is not explicitly present. The first step to perform Monte Carlo kinetic simulations is to add a new label $t$ to the particles, serving as a \textit{clock} with value updated using the kinetic energy and the distance traveled by the neutron between events. This time is set to zero ($t\!=\!0$) at the beginning of the simulation and is updated as the particle is transported in the simulation.  

\subsection{Simulation time boundary}
\label{subsection:time_boundary}
To simulate transient events in fissile systems the evolution was divided in discrete time intervals. There are two reasons for this: First, the variance reduction and population control techniques require a time grid to be applied. The second reason is that changes in the geometry or reactivity of the system can take place in a transient simulation, changes that can be introduced at the end of a time interval. It is important to notice that the size of the time intervals can be choosen freely and they do not affect the validity or accuracy of the results obtained from the simulation.
When a particle crosses a time boundary, its trajectory is stopped exactly at the boundary, with the spacial position that corresponds to the time boundary, then the particle is stored to continue the simulation at the next time step.

\subsection{Time tally}
\label{subsection:time_tally}
In order to tally the measured quantities in time, the tallies in OpenMC were modified and a new filter was added. This \textit{time} filter added the capability to monitor the time evolution of any of the tallies already present in the code.

\section{Delayed neutron precursors}
\label{section:delayed_neutron_precursors}
In this section the time delay of the $\beta$-delayed neutron emission and its consequences in the context of a Monte Carlo simulation are explored. A key point of this discussion is the large variance in the simulation results if an \textit{analog} Monte Carlo method was used, issue that will be dealt with at the end of the section. Then the inclusion of $\beta$-delayed neutron emission from individual precursors, one of the main objectives of this work, is addressed. Following this discussion, the precursor particle defined in this work, for the simulation was presented. In the last part of this section, techniques chosen to solve the problem caused by time differences between prompt and delayed emission of neutrons are described.    
\subsection{The time delay of the precursors and its consequences}
\label{subsection:precursors_and_consequences}
The delayed neutron precursor decay is a stochastic process, which can be described by
\begin{equation}
p_i(t) = \lambda_i \, e^{-\lambda_i(t-t_0)} \, \theta(t-t_0),
\label{ec:decay_probability}
\end{equation}
where $p_i(t)$ is the probability the $i$-th precursor decay at a given time $t$, $\lambda_i$ is the decay constant, $t_0$ is the time when the precursor was created and $\theta$ the Heavyside function.
Given this probability, an \textit{analog} Monte Carlo simulation could be performed to, in principle, describe what happens in a fissile system: at time $t_f$ of the fission event, $\nu_p$ prompt neutrons are produced and then, at a time $t\!=\!t_f\!+\!t_d$, $\nu_d$ delayed neutrons are inserted into the simulation. Time $t_d$ is sampled from Eq.~\eqref{ec:decay_probability} and the energy is determined from the precursor delayed neutron energy distribution (see Section~\ref{subsection:quantities_of_interest}).

\begin{figure}[h]
	\centering
	\includegraphics[width=15cm]{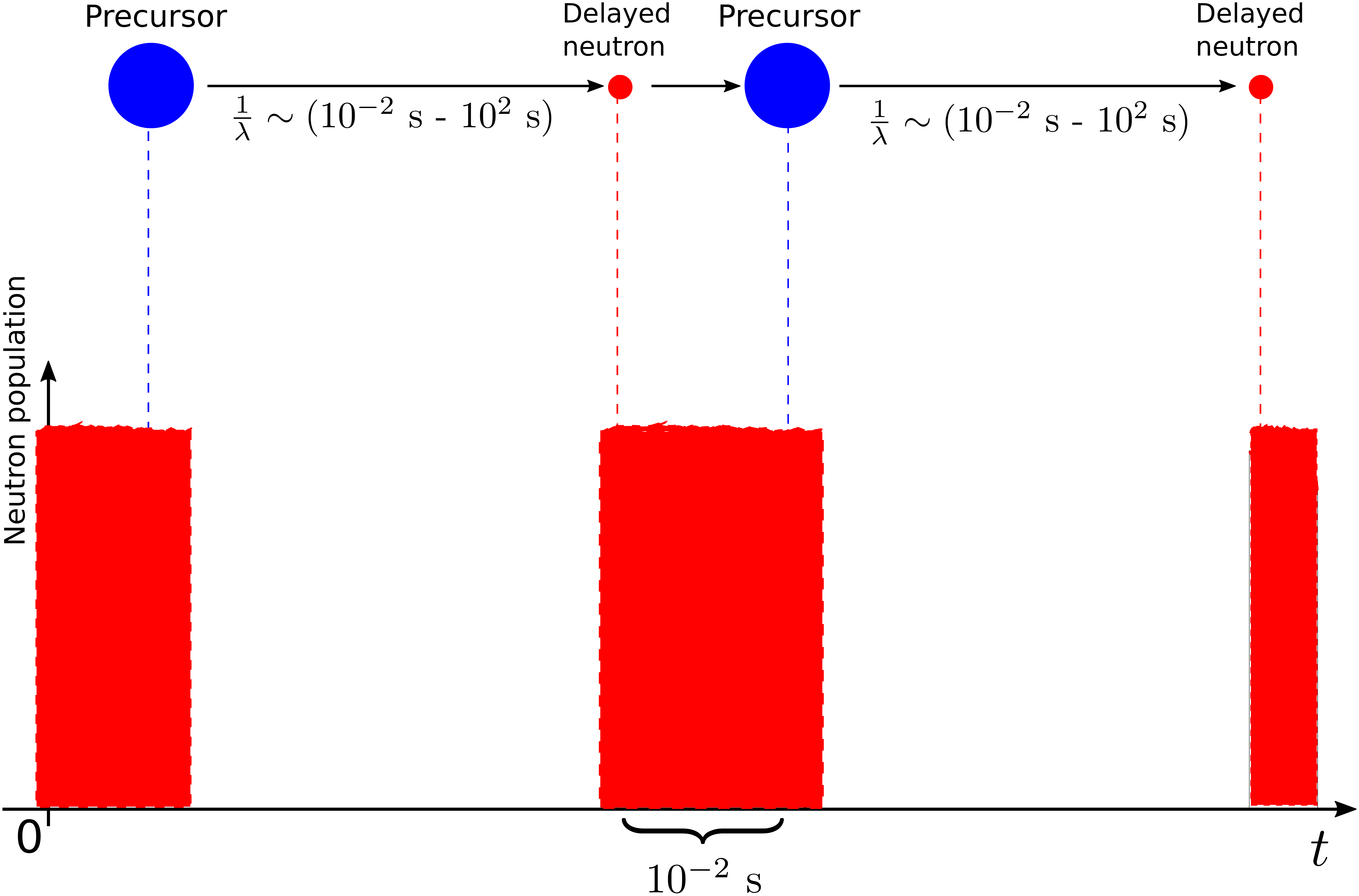}
	\caption{Schematic representation of the time scales associated to the delayed neutron emission and the lifetime of the prompt chains. This different time scales produce large variance in the quantities scored.}
	\label{fig:delayed_emission_consequences}
\end{figure}

Although this strategy emulates what happens in a nuclear reactor, a large variance in the results obtained due to the difference in the time scales associated with prompt and delayed events. Indeed, as it was mentioned in Section~\ref{subsection:importance_of_delayed_neutrons} there exists a time delay between the nuclear fission event and the emission of a delayed neutron from the decay of a precursor. The average lifetime of a prompt neutron in a light water reactor is $\sim\!10^{-4}$~s and the average length of a fission chain in a system close to critical is $\sim\!150$~neutrons~\cite{doi:10.13182/NSE12-44}. This implies that the average lifetime of a neutron chain is $\sim\!10^{-2}$~s. At the same time, a prompt fission chain will produce on average one precursor, which in turn will decay to a delayed neutron and then produce a new fission chain in a few seconds. During this time there would be no new neutrons produced in an analog Monte Carlo simulation, as it is shown in Fig.~\ref{fig:delayed_emission_consequences}. This lack of particles would in turn lead to large variance in the quantities scored. In an actual fissile system this does not happen because of the large number of neutrons produced so the effect is averaged out. Of course, due to limitations imposed by computer calculation power and memory, it is not possible to simulate this many fission chains. Due to this fact, and in order to obtain results with acceptable statistics, delay of precursors decay must be simulated in another way.

\subsection{Individual precursors}
\label{subsection:individual_precursor}
In $1957$, Keepin measured the periods, relative abundances and yields of delayed neutrons from fission. He had the idea of grouping the $\beta^{-}$ delayed neutron emitters into groups according to their half-lives and assuming that the total emission rate could be represented as a sum of exponential functions~\cite{KEEPIN1957IN2}. It is important to note that the number of groups is arbitrary, considering that the total number of precursors produced in ${}^{235}$U fission is more than $270$. Nonetheless, Keepin found that a six group representation properly fitted the measured experimental activity. 

At the time of the writing of this thesis, there were no published Monte Carlo codes for neutron transport in fissile systems which include the delayed neutron emission from individual precursors, i.e. all of the existing codes use the group structure to take into account $\beta$-delayed neutron emissions and insert a delayed neutron directly into the simulation. 
In the case of this work a precursor is created, and then this precursor can decay, emitting a delayed neutron which is inserted into the system. To further illustrate this point, the steps needed to take into account $\beta$-delayed neutron emissions will be outlined using, i) the group structure or, ii) the individual precursors.    

\noindent \textbf{i) $\boldsymbol{\beta}$-delayed neutron emission with group structure} 

If a delayed fission is sampled, the next step is to choose which precursor group will be sampled. Here, the relative yield, is utilized. If the $j$-th group is chosen, then the delayed time associated with the delayed emission will be sampled using the {group decay constant} and Eq.~\eqref{ec:decay_probability}. Finally, the delayed neutron energy will be chosen from the $j$-th {delayed neutron energy group spectra}~\cite{BROWN20181}. 
\newpage 
\noindent \textbf{ii) $\boldsymbol{\beta}$-delayed neutron emission with individual precursors} 

On the other hand, if a delayed fission is sampled and the delayed neutron emission from individual precursors is being simulated, instead of directly inserting a delayed neutron, a precursor is produced. The next step is to choose which precursor nuclide will decay. In order to do this, the {precursor importance} or {relative yield} of the individual precursor $i$-th, $I_i$, is defined as~\cite{foligno_thesis}:
\begin{equation}
I_i \equiv \frac{CY_i \, P_{n,i}}{\nu_d},
\label{eq:precursor_importance}
\end{equation}
with $CY_i$ the cumulative fission yield, $P_{n,i}$ the precursor delayed neutron emission probability, and $\nu_d$ the average delayed neutron yield.
Once the precursor has been chosen, the delayed time associated with this emission is sampled using the {precursor decay constant} and Eq.~\eqref{ec:decay_probability}. Finally, the delayed neutron energy will be the average energy from the corresponding {precursor delayed neutron spectrum}. 

Another point to consider is the number of precursors to include in the simulation, for which the precursor importance is useful because it shows the fraction of the total delayed neutron yield that the precursor represents (i.e. how important it is). As an example, when using the cumulative yields from JEFF-$3.1.1$ library and the $\beta$-delayed neutron emission probabilities from ENDF/B-VIII.$0$, the average delayed neutron yield obtained is $1.57 \times 10^{-2}$. Then from the values presented in Table~\ref{tab:yields_and_probs_example}, the importance for any given precursor can be calculated. For example, for ${}^{137}$I, the precursor importance obtained is $16.26 \%$, which means that the delayed neutrons emitted from the ${}^{137}$I decay account for $16.26 \%$ of the total $\beta$-delayed neutron emission. 

So, although there are data for $269$ precursors, in this work only $50$ will be included in the simulation. To justify this choice the precursors were ordered by importance using Eq.~\eqref{eq:precursor_importance} and then the cumulative importance ($\sum_i I_i$), was calculated. It was found that the first $50$ precursors account for $99.16 \%$ of the total delayed neutron yield, which means that the remaining $219$ precursors have a combined importance of $0.84 \%$. This small contribution in comparison to the contribution of the first $50$ precursors was judged to be negligible for the purpose of this work.

Table~\ref{tab:precursors_comparison} summarizes the differences when comparing between the simulation of the $\beta$-delayed neutron emission using the group structure and the delayed neutron emission when using individual precursors. 

\begin{table}[h!]
	\footnotesize
\centering
\begin{tabular}{ccc}\toprule
{\textbf{Quantity}} & \textbf{N-group structure} & \textbf{This work} \\
\midrule
\textbf{Relative abundance} & ${\beta_i}/{\beta} \quad \text{with} \quad 1<i<6 \; \text{or} \; 8$ & ${(CY_i \, P_{n,i})}/{\nu_d} \quad \text{with} \quad 1<i<50$\\
\textbf{Decay constants} & Precursors in $6$- or $8$- groups  & $50$ individual precursors \\
\textbf{Energy spectra} &  Precursors in $6$- or $8$- groups   & $50$ individual precursors \\
\bottomrule	
\end{tabular}
\caption{Summary of the differences when including the $\beta$-delayed neutron emission using the precursor group structure or the individual precursors \label{tab:precursors_comparison}}
\end{table}
Finally, it is worth mentioning that the choice of including $50$ out of the $269$ precursors was made taking into account the calculation time and the cumulative importance of these $50$ precursors, but should need arise, the code developed can handle the whole set of precursors.  
\subsection{The precursor particle}
\label{subsection:precursor_particle}
The first step to include the precursor decay in the simulation involves adding the precursors in the simulation. So a new particle type is defined in the code, the \textit{precursor particle}. All precursors (or precursor groups) are combined into a single precursor particle~\cite{doi:10.13182/NSE12-44}. The decay probability for this particle is given by
\begin{equation}
p_{\mathit{combined}}(t) = \sum_i \Upgamma_i \, \lambda_i \, e^{-\lambda_i (t-t_0)} \, \theta(t-t_0),  
\label{eq:probability_combined}
\end{equation}
with, $t_0$ the time when the precursor was created, and $\Upgamma_i$ a factor that depends on whether precursor groups or individual precursors are being considered:
\begin{equation}
\Upgamma_i =
\begin{cases}
\frac{\beta_i}{\beta}, & \text{for precursor group} \\
I_i, & \text{for individual precursor}.
\end{cases}
\label{eq:upgamma_t}
\end{equation}
Here $\beta_i$ is the delayed fraction for the $i$-th precursor and $\sum_i \beta_i = \beta$, with $\beta$ the total delayed neutron fraction. $I_i$ is the precursor importance for the $i$-th precursor. 
Fractions $\Upgamma_i$ must be defined differently in some cases, as will be shown in Section~\ref{section:initial_particle_source}.
In principle, the statistical weight of a delayed neutron emitted from the $\beta$-decay of a precursor is given by
\begin{equation}
w_d(t) = w_c \, \sum_i \Upgamma_i \, \lambda_i \, e^{-\lambda_i (t-t_0)} \, \theta(t-t_0),
\label{eq:weight_delayed_analog}
\end{equation}
with $w_c$ the main precursor weight. This weight is the number of \textit{physical} precursors that this precursor particle represents at the time of its creation in the simulation. It must be noted that this weight does not change with time and it can only be altered by means of variance reduction techniques, as it will be explained in Section~\ref{subsection:pop_control}. After the precursor decay is produced, the energy of the emitted delayed neutron must be chosen from the corresponding precursor or precursor group. The probability of choosing the $i$-th group or precursor is a function of time given by
\begin{equation}
P_i(t) = \frac{\Upgamma_i \, \lambda_i \, e^{-\lambda_i (t-t_0)}}{\sum_i \Upgamma_i \, \lambda_i \, e^{-\lambda_i (t-t_0)}}.
\label{eq:probability_choose}
\end{equation}
This means that this probability must be evaluated at the time of decay to select the correct group or precursor for the energy spectrum.

Aside from the main precursor weight $w_c$, there is another statistical weight which will be utilized during this work. This is the weight of the precursor at a time $t$ and it represents the number of physical precursors that a precursor particle represents at a given time $t$ and is given by
\begin{equation}
w_p(t) = w_c \sum_i \Upgamma_i e^{-\lambda_i (t-t_0)}.
\label{eq:timed_weight}
\end{equation}
The last statistical weight that can be utilized is the expected delayed neutron weight. The precursor interacts with the system through delayed neutrons, so the weight of the delayed neutrons can be used for variance reduction. The problem is that the decay time is not known \textit{a priori}, so this weight is defined as~\cite{doi:10.13182/NSE12-44}
\begin{equation}
	w_{d,av} = \frac{1}{\Delta t} \int_{t_1}^{t_1 + \Delta t} w_d(t) dt,
\label{eq:expected_delayed_w_definition}
\end{equation}
where $t_1$ is the start of the \textit{next} interval. Using Eq.~\eqref{eq:weight_delayed_analog}, the expected delayed neutron weight becomes
\begin{equation}
	w_{d,av} = w_c \sum_i \Upgamma_i (e^{-\lambda_i(t-t_0)} - e^{-\lambda_i(t_1 + \Delta t - t_0)}).
	\label{eq:expected_weight_calculated}
\end{equation}
\subsection{Precursor forced decay}
\label{subsection:forced_decay}
As explained in Section~\ref{subsection:precursors_and_consequences}, a direct simulation of delayed neutron precursor decay would lead to significant variance in the system, so that another way to simulate the precursors must be utilized. Since this variance is caused by the fact that there are too few fission chains per unit of time caused by delayed neutron decays, one strategy would be to modify the precursor decay probability, forcing the decay of all the precursors in each interval and thus having more delayed neutrons present. In this technique, called ``forced decay''~\cite{LEGRADYPHYSOR2008}, the sampling of the delayed neutrons is biased and the Monte Carlo fair game is preserved by altering the statistical weight of the emitted delayed neutrons.
Regarding the biased decay probability, the simplest choice would be a uniform decay probability, forcing the decay of all of the precursors in each one of the time intervals defined in Section~\ref{subsection:time_boundary}. With this choice the biased decay probability is
\begin{equation}
\bar{p}(t) = \frac{1}{t_{j+1} - t_{j}} = \frac{1}{\Delta t},
\label{eq:biased_decay_probability}
\end{equation}
where $t$ is the time when the forced decay happens and $\Delta t$ is the size of the time bin. To ensure an unbiased result the weight of the delayed neutrons produced by forced decay is adjusted to
\begin{equation}
w_d(t) = \frac{p(t)}{\bar{p}(t)} = w_c \, \Delta t \, \sum_i \Upgamma_i \lambda_i e^{-\lambda_i(t - t_0)} \quad  \text{with } \quad  t_j<t<t_{j+1} 
\label{ec:weight_precursor}
\end{equation}
where $w_c$ is the statistical weight of the precursor. The delayed neutron produced will be transported and may in turn cause new fissions.
Once the delayed neutron of weight $w_d(t)$ has been created during the corresponding time interval between $t_j$ and $t_{j+1}$, the precursor is not eliminated from the simulation. Instead, it is added to a precursor bank with weight
\begin{equation}
	w_p(t) = w_c \sum_i \Upgamma_i e^{-\lambda_i(t_{i+1} - t_i)},
	\label{eq:precursor_weight_in bank}
\end{equation}
where it will undergo forced decay, producing more delayed neutrons. It is important to note that the precursor is not being transported in the simulation and only affects the simulation through the delayed neutrons that emits.
\begin{figure}[h!]
	\centering
	\includegraphics[width=14cm]{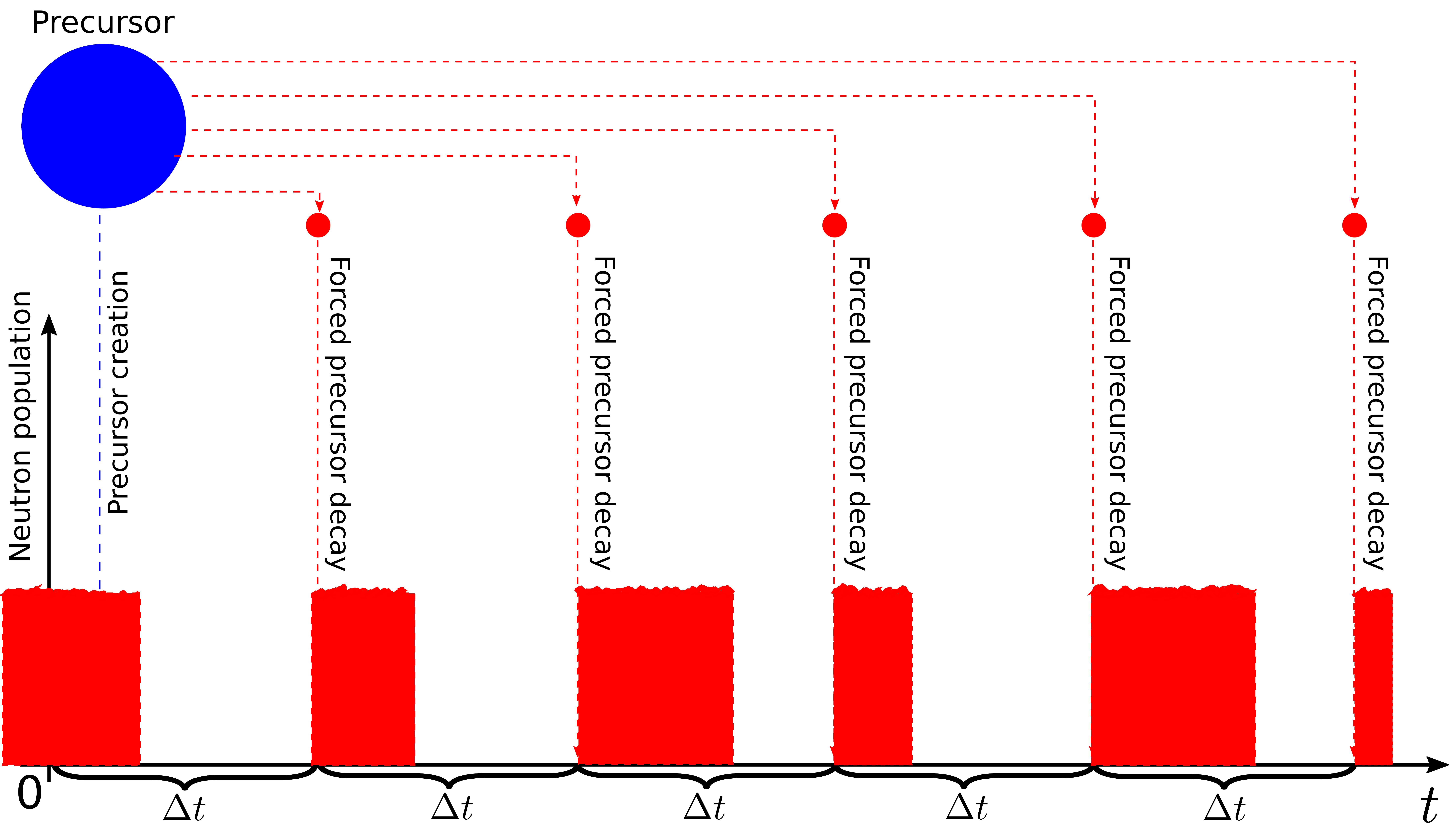}
	\caption{Schematic representation of the forced decay scheme for the precursors, where the precursor is forced to decay at the beginning of each time bin, so there are scored in every time interval.}
		\label{fig:forced_decay_scheme}
\end{figure}
\section{Initial transient particle source}
\label{section:initial_particle_source}
To begin a transient Monte Carlo, an initial transient source distribution will be constructed using the converged source distribution from an eigenvalue calculation, with $k_{\mathit{eff}} \sim 1$. To assess the convergence of the source distribution in the criticality calculation, OpenMC code has the capability of define a suitable spatial mesh and monitor the Shannon entropy. 
There are two methods to create an initial particle source. The first method is to transform the converged neutron source into a mix source comprised of neutrons and precursors. The second method consists of sampling the initial neutrons and precursors using appropriate tallies after the eigenvalue calculation.
For the first stage in the development of this work, when a $1$-group and monoenergetic system was studied, the first method is an acceptable choice. 
As it was shown in Section~\ref{subsection:nte_with_fission_neutrons} and according to Eq.~\eqref{eq:precursor_conc.}, the precursor concentration for one group at stationary state given by (${\partial C}/{\partial t}=0$) is
\begin{equation}
C_{0}(\mathbf{r}) = \frac{\beta_i}{\lambda_i} \nu \Sigma_f \psi(\mathbf{r}).
\label{eq:precursor_conc_t0}
\end{equation}
To determine the fraction of neutrons in position $\mathbf{r}$ the following relation is useful
\begin{equation}
	\frac{n_0(\mathbf{r})}{n_0(\mathbf{r}) + C_0(\mathbf{r})} = \frac{\frac{1}{v} \psi(\mathbf{r})}{\frac{1}{v} \psi(\mathbf{r}) + \frac{\beta_i}{\lambda_i} \nu \Sigma_f \psi(\mathbf{r})} = \frac{1}{1 + \frac{\beta_i}{\lambda_i} \nu v \Sigma_f}.
	\label{eq:neutron_fraction}
\end{equation}
This relation is valid only for constant neutron energy. For the mono-energetic system studied in this work and using the parameters shown in Table~\ref{tab:parameters_mono_system}, it is obtained that for every neutron there are about $10^4$ precursors and that the fraction of prompt neutrons in steady state is $0.08~\%$.

For the second method~\cite{doi:10.13182/NSE12-44} the energy dependent initial source is sampled from an eigenvalue calculation. For the number of neutrons, the estimator used was 
\begin{equation}
	n_0(\mathbf{r}, E) = \int_{4 \pi} \frac{\psi_0(\mathbf{r}, \mathbf{{\Omega}}, E)}{v(E)} d \mathbf{{\Omega}},
	\label{eq:tally_neutrons_source}
\end{equation}
while for the precursors the estimator utilized was
\begin{equation}
	C_{i,0}(\mathbf{\mathbf{r}}) = \int_{4 \pi} \int_0^{\infty} \frac{\beta_i(\mathbf{\mathbf{r}},E) \nu(\mathbf{\mathbf{r}},E) \Sigma_f(\mathbf{\mathbf{r}},E)}{\lambda_i} \psi_0(\mathbf{\mathbf{r}},\mathbf{{\Omega}},E) \, dE d\mathbf{{\Omega}},
	\label{eq:tally_precursors_source}
\end{equation}
where $\psi_0$ is the flux sampled by an already existing flux tally in OpenMC. 

It is important to mention that the probability distribution for a precursor created in a fission event (shown in Sec.~\ref{subsection:precursor_particle}) is different than the one for a precursor created from the steady state distribution. 
This is because the precursors have undergone a portion of its decay before $t=0$. The different precursors with different decay constants result in a steady state group distribution given by
\begin{equation}
	P_i = 
	\begin{cases}
	\dfrac{\lambda^b}{\lambda_i} \dfrac{\beta_i}{\beta}, & \text{for precursor group} \\[12pt]
    \dfrac{\lambda^b}{\lambda_i}  I_i, & \text{for individual precursor}, \\
	\end{cases}
\label{eq:correct_prob_at_zero}
\end{equation}
where $\lambda^b$ is the inversely weighted decay constant defined as
\begin{equation}
	\lambda^b = 
	\begin{cases}
\dfrac{\beta}{\sum_i \dfrac{\beta_i}{\lambda_i}}, & \text{for precursor group} \\[12pt]
\dfrac{1}{\sum_i \dfrac{I_i}{\lambda_i}}  , & \text{for individual precursor},
\end{cases}
\label{eq:lambda^b}	
\end{equation}
This difference in the probability distributions is implemented in the code according to the time of creation of the precursor. 
\section{Population control}
\label{subsection:pop_control}
When using the ``forced decay'' method the precursors always survive after they decay into delayed neutrons. This means that the population of precursors is continuously increasing, so population control for precursors must be implemented. The method implemented in the OpenMC code in this work is the \textit{Combing method}~\cite{BOOTH1996}, which was originally developed for stationary Monte Carlo simulations. The idea of this method is to preserve the total statistical weight while mantaining a fixed number of particles. In the context of this work, keeping constant the number of particles serves for two purposes: i) variance reduction and ii) reduced computing time by keeping the population size approximately constant. If the system is super-critical, combing prevents the unlimited growth of the population, while if the system is sub-critical, keeps the simulation running by preventing the population from dying. If the system is critical, combing prevents the divergence of the population due to fluctuations of fission chains~\cite{de_Mulatier_2015}.

\begin{figure}[h!]
	\centering
	\includegraphics[width=14cm]{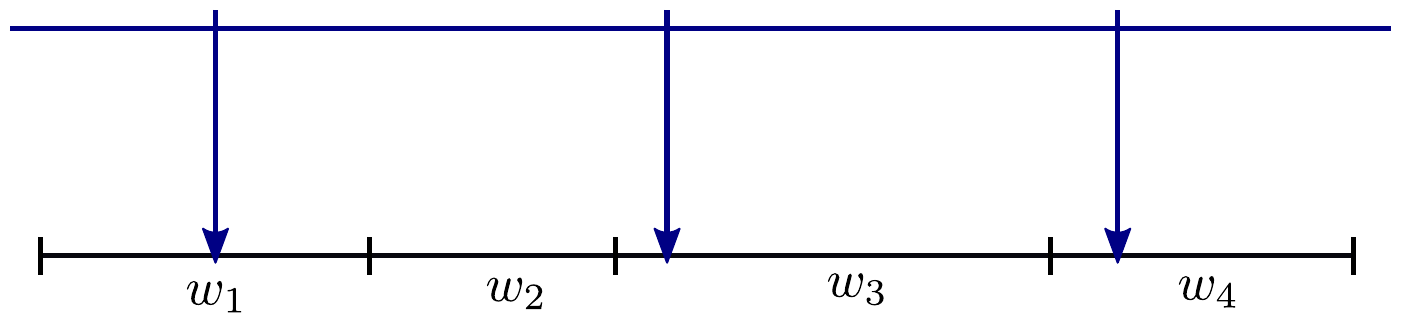}
	\caption{Diagram of the application of the combing method for $4$ particles of total weight $W$ combed into $M=3$. The particles kept by the comb are particle $1$, particle $3$ and particle $4$, each with weight $W/3$.}
	\label{fig:combing_method_scheme}
\end{figure}

If there are $K$ particles at the end of a time interval and the objective is to comb them to $M$ particles. These $K$ particles will be combed into $M$ using a comb with $M$ teeth. Figure~\ref{fig:combing_method_scheme} shows an example situation with $K=6$ and $M=4$. 
The length of the comb is the sum of the particle weights
\begin{equation}
	W = \sum_{i=1}^K = w_i.
	\label{eq:comb_W_total}
\end{equation}
The comb teeth are equally spaced with the position of the teeth randomly selected as
\begin{equation}
	t_m = \xi \frac{W}{M} + (m-1)\frac{W}{M}.
	\label{eq:teeth_spacing}
\end{equation}
Each time a tooth hits interval $i$, the $i$-th article is duplicated ad assigned a weight
\begin{equation}
	w'_i = \frac{W}{M},
	\label{eq:postcombed_weight}
\end{equation}
where $w_i'$ is the weight after combing.
Defining the integer $j$ by
\begin{equation}
	j < \frac{w_i}{W/M} \leq j +1,
	\label{eq:combing_j}
\end{equation}
it can be seen that either $j$ or $(j+1)$ teeth of a comb with a pitch of $W/M$ will hit an interval of length $w_i$. In particular, the probability of $j$ teeth fall in an interval $i$ is
\begin{equation}
p_{i,j} = j + 1 -w_i\frac{M}{W},
\label{eq:comb_prob_j}
\end{equation} 
while the probability that $j+1$ teeth fall in interval $i$ is
\begin{equation}
p_{i,j+1} = w_i\frac{M}{W} - j.
\label{eq:comb_prob_j+1}
\end{equation} 
The expected weight for a single particle after combing is 
\begin{equation}
w_i' = p_{i,j} j \frac{W}{M} + p_{i,j+1}(j+1)\frac{W}{M} = \frac{w_i}{W/M} \frac{W}{M} = w_i,
\label{eq:combing_expected}
\end{equation}
this implies that the combing preserves the total weight because after combing each particle is asigned a weight $w_i'=W/M$ and since there are $M$ particles, the total weight is preserved.
In this work both the neutron and precursor populations are combed separately and for the monitoring of the precursor population the timed precursor weight (Eq.~\eqref{eq:timed_weight}) or the expected delayed neutron weight (Eq.~\eqref{eq:expected_weight_calculated}) can be used.

\chapter{Results and discussion}
\label{ch:results}
The time dependence in neutron transportation, including $\beta$-delayed neutron emission from fission products, added in this work to the original OpenMC code were tested and the results are discussed in this chapter. This modified version of the mentioned code will be denoted as Time-Dependent OpenMC or OpenMC(TD).

In Section~\ref{section:preliminar_work}, the inclusion of time dependence and individual precursors in OpenMC(TD) was evaluated. Related to time dependence, the tests made were: i) time tally (see Sec.~\ref{subsection:time_tally}), by scoring time dependent quantities in a fixed source calculation in a subcritical configuration for the RA-$6$ reactor (see Sec.~\ref{subsection:time_tally_results}), ii) time boundary Monte Carlo simulation (see Sec.~\ref{subsection:time_boundary} and Sec.~\ref{subsection:neutron_time_evolution}), by transporting neutrons in a fixed source calculation and in a Monte Carlo simulation divided in time intervals (see Sec.~\ref{subsection:transport_checks}), and iii) scoring of time dependent quantities in a simulation divided in time intervals (see Sec.~\ref{subsection:test3}). The inclusion of individual precursors lead to a discussion about the $\beta$-delayed neutron activity comparing the standard $6$-group precursor structure and the $50$ individual precursor structure studied in this work (see Sec.~\ref{subsection:activity_individual_precursors}). Likewise, the $\beta$-delayed average neutron energy for the $8$-group precursor structure in OpenMC(TD) was compared with the neutron spectrum energy for the JEFF-$3$.$1$.$1$ $8$-group precursor structure (see Sec.~\ref{subsection:average_energies}). 

In Section~\ref{section:mono-energetic_system} a monoenergetic fissile system was simulated considering $1$-group precursor structure. Three configurations were studied and discussed: subcritical (See Sec.~\ref{subsection:mono_subcritical_state}), critical (Sec.~\ref{subsection:mono_steady_state}) and reactivity insertion (See Sec.~\ref{subsection:mono_reactivity}).

In Section~\ref{section:energy_dependent} an energy dependent system using ${}^{235}$U was simulated considering different precursor structures. Two configurations were studied and discussed: subcritical (see Sec.~\ref{subsection:energy_u235_vacuum_sub}) and supercritical (see Sec.~\ref{subsection:energy_u235_vacuum_super}).

In Section~\ref{section:energy_individual_precursors} an energy dependent and light-water moderated system using ${}^{235}$U was simulated using different precursor structures and criticality configurations. Afterwards, the $6$-group precursor structure effective multiplication factor was compared to the $50$ individual precursor structure effective multiplication factor (see Sec.~\ref{subsection:criticality_50precursors}).
Finally, the following cases were studied and discussed: i) comparison between $6$-group and $50$ individual precursor structure in a critical configuration (see Sec.~\ref{subsection:critical_state_50precursors}), and ii) comparison between $6$-group, $50$ individual and $40$ individual precursor structure in a critical configuration (see Sec.~\ref{subsection:only_10_precursors}). 

Simulations were run at CSICCIAN (spanish acronym for Simulation and Calculation Center in Nuclear Sciences and Applications) clusters from the Chilean Nuclear Energy Commission, its specifications are shown in Appendix~\ref{app:summary_calculations}, along with a summary of the simulations presented in this work. 

\section{Inclusion of time dependence and individual precursors in OpenMC(TD)}
\label{section:preliminar_work}
As explained in Sec.~\ref{section:details_time_dependence}, there were some starting points that needed to be addressed in order to include time dependency in a Monte Carlo transport code. In short: i) time is explicitly added by means of time label to the particles, ii) the total simulation time is divided in discrete time intervals and, iii) a new filter is added, so the code has the capability to score time-dependent quantities. In order to check the correct implementation of these characteristics into the code, three tests were conducted prior to the inclusion of the precursors and delayed neutrons.   

At the time of the writing of this thesis, measured $\beta$-delayed neutron energy spectra in databases~\cite{jeff311,BROWN20181} were available only for $34$ precursors~\cite{osti_6187550}. In this work the average energy of the $\beta$-delayed neutron was used for each individual precursor (see Sec.~\ref{subsection:average_energies}).  

\subsection{Scoring of time dependent quantities in a fixed source calculation}
\label{subsection:time_tally_results}
\begin{figure}[h!]
	\centering
	\includegraphics[width=\textwidth]{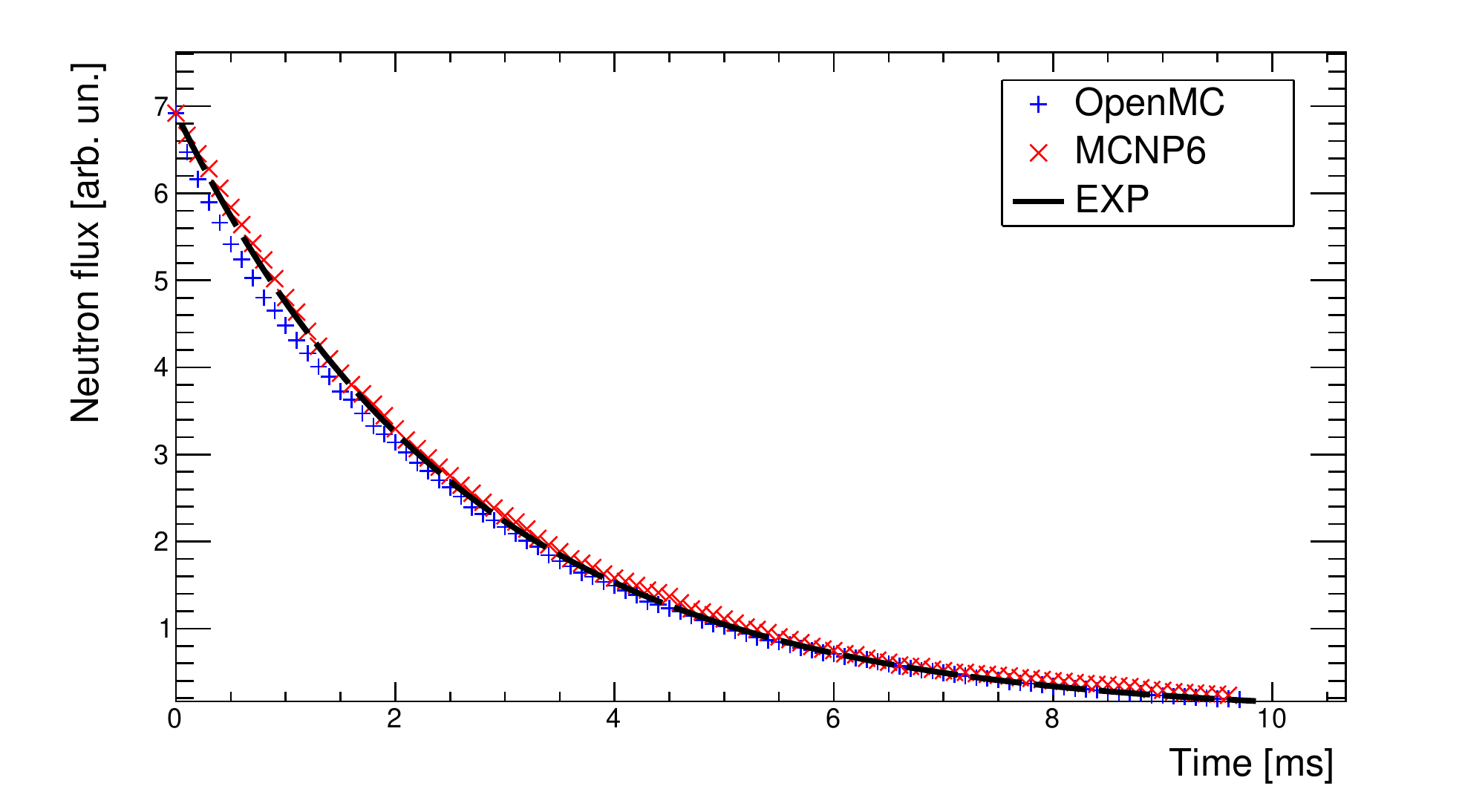}
	\caption{Time evolution of the neutron flux in a subcritical configuration for the RA-$6$ reactor, obtained from running a fixed source simulation in both OpenMC(TD) and MCNP. Results were scored for $10$~ms. Both codes are in good agreement with the experimental benchmark result.}
	\label{fig:flux_subcritical_RA6}
\end{figure}

The tallying capabilities of OpenMC were expanded and a time filter was added to monitor the time evolution of any of the tallies already present in the code. In order to examine the proper functioning of this filter, MCNP and OpenMC(TD) were used to estimate the time evolution of the neutron flux in the RA-$6$ reactor using the pulsed method~\cite{doi:10.13182/NSE3-595-608}. In this method, a burst of neutrons is injected into a subcritical system and then the decay of the prompt neutron flux as a function of time is observed. Since the phenomena being studied is the prompt neutron decay the contribution from delayed neutrons can be neglected from point kinetics Eq.~\eqref{eq:neutrons_point_kinetic}, which in that case reads, 
\begin{equation}
\frac{\partial}{\partial t} s(t) = \frac{\rho - \beta_{\mathit{eff}}}{\Lambda_{\mathit{eff}}} \,s(t).
\label{eq:neutrons_pulsed}
\end{equation} 
The solution to Eq.~\eqref{eq:neutrons_pulsed} is given by
\begin{equation}
	s(t) = s_0 e^{\alpha \,t}, \quad \text{with} \quad \alpha= \frac{\rho - \beta_{\mathit{eff}}}{\Lambda_{\mathit{eff}}}
	\label{eq:prompt_decay}
\end{equation}
where $s_0$ is the initial flux density and the decay constant $\alpha$ is the decay constant of the neutron population.

MCNP and OpenMC(TD) were used to simulate the neutron source and then the prompt neutron decay was scored during $10$~ms. The flux as a function of time obtained with both codes was then compared with the experimental results for the decay constant from the graphite reflected RA-$6$ benchmark from ~\cite{ICSBEP,bazzana}.
Fig.~\ref{fig:flux_subcritical_RA6} shows results obtained, where blue (red) crosses (x marks) denote OpenMC(TD) (MCNP) results. Dashed curve denote the benchmark value. It can be observed good agreement between the decay constants from fit parameters and the result from the benchmark.  
\begin{table}[h!]
\centering
\begin{tabular}{@{}ccccc@{}}
\toprule
      & \textbf{OpenMC}      & \textbf{MCNP}        &   \textbf{Benchmark}         \\ \midrule
$\alpha$ [s${}^{-1}$]    & $-370(1)$ & $-354(3)$ &   $-378(3)$ \\ \bottomrule
\end{tabular}
\caption{Decay constants obtained for the time evolution of the neutron flux obtained using the pulsed method in the RA-6 reactor.}
\label{table:alpha_ra6}
\end{table}

As it can be seen in Table~\ref{table:alpha_ra6}, the values obtained for the decay constant are in reasonable agreement between each other. In conclusion, the \textit{time} filter implemented works as expected and OpenMC(TD) can score time dependent quantities in fixed source calculations.     

\subsection{Transport logic in a simulation divided in time intervals}
\label{subsection:transport_checks}
Another modification needed to be the implemented in the code is the division of the total simulation time in discrete time intervals. It is important to check that there are no errors in the crossing of time intervals. To do this, neutron transport in the monoenergetic fissile system described in Appendix~\ref{app:specifications_monoenergetic_system} was studied.
\begin{figure}[h!]
	\centering
	\includegraphics[width=\textwidth]{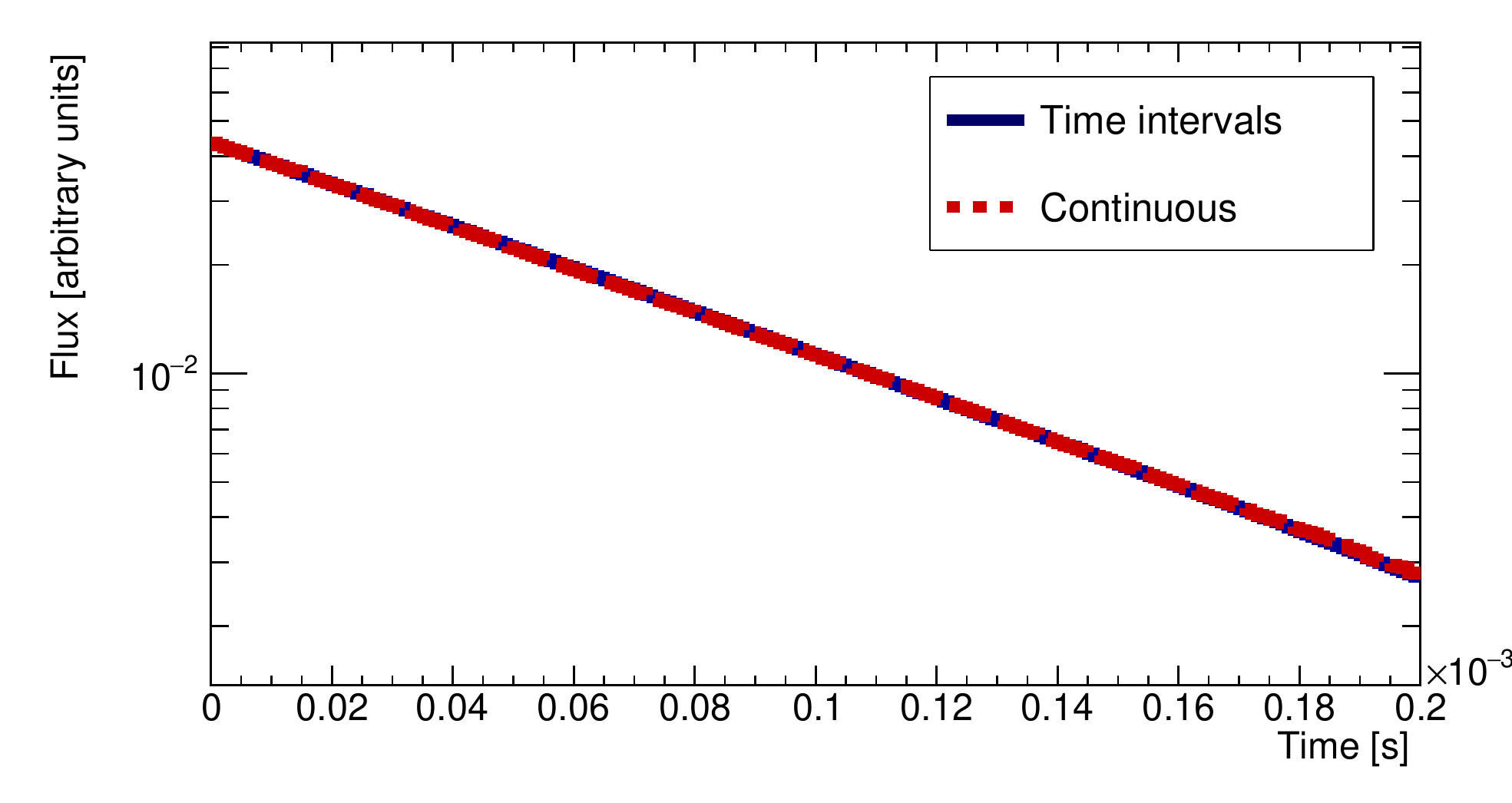}
	\caption{Neutron flux as a function of time in a simple transport problem. In red, the neutron flux obtained for a non-transient fixed source simulation is shown. In blue the neutron flux obtained from a transient simulation divided in time intervals is shown. Both results are equivalent.}
	\label{fig:test2_transport_logic}
\end{figure}
Since the purpose of this test was only to check for errors in the particle transport when dividing the simulation in time intervals, fission reactions were not considered. Results obtained are shown in Fig.~\ref{fig:test2_transport_logic}, where it can be seen that the neutron flux obtained when the simulation is divided in discrete time intervals is the same when a regular fixed source calculation is performed, thus, the transport logic is correct and OpenMC(TD) correctly transports neutrons across time intervals.

\subsection{Scoring of time dependent quantities in a simulation divided in time intervals}
\label{subsection:test3}
Test $3$ was a combination of tests $1$ and $2$, i.e., flux scoring as a function of time in a subcritical configuration when the simulation was divided in time intervals. The transport problem studied was the monoenergetic fissile system detailed in Appendix~\ref{app:specifications_monoenergetic_system}. The advantage of studying a system like this one is that the time evolution of the neutron flux can be described by an analytical expression, making direct its validation. 

\begin{figure}[h!]
	\centering
	\includegraphics[width=\textwidth]{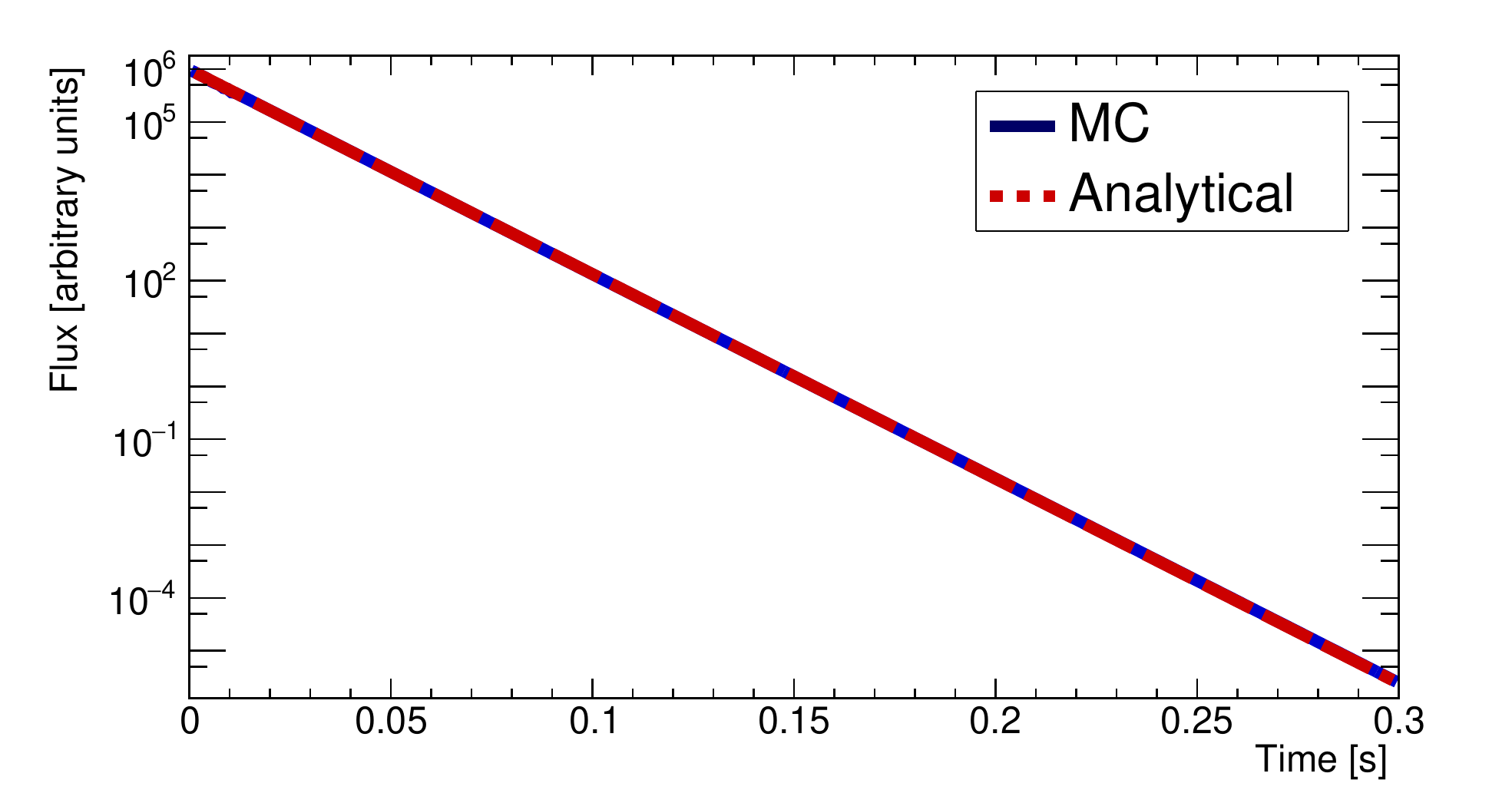}
	\caption{Time evolution of the neutron flux for the monoenergetic system studied. Results obtained using OpenMC(TD) are shown in blue, while the fit to the point kinetics solution given by Eq.~\eqref{eq:prompt_decay} is shown in red.}
	\label{fig:test3_time_intervals_check}
\end{figure}

First, the configuration was made subcritical by increasing the absorption cross section to $\Sigma_a\!=\!0.5854$~cm${}^{-1}$, while mantaining the total cross section constant. A criticality calculation with $10^{6}$~neutrons, $5000$~batches and $300$~skipped cycles for this configuration gives $k_{\mathit{eff}}\!=\!0.99344\!\pm\!0.00003$. Then test $3$ was conducted transporting $10^6$~neutrons for $300$~ms using a time interval of $1$~ms and $10$~batches.  Fig.~\ref{fig:test3_time_intervals_check} shows a comparison between prompt neutron flux obtained from the Monte Carlo simulation and the analytical solution obtained using Eq.~\eqref{eq:prompt_decay} and the parameters for this system given in Table~\ref{tab:parameters_mono_system}. 
The fitted time constant parameter obtained for the decay of the prompt neutron flux is $\alpha_{\mathit{fitted}}\!=\!-89.08(3)$~s${}^{-1}$, while the calculated value is given by $\alpha_{\mathit{teo}}\!=\!-90.20(1)$~s${}^{-1}$. A summary of the obtained results is shown in Table~\ref{table:test3}. Both values are in excellent agreement with each other ($1.2 \%$ difference). Therefore, the scoring of time dependent quantities when the simulation is divided in time intervals works correctly.

\begin{table}[h!]
	\centering
	\begin{tabular}{@{}ccccc@{}}
		\toprule
		& \textbf{Calculated}      & \textbf{Fitted}        &   $\Delta$         \\ 
		&  \textbf{Decay constant}  &  \textbf{Decay constant}                     &                  \\ \midrule
		$\alpha$ [s${}^{-1}$]    & $-90.20(1)$ & $-89.08(3)$ &   $-1.12(3)$ \\ \bottomrule
	\end{tabular}
	\caption{Decay constants obtained for the time evolution of the neutron flux using the RA-6 reactor.}
	\label{table:test3}
\end{table}

\subsection{Activity of individual precursors}
\label{subsection:activity_individual_precursors}
The purpose behind the activity calculation for individual precursors and its comparison to the $6$-group activity was to verify the suitability of the $50$ individual precursors chosen for the emission of the $\beta$-delayed neutrons as part of the new capabilities OpenMC(TD) code. This test was necessary because if there were differences in results obtained for the time evolution of the neutron flux using the $N$-group structure or in the individual precursors, it was relevant to know if the activity of the $\beta$-delayed neutron emission was the cause of these eventual discrepancies.

\begin{figure}[h!]
	\centering
	\includegraphics[width=\textwidth]{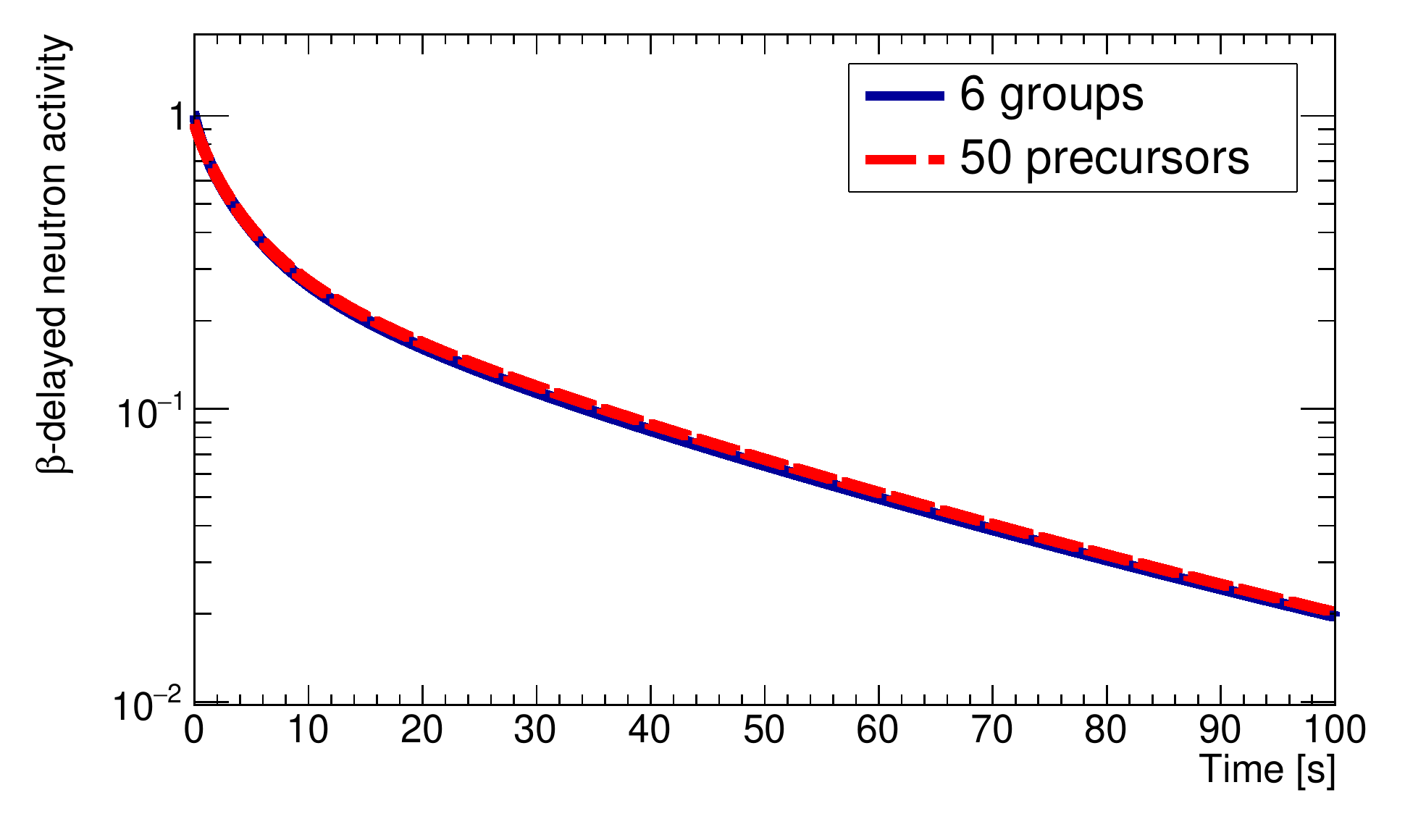}
	\caption{$\beta$-delayed neutron activity for $6$ precursor groups and $50$ individual precursors is shown. In blue, $A_6(t)$ is shown, while $A_{50}(t)$ is shown in red. As it can be seen, both activities are equivalent.}
	\label{fig:activity_decay_50precs_vs_6_groups}
\end{figure}

The calculated activity for the $6$-precursor groups, denoted by $A_6(t)$, is given by
\begin{equation}
A_6(t) = \sum_{i=1}^{6} a_i \exp(-\lambda_i t),
\label{eq:6group_activity}
\end{equation}
where $a_i\!=\!\beta_i/\beta$ is the $i$-th group relative abundance and $\lambda_i$ is the $i$-th group decay constant (see Table~\ref{table:6-group_precursor_structure}). Conversely, the calculated activity for the $50$ individual precursors, denoted by $A_{50}(t)$ reads
\begin{equation}
A_{50}(t) = \sum_{i=1}^{50} I_i \exp(-\lambda_i t),  \quad \text{with} \quad I_i = \frac{CY_i \, P_{n,i}}{\nu_d},
\label{eq:50group_activity}
\end{equation}
where $I_i$ is the $i$-th precursor importance as defined in Eq.~\eqref{eq:precursor_importance} (see Table~\ref{table:50-individual_precursor_structure}) and $\lambda_i$ is the $i$-th precursor decay constant (See Sec.~\ref{subsection:quantities_of_interest}).
The calculated activity for $6$ precursor groups and $50$ individual precursors is shown in Fig.~\ref{fig:activity_decay_50precs_vs_6_groups}. In blue, $A_6(t)$ is shown, while $A_{50}(t)$ is shown in red. As it can be seen, both activities are equivalent. Quantitatively, comparing $\beta$-delayed neutron emission for $A_{50}(t)$ and $A_{6}(t)$ up to $100$~s, it is obtained that 
\begin{equation}
\frac{\int_{0}^{100} A_{50} (t) dt}{\int_{0}^{100} A_{6} (t) dt} = 0.9916.
\label{eq:quant_a6_a50`} 
\end{equation}
This indicates that adding the remaining $219$ precursors only contributes to $0.84 \%$ of delayed neutron emission.  

\subsection{Discussion about the use of average energies from precursor delayed neutron spectra}
\label{subsection:average_energies}
At the time of the writing of this work there exists experimental measurements for only $34$ $\beta$-delayed neutron energy spectra. This data was compiled and completed by Brady in $1989$~\cite{osti_6187550}. The remaining $\beta$-delayed neutron energy spectra present in ENDF/B-VIII.$0$ comes from QRPA calculations~\cite{BROWN20181}. Given that the capabilities added to the OpenMC code allows to run simulations using up to $269$ individual precursors and with the intention of having these precursors on the same footing regarding the $\beta$-delayed neutron emission energies, it was decided that the average energy for the delayed neutron emission would be used. Nevertheless, if the $\beta$-delayed neutron energy spectra databases were updated in the future, its inclusion could be easily implemented in the code.  

Since the average energy for the $\beta$-delayed neutron emission was used, it was important to verify that the results obtained for the time evolution of the neutron flux when using the delayed neutron average energies were equivalent to sampling the delayed neutron energy from the corresponding spectra. To this end, a transient simulation using OpenMC(TD) in a subcritical configuration was run. Subcriticality was achieved by decreasing the ${}^{235}$U density from $\delta_{U235}\!=\!3.2675\!\times\!10^{-2}$~(atoms/b cm) to $\delta_{U235}\!=\!3.19\!\times\!10^{-2}$~(atoms/b cm), while mantaining the dimensions of the box constant, obtaining an effective multiplication system of $k_{\mathit{eff}}\!=\!0.98663\!\pm\!0.00004$ for the system. The simulation was run using $3$ batches and the total simulation time was $10$~ms divided in $10000$ time intervals of $1$~$\mu$s each. Population control was applied at the end of each interval. 

\begin{figure}[h!]
	\centering
	\includegraphics[width=\textwidth]{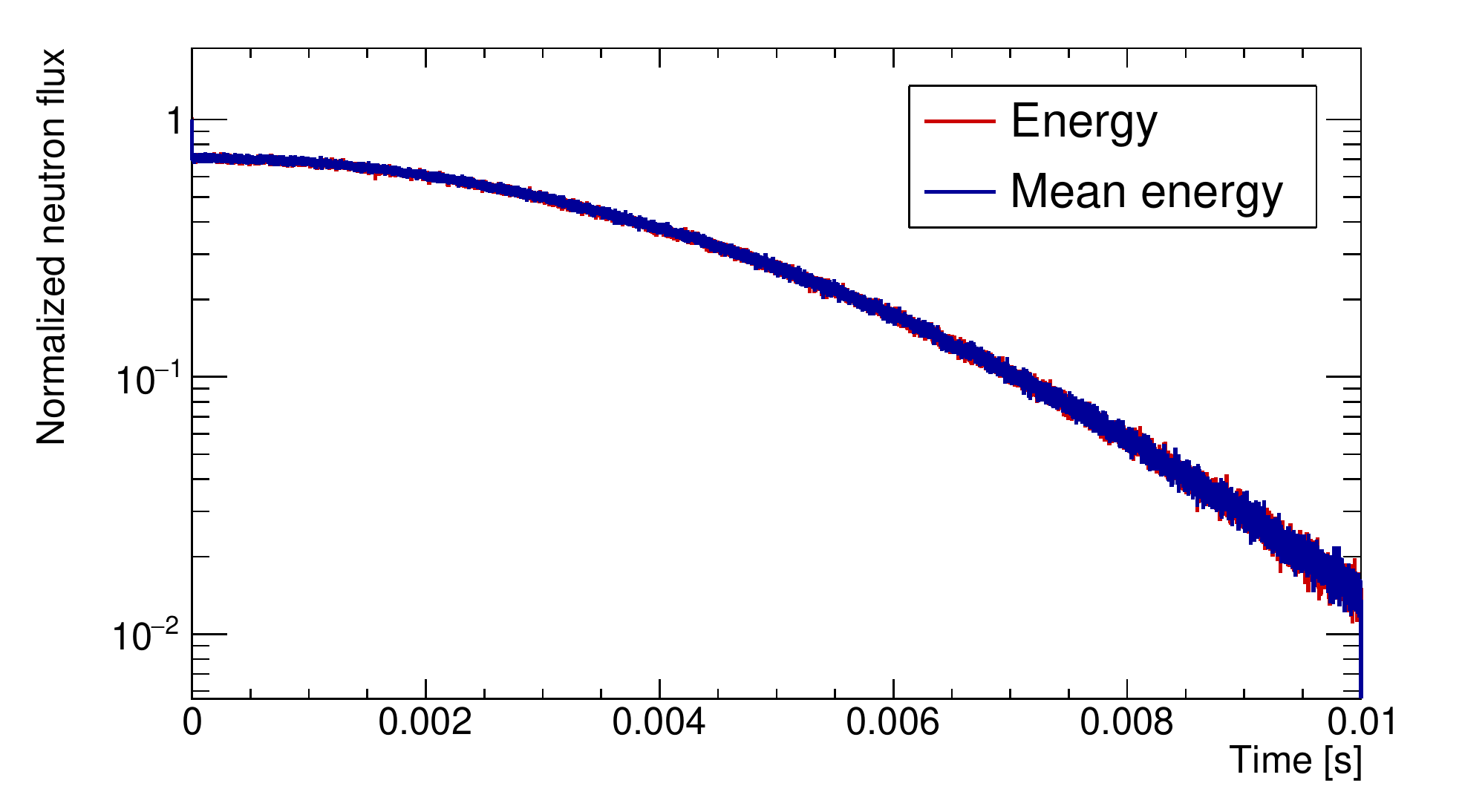}
	\caption{Time evolution of neutron flux for a subcritical configuration with \texorpdfstring{$k_{\mathit{eff}}\!=\!0.98663\!\pm\!0.00003$}. Both simulations were run for $3$ batches, total simulation time was $10$~ms divided in $10000$ time intervals of $1$~$\mu$s each. Results of neutron flux when delayed neutron energy was sampled from spectra are shown in red (behind the blue curve), while results obtained when using the average energy for delayed neutron emission are shown in blue. It can be seen that both results are equivalent.}
	\label{fig:comparison_mean_energy_vs_energy}
\end{figure}

Results obtained from transient Monte Carlo simulation using OpenMC(TD) for the time evolution of the neutron flux when the delayed neutron energy was sampled from spectra are shown in red in Fig.~\ref{fig:comparison_mean_energy_vs_energy}, while results obtained when using the average energy for the delayed neutron emission are shown in blue. From Fig.~\ref{fig:comparison_mean_energy_vs_energy} it can be seen that the time evolution of the neutron flux obtained with transient Monte Carlo code OpenMC(TD) using the average delayed neutron energy and delayed energy sampled from spectra are equivalent.

\section{Monoenergetic fissile system with 1-group precursor structure}
\label{section:mono-energetic_system}
Once the preliminary work described in Section~\ref{section:preliminar_work} was completed, the new capabilities added to the OpenMC code, namely division of the simulation in discrete time intervals, scoring of time dependent quantities, forced decay of precursors and population control, were tested in the monoenergetic fissile system described in Appendix~\ref{app:specifications_monoenergetic_system}. The objective of this section is to lay the groundwork for the study of the delayed neutron emission from individual precursors for transient calculations using OpenMC. 
 
Prior to transient simulations with Monte Carlo code OpenMC(TD) presented in this section, a non-transient standard steady state criticality calculation was done with $10^6$~neutrons, $3000$~batches and $200$ skipped cycles, using OpenMC. The effective multiplication factor obtained was $k_{\mathit{eff}}\!=\!1.00010 \pm 0.00003$. Afterwards, the initial transient source was created as described in Sec.~\ref{section:initial_particle_source}, with $10^5$ neutrons and $9\times 10^{5}$ precursors. This initial transient source was used in subcritical (See Sec.~\ref{subsection:mono_subcritical_state}), critical (See Sec.~\ref{subsection:mono_steady_state}) and reactivity insertion (See Sec.~\ref{subsection:mono_reactivity}) configurations, presented in the following subsections, in order to start the transient simulation.

In this section, the code input was the macroscopic absorption cross section, $\Sigma_a$, which was suitably modified in order to produce reactivity changes in the monoenergetic fissile system (critical, subcritical or supercritical configurations). Thus, the output values (observables) were the effective multiplication factor $k_{\mathit{eff}}$ and the time evolution of the neutron flux $\phi(t)$, which were compared with point kinetics calculations.

\subsection{Subcritical configuration}
\label{subsection:mono_subcritical_state}
The code was firstly tested in a subcritical configuration. Subcriticality was achieved by increasing the absorption cross section from $\Sigma_a\!=\!0.5882$~cm${}^{-1}$ to $\Sigma_a\!=\!0.5952$~cm${}^{-1}$. Total cross section $\Sigma_t$ was kept constant, then the effective multiplication of the system $k_{\mathit{eff}}\!=\!0.98821 \pm 0.00003$. This increasing in the absorption cross section $\Sigma_a$ is equivalent to decrease the density of the fissile material $\delta_f$ of the system. 

\begin{figure}[h!]
	\centering
	\includegraphics[width=\textwidth]{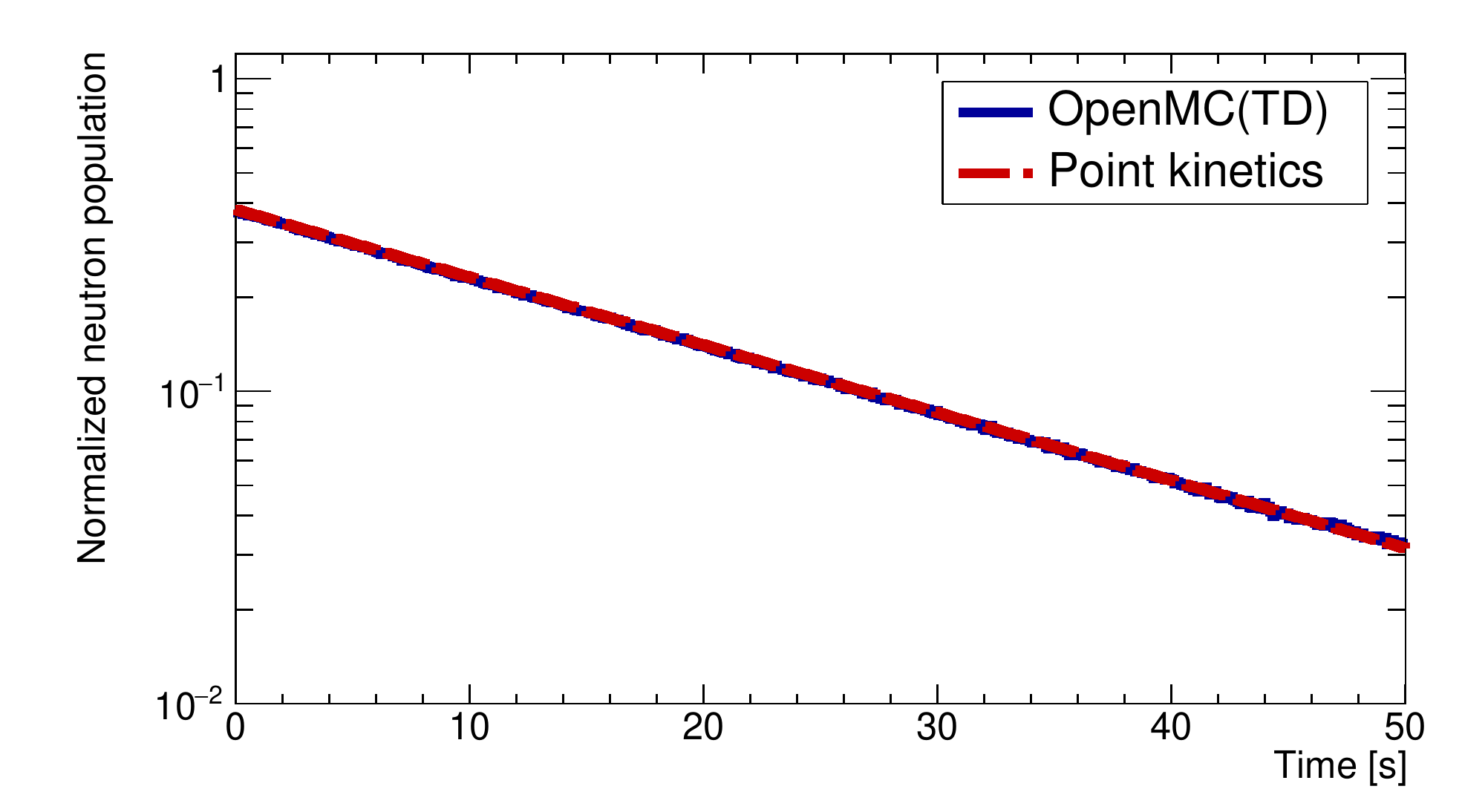}
	\caption{Time evolution of the neutron population for a monoenergetic system in a subcritical configuration (\texorpdfstring{$k_{\mathit{eff}}\!=\!0.98821 \pm 0.00003$}) obtained using OpenMC(TD) code. The initial transient source is prepared in a critical configuration and at the beginning of the transient simulation the system is made subcritical. The result is compared to the analytical solution from the point kinetics equations.}
	\label{fig:mono_subcritical_50s}
\end{figure}

The simulation was run using $60$ batches and the total simulation time was $50$~s, divided in $500$ time intervals of $100$~ms each one. At the end of each time interval population control was applied, using the technique explained in Sec.~\ref{subsection:pop_control}. 
Results obtained from transient Monte Carlo simulation using OpenMC(TD) are shown in blue in Fig.~\ref{fig:mono_subcritical_50s}, meanwhile the point kinetic solution of the neutron population as a function of time is shown in red. From Fig.~\ref{fig:mono_subcritical_50s} it can be seen that the time evolution of the neutron population calculated using transient Monte Carlo code OpenMC(TD) and point kinetics solution using Eq.~\eqref{eq:solution_neutron_1precursor} are equivalent.

Quantitatively, from Fig.~\ref{fig:mono_subcritical_50s} the reactivity value $\rho$ can be obtained as a fitted parameter of Eq.~\eqref{eq:neutron_population_1precursor}. This $\rho_{\mathit{fit}}\!=\!-0.01193(626)$ was compared to the reactivity from the criticality calculation using OpenMC(TD), $\rho \!=\! (k_{\mathit{eff}}-1)/k_{\mathit{eff}} \!=\! -0.01193(3)$. 

It is important to notice that in this case the population control prevents the dying out of the neutron population\footnote{See for instance Fig.~\ref{fig:test3_time_intervals_check} where in a non-transient standard Monte Carlo fixed source calculation, using both MCNP or OpenMC, the neutron population extinguishes in $\sim 50$~ms.}. This new time dependent capability added to OpenMC allows the observation of the slow decay of the neutron population due to the $\beta$-delayed neutron emission.

The reactivity value is usually obtained by running a criticality Monte Carlo calculation. In this work, using OpenMC(TD), this value can be obtained by fitting  Eq.~\eqref{eq:solution_neutron_1precursor} to the time evolution of the neutron population. A summary of the results obtained is shown in Table~\ref{table:summary_mono_sub_reactivity}.   

\begin{table}[h!]
	\centering
	\begin{tabular}{@{}ccccc@{}}
		\toprule
		\textbf{}  & \textbf{Calculated}   & \textbf{Fitted}               \\ 
		  &     \textbf{reactivity}     &    \textbf{reactivity}      \\ \midrule
		$\rho$ [pcm]  & $-1193(3)$ & $-1193(626)$ \\  \bottomrule
	\end{tabular}
	\caption{Results obtained for the reactivity of the monoenergetic simulated system in a subcritical configuration using $1$-group precursor structure.}
	\label{table:summary_mono_sub_reactivity}
\end{table}

\subsection{Critical configuration}
\label{subsection:mono_steady_state}
\begin{figure}[h!]
	\centering
	\includegraphics[width=\textwidth]{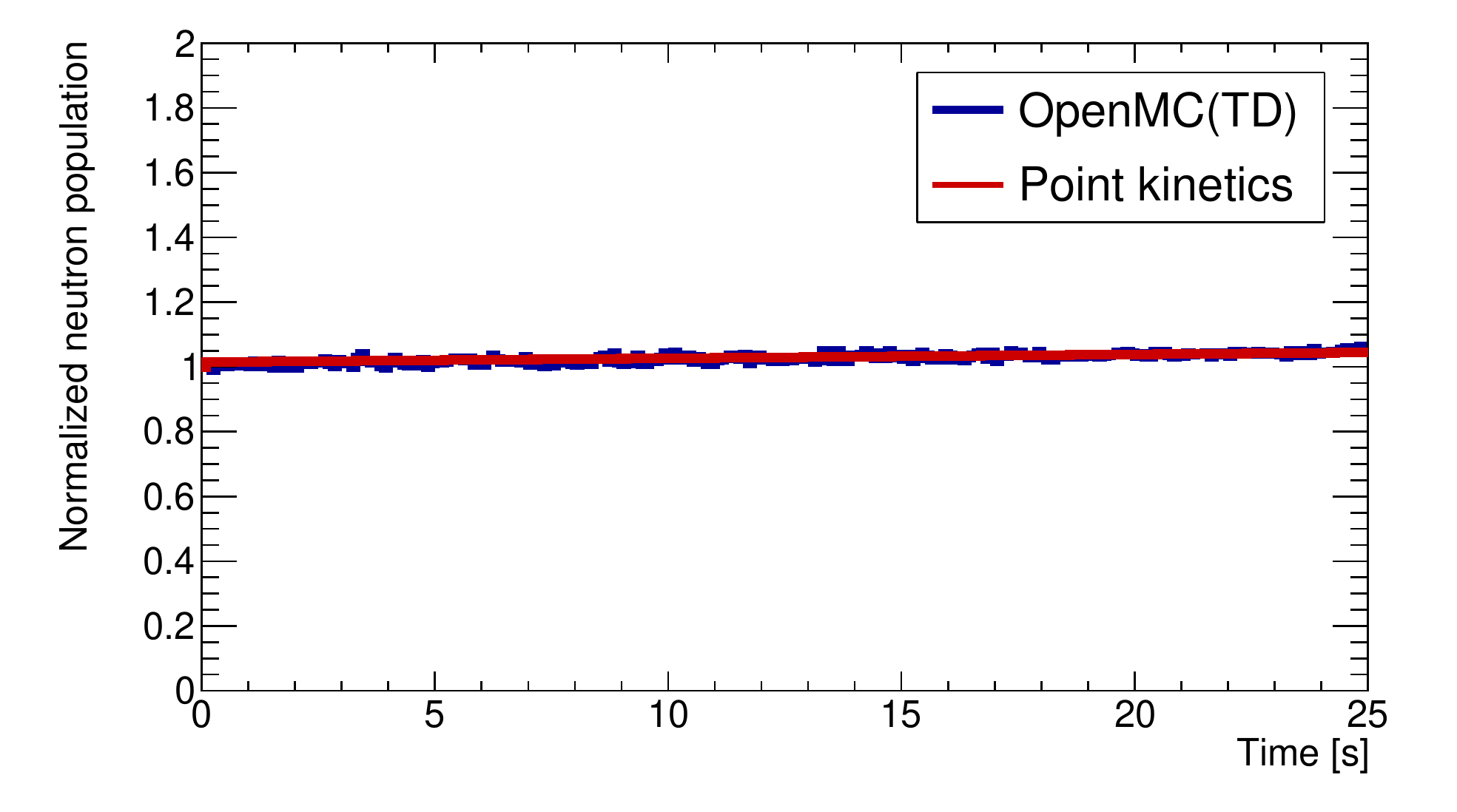}
	\caption{Time evolution of the neutron population for a monoenergetic system in a critical configuration (\texorpdfstring{$k_{\mathit{eff}}\!\!=\!\!1.00010\!\pm\!0.00003$}) obtained using the OpenMC(TD) code. The result is compared to the analytical solution from the point kinetics equations.}
	\label{fig:mono_critical_25s}
\end{figure}
In the second test, transient analysis was done in a critical configuration, with $k_{\mathit{eff}}\!=\!1.00010\!\pm\!0.00003$. The simulation was run using $4$ batches and the total simulation time was $25$~s, divided in $250$ time intervals of $100$~ms each. Population control was applied at the end of each interval. 

Results obtained from transient Monte Carlo simulation using OpenMC(TD) are shown in blue in Fig.~\ref{fig:mono_critical_25s}, meanwhile the point kinetic solution of the neutron population as a function of time is shown in red. From Fig.~\ref{fig:mono_critical_25s} it can be seen that the time evolution of the neutron population calculated using transient Monte Carlo simulation using OpenMC(TD) and point kinetics solution using Eq.~\eqref{eq:solution_neutron_1precursor} are equivalent. 

Neutron population remains practically constant in time, as it is expected for a critical system. It can also be noted that since this is a critical configuration, the fission chains tend to diverge as it was mentioned in Sec.~\ref{subsection:pop_control}, but population control prevents this from happening and the simulation remains stable.

Quantitatively, from Fig.~\ref{fig:mono_critical_25s} the reactivity value $\rho$ can be obtained as a fitted parameter of Eq.~\eqref{eq:neutron_population_1precursor}. This $\rho_{\mathit{fit}}\!=\!0.00013(70)$, was compared to the reactivity from the criticality calculation using OpenMC, $\rho\!=\!(k_{\mathit{eff}}-1)//k_{\mathit{eff}}\!=\!0.00010(3)$. 

The reactivity value is usually obtained by running a criticality Monte Carlo calculation. In this work, using OpenMC(TD), this value can be obtained by fitting  Eq.~\eqref{eq:solution_neutron_1precursor} to the time evolution of the neutron population. A summary of the results obtained is shown in Table~\ref{table:summary_mono_critical_reactivity}. 

\begin{table}[h!]
	\centering
	\begin{tabular}{@{}ccccc@{}}
		\toprule
		\textbf{}  & \textbf{Calculated}   & \textbf{Fitted}               \\ 
		&     \textbf{reactivity}     &    \textbf{reactivity}      \\ \midrule
		$\rho$ [pcm]  & $10(3)$ & $13(70)$ \\  \bottomrule
	\end{tabular}
	\caption{Results obtained for the reactivity of the monoenergetic simulated system in a critical configuration using $1$-group precursor structure.}
	\label{table:summary_mono_critical_reactivity}
\end{table}

\subsection{Reactivity insertion}
\label{subsection:mono_reactivity}
\begin{figure}[h!]
	\centering
	\includegraphics[width=\textwidth]{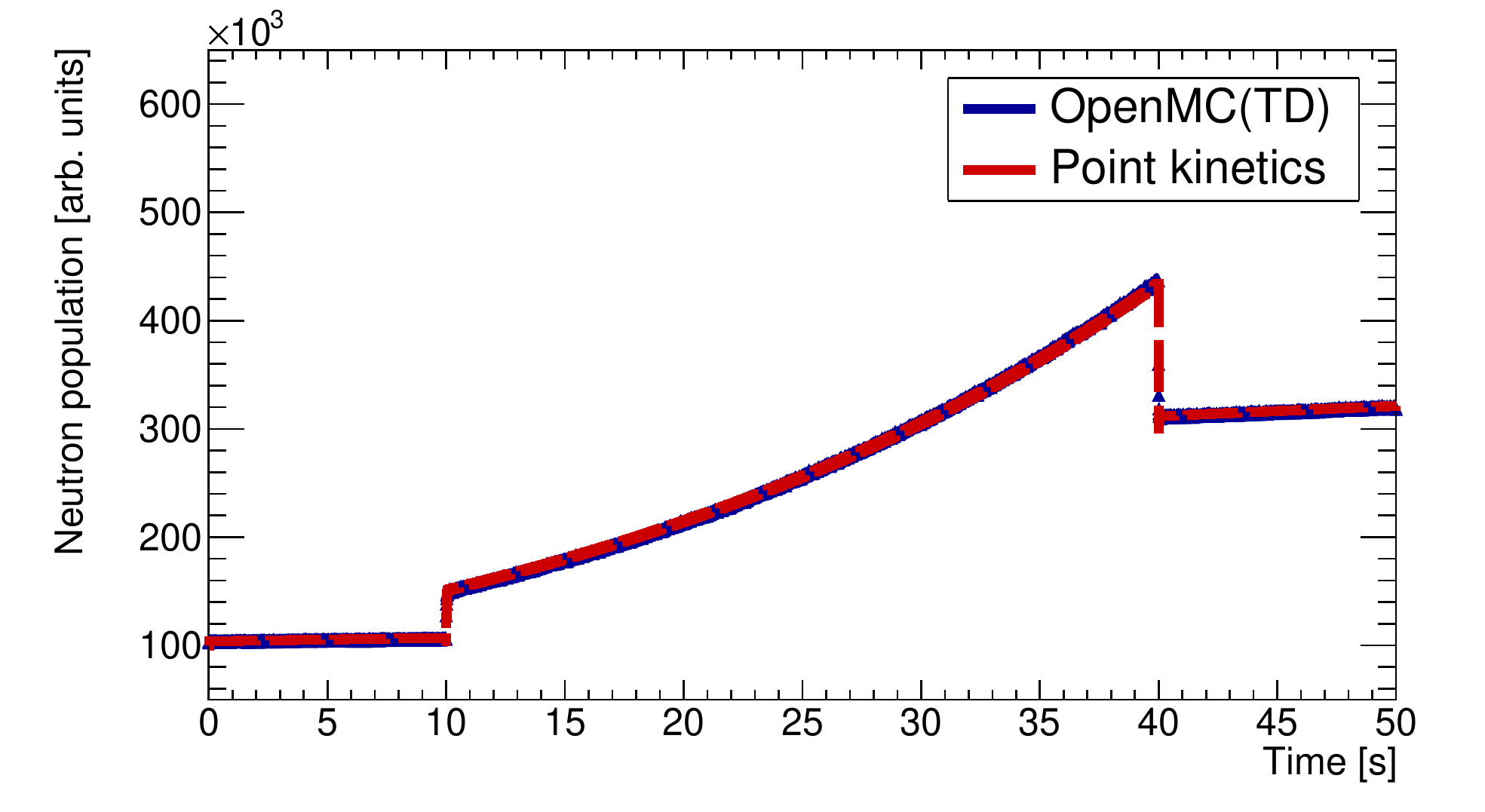}
	\caption{Time evolution of the neutron population for a monoenergetic system obtained using OpenMC(TD). The system is initially in a critical configuration, then, at $t\!=\!10$~s a reactivity of $211$~pcm is inserted. After $30$~s the system is brought back to critical configuration.}
	\label{fig:mono_reactivity_insertion_25s}
\end{figure}

The last case studied was a mixture of the two cases previously presented. First, the configuration was critical, then reactivity is inserted and afterwards, the configuration is brought back to critical by a negative reactivity insertion. This system is a first approximation to simulate the operation of a nuclear reactor.
Concretely, these reactivity insertions were simulated by changing the absorption cross section while keeping the total cross section constant.

In Fig.~\ref{fig:mono_reactivity_insertion_25s} the reactivity insertion case is shown. For the first $10$~s the configuration was kept critical with $\Sigma_a \!=\! 0.5882$~cm${}^{-1}$. Then, for $10$~s $< t < $ $40$~s the absorption cross section was reduced from $\Sigma_a \!=\! 0.5882$~cm${}^{-1}$ to $\Sigma_a \!=\! 0.5870$~cm${}^{-1}$, inserting a positive reactivity of $211$~pcm, thus making the configuration supercritical. This fast increase in neutron population is known as \textit{prompt jump}. In $t\!=\!40$~s the system is brought back to $\Sigma_a\!=\!0.5882$~cm${}^{-1}$. The neutron population stops growing and decreases rapidly. This fast change in neutron population is known as \textit{prompt drop}. In $t\!=\!40$~s the neutron population is almost three times the initial neutron population. The final state of the configuration is slightly supercritical.

The simulation was run for $25$ batches and the simulation time was divided in $5000$ time intervals of $10$~ms each and population control was applied at the end of every interval. Results obtained are shown in Fig.~\ref{fig:mono_reactivity_insertion_25s}, where the time evolution of the neutron population calculated from point kinetic equations is also shown. From Fig.~\ref{fig:mono_reactivity_insertion_25s} it can be seen that the time evolution of the neutron population calculated using transient Monte Carlo simulation using OpenMC(TD) and point kinetics solution using Eq.~\eqref{eq:solution_neutron_1precursor} are equivalent.

It is important to notice that in the reactivity insertion case, both prompt jump and prompt drop can be studied in detail given that short time intervals of $10$~ms were used in the simulation. This new Monte Carlo capabilitity, implemented in this work, allows to reduce time windows as much as desired, so parameters as the Rossi-$\alpha$~\cite{doi:10.1080/00223131.2016.1274686} can be calculated. 

\section{Energy-dependent \texorpdfstring{${}^{235}$U}{U235} system}
\label{section:energy_dependent}
After the new capabilities added to the code were successfully tested for the monoenergetic system described in the previous Section~\ref{section:mono-energetic_system}, the following study involved testing the code in a system with continuous, energy-dependent cross sections (i.e. not monoenergetic). The objective of this section is to simulate a more realistic system, but at the same time keeping it simple enough to compare to the point kinetics model, whenever is possible. In order to do this, the material of the box from the preceding section was made of pure ${}^{235}$U, using the continuous energy cross sections from JEFF-$3$.$1$.$1$~\cite{jeff311} nuclear data library and the geometry was surrounded by vacuum. 

Prior to the transient simulations with Monte Carlo code OpenMC(TD) presented in this section, a non-transient standard steady state criticality calculation was done with $10^6$~neutrons, $5000$~batches and $300$ skipped cycles, using OpenMC. The effective multiplication factor obtained was $k_{\mathit{eff}}\!=\!1.00015 \pm 0.00003$. Afterwards, the initial transient source was sampled as described in Sec.~\ref{section:initial_particle_source}, with $10^5$ neutrons and $9\times 10^{5}$ precursors. This initial transient source was used in subcritical and supercritical tests, presented in the following subsections, in order to start the transient simulation. Intentionally, the critical configuration was not considered in this set of tests because the main objective of this part of the work was to examine whether the code had the capability to resolve fast changes in the neutron flux. 

In this section, the code input was the density of the fissile material, which will be denoted as $\delta_{U235}$. Since the box made of pure ${}^{235}$U, the only two ways to insert reactivity to the system are: i) by changing the box dimensions or, ii) by changing the density of the fissile material. The latter method was chosen and the dimensions of the box were kept constant throughout the different cases. Thus, the output value (observable) was the effective multiplication factor $k_{\mathit{eff}}$ and the time evolution of the neutron flux $\phi(t)$, like in the previous section. Since this is not a monoenergetic system, Eq.~\eqref{eq:Lambda_simple} for the effective generation time no longer holds. In consequence, for the calculation of $\Lambda$ and $\beta_{\mathit{eff}}$, a simulation of the system in MCNP was made, given that this code can estimate these parameters using the weighted adjoint flux. These two quantities were then compared with the fitted parameters from Eq~\eqref{eq:solution_neutron_1precursor}, which is the solution to the point kinetics equations.

Different group structures were simulated in this section. When it was possible, the energy of the $\beta$-delayed neutrons was taken from a distribution (JEFF-$3$.$1$.$1$). Otherwise, the average energy was used for each precursor or group (ENDF-B/VIII.$0$). For comparison purposes, a simulation using the energy distribution and the average energy from the first group were also studied.  

\subsection{Subcritical configuration}
\label{subsection:energy_u235_vacuum_sub}
The first case studied was a subcritical configuration. The system was made subcritical by decreasing the ${}^{235}$U density from $\delta_{U235}\!=\!4.496 \times 10^{-2}$~(atoms/b cm)~\cite{doi:10.13182/NSE12-44} to $\delta_{U235}\!=\!4.4362 \times 10^{-2}$~(atoms/b cm), while mantaining the dimensions of the box constant, making the effective multiplication factor of the system $k_{\mathit{eff}}\!=\!0.98956 \pm 0.00003$. 

\subsubsection{i) First group with energy distribution from JEFF-3.1.1}
\label{subsub:subcritical_first_group_energy_real}

The first precursor group, characterized by a half-life $T_{1/2}\!=\!55.6$~s, was simulated. The delayed neutron energy was sampled from its neutron energy distribution, reported from JEFF-$3$.$1$.$1$. Group $1$ $\beta$-delayed neutron energy spectrum from JEFF-$3$.$1$.$1$. is shown in Fig.~\ref{fig:jeff311_group1_spectrum_1}. In Appendix~\ref{app:group_spectra}, the $\beta$-delayed neutron energy spectra for all $8$ groups can be found.

\begin{figure}[h!]
	\centering
	\includegraphics[width=0.8\textwidth]{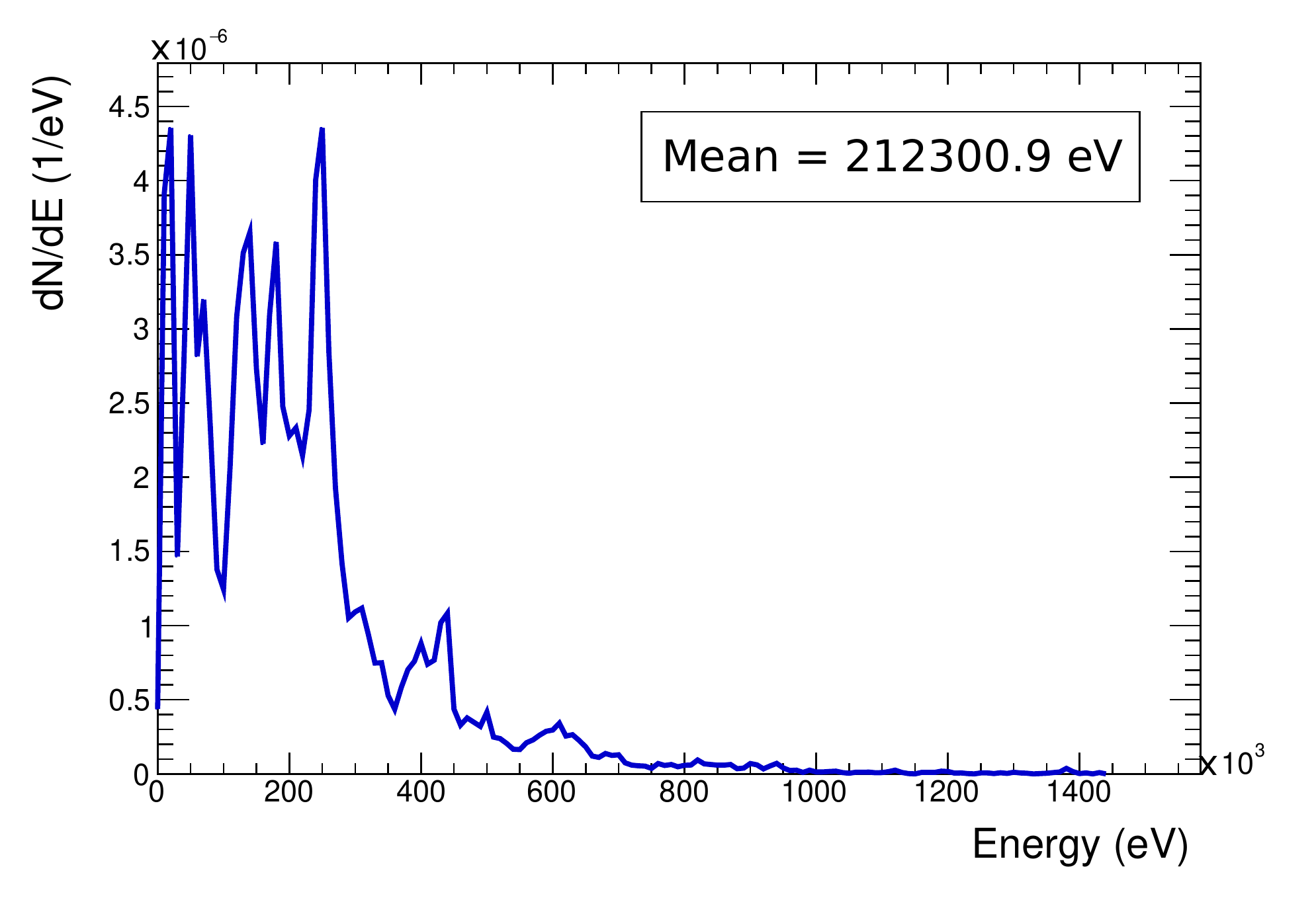}
	\caption{Group $1$ $\beta$-delayed neutron energy spectrum from JEFF-$3$.$1$.$1$.}
	\label{fig:jeff311_group1_spectrum_1}
\end{figure}

This simulation was run using $22$ batches and the total simulation time was $0.1$~ms divided in $1000$ time intervals of $100$~ns each. Population control (see Sec.~\ref{subsection:pop_control}) was applied at the end of each interval. 

\begin{figure}[ht!]
	\centering
	\includegraphics[width=\textwidth]{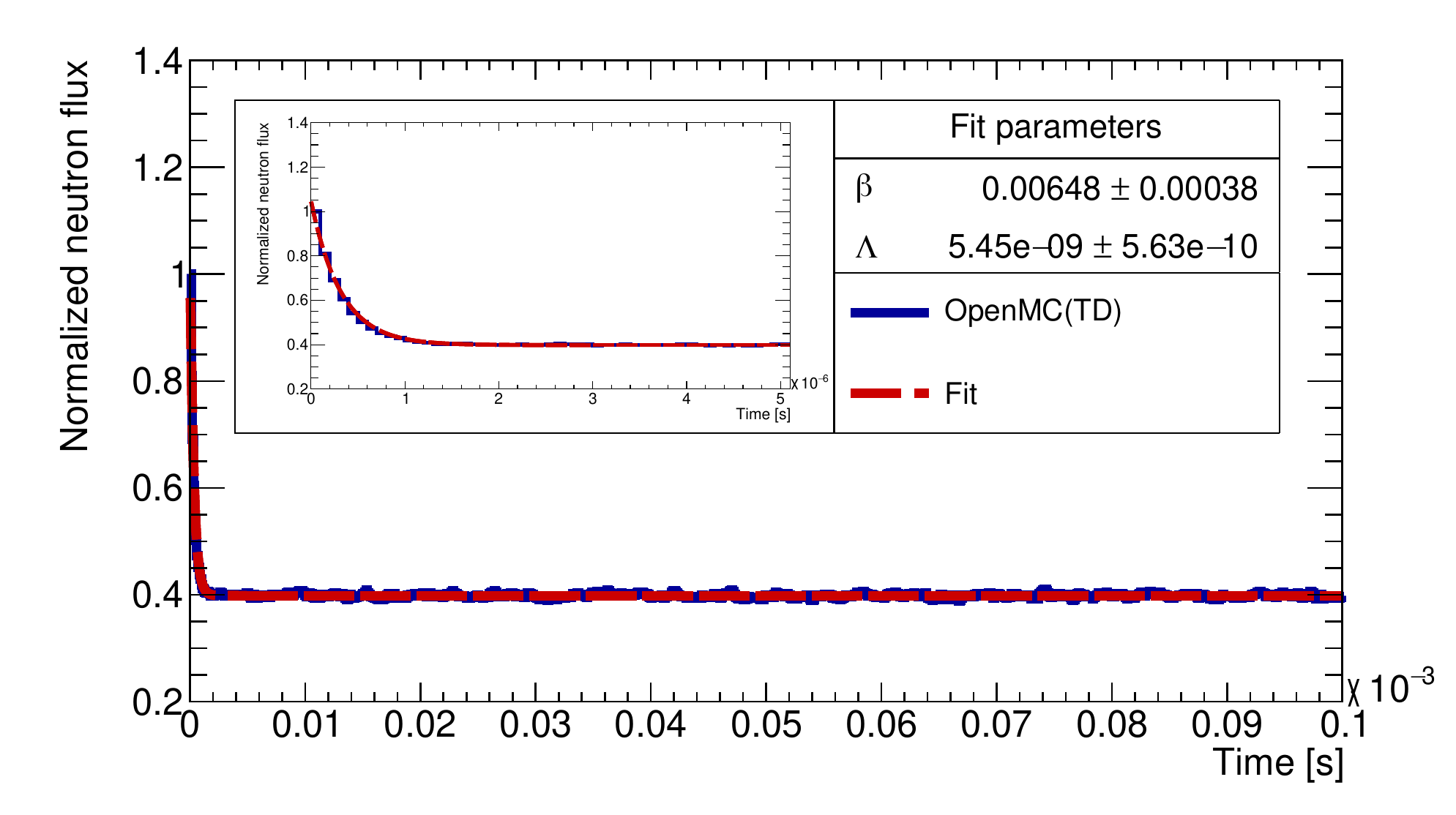}
	\caption{Study i). Time evolution of the neutron flux in the studied subcritical configuration. The initial transient source is prepared in a critical state and at the beginning of the transient Monte Carlo simulation using OpenMC(TD), the system is made subcritical by decreasing $\delta_{U235}$. The time evolution of the neutron flux is shown in blue, while the fit obtained is shown in red. Between $0 < t < 5 \; \mu\text{s}$ the prompt drop can be observed, and then the decay of the neutron population slows. Inset figure shows the prompt drop zoomed for the first $5$ $\mu$s.}
	\label{fig:real_subcritical_vacuum}
\end{figure}

\begin{sloppypar}
Results obtained from the transient Monte Carlo simulation using OpenMC(TD) are shown in blue in Fig.~\ref{fig:real_subcritical_vacuum}, while the fit obtained by adjusting the results to Eq~\eqref{eq:solution_neutron_1precursor} are shown in red. In Fig.~\ref{fig:real_subcritical_vacuum} the prompt drop can be seen for the first $5 \; \mu\text{s}$, and then for $t>5 \; \mu\text{s}$ the decay of the neutron flux stabilizes.
\end{sloppypar}

Quantitatively, the fitted effective neutron generation time obtained was $\Lambda_{\mathit{fitted}} \!=\! 5.45(56)$~ns. By comparison, MCNP obtained value was $\Lambda_{\mathit{MCNP}}\!=\!5.74(1)$~ns, giving a $\sim 5.1 \%$ difference between both quantities.
The fitted effective delayed neutron fraction obtained was $\beta_{\mathit{eff}}^{(\mathit{fitted})}\!=\!0.00648(38)$. By comparison, MCNP obtained value was $\beta_{\mathit{eff}}^{(\mathit{MCNP})}\!=\!0.00644(6)$. Table~\ref{table:sub_real_first_group_realE} shows a summary of results obtained in this section.

\begin{table}[h!]
\centering
\begin{tabular}{@{}ccccc@{}}
\toprule
\textbf{Parameter}  & \textbf{Calculated}   & \textbf{Fitted}        &   $\Delta$         \\ 
\textbf{Unit}  &     \textbf{MCNP}      &   \textbf{OpenMC}       &              \\ \midrule
$\Lambda$ [ns]  & $5.74(1)$ & $5.45(56)$ &   $5.1$$\%$ \\ 
$\beta_{\mathit{eff}}$ [pcm]  & $644(6)$ & $648(38)$ &   $1$$\%$ \\  \bottomrule
\end{tabular}
\caption{Study i). Values of the parameters obtained from running an OpenMC(TD) transient simulation in a subcritical configuration. The calculated values for the parameters were calculated using MCNP, while the OpenMC(TD) parameters were obtained by fitting Eq.~\eqref{eq:solution_neutron_1precursor} to the time evolution of the neutron flux.}
\label{table:sub_real_first_group_realE}
\end{table}

\subsubsection{ii) First group with average energy from JEFF-3.1.1}
\label{subsub:subcritical_first_group_energy_average}
The first precursor group was simulated, but in this case the delayed neutron was emitted with the average energy of the first group energy distribution reported from JEFF-$3$.$1$.$1$. This energy is $\bar{E}_{1g} \!=\! 212.31$~keV and it was calculated as the weighted average per eV from the distribution shown in Fig.~\ref{fig:jeff311_group1_spectrum_1}.

This simulation was run using $3$ batches and the total simulation time was $0.05$~ms divided in $500$ time intervals of $100$~ns each. Population control (see Sec.~\ref{subsection:pop_control}) was applied at the end of each interval. 

\begin{figure}[ht!]
	\centering
	\includegraphics[width=\textwidth]{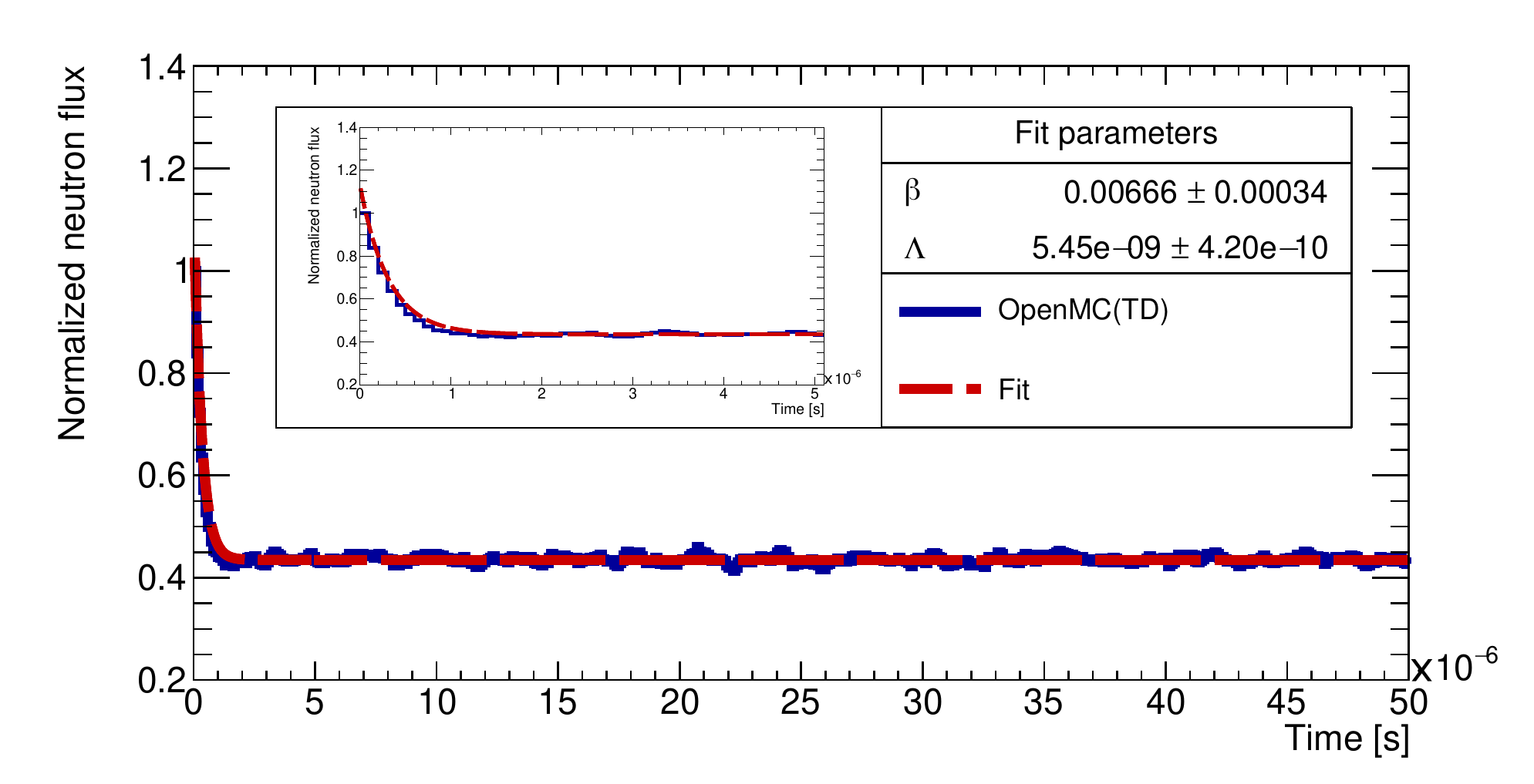}
	\caption{Study ii). Time evolution of the neutron flux in the studied subcritical configuration. The initial transient source is prepared in a critical state and at the beginning of the transient Monte Carlo simulation using OpenMC(TD), the system is made subcritical by decreasing $\delta_{U235}$. The time evolution of the neutron flux is shown in blue, while the fit obtained is shown in red. Between $0 < t < 5 \; \mu\text{s}$ the prompt drop can be observed, and then the decay of the neutron population slows. Inset figure shows the prompt drop zoomed for the first $5$ $\mu$s.}
	\label{fig:subcritical_ii}
\end{figure}

\begin{sloppypar}
Results obtained from the transient Monte Carlo simulation using OpenMC(TD) are shown in blue in Fig.~\ref{fig:subcritical_ii}, while the fit obtained by adjusting the results to Eq~\eqref{eq:solution_neutron_1precursor} are shown in red. In Fig.~\ref{fig:subcritical_ii} the prompt drop can be seen for the first $5 \; \mu\text{s}$, and then for $t>5 \; \mu\text{s}$ the decay of the neutron flux stabilizes.
\end{sloppypar}

Quantitatively, the fitted effective neutron generation time obtained was $\Lambda_{\mathit{fitted}} \!=\! 5.45(42)$~ns. By comparison, MCNP obtained value was $\Lambda_{\mathit{MCNP}}\!=\!5.74(1)$~ns, giving a $\sim 5.1 \%$ difference between both quantities. The fitted effective delayed neutron fraction obtained was $\beta_{\mathit{eff}}^{(\mathit{fitted})}\!=\!0.00666(34)$. MCNP obtained value was $\beta_{\mathit{eff}}^{(\mathit{MCNP})}\!=\!0.00644(6)$. Table~\ref{table:subcritical_ii} shows a summary of results obtained in this section.
\vspace{1cm}
\begin{table}[h!]
	\centering
	\begin{tabular}{@{}ccccc@{}}
		\toprule
		\textbf{Parameter}  & \textbf{Calculated}   & \textbf{Fitted}        &   $\Delta$         \\ 
		\textbf{Unit}  &     \textbf{MCNP}      &   \textbf{OpenMC}       &              \\ \midrule
		$\Lambda$ [ns]  & $5.74(1)$ & $5.45(42)$ &   $5.1$$\%$ \\ 
		$\beta_{\mathit{eff}}$ [pcm]  & $644(6)$ & $666(34)$ &   $3.4$$\%$ \\  \bottomrule
	\end{tabular}
	\caption{Study ii). Values of the parameters obtained from running an OpenMC(TD) transient simulation in a subcritical configuration. The calculated values for the parameters were calculated using MCNP, while the OpenMC(TD) parameters were obtained by fitting Eq.~\eqref{eq:solution_neutron_1precursor} to the time evolution of the neutron flux.}
	\label{table:subcritical_ii}
\end{table}

\subsubsection{iii) $1$-group with average energy from ENDF-B/VIII.0}
\label{subsub:subcritical_one_group_energy_average}

A $1$-group precursor structure was simulated. Delayed neutrons were emitted with the average energy of the $6$-group precursor structure from ENDF-B/VIII.$0$. This energy is $\bar{E}_{6g} = 501.31$~keV and was calculated as the weighted average of the reported average energies per group, according to
\begin{equation}
	\bar{E}_{6g} = \sum_{i=1}^{6} \frac{\beta_i}{\beta} \bar{E}_i,
	\label{eq:E6g}
\end{equation}
where $E_i$ is the average energy for $i$-th group, see Table~\ref{table:6groups_precursors}.

This simulation was run using $3$ batches and the total simulation time was $0.05$~ms divided in $500$ time intervals of $100$~ns each. Population control (see Sec.~\ref{subsection:pop_control}) was applied at the end of each interval. 

\begin{figure}[h!]
	\centering
	\includegraphics[width=\textwidth]{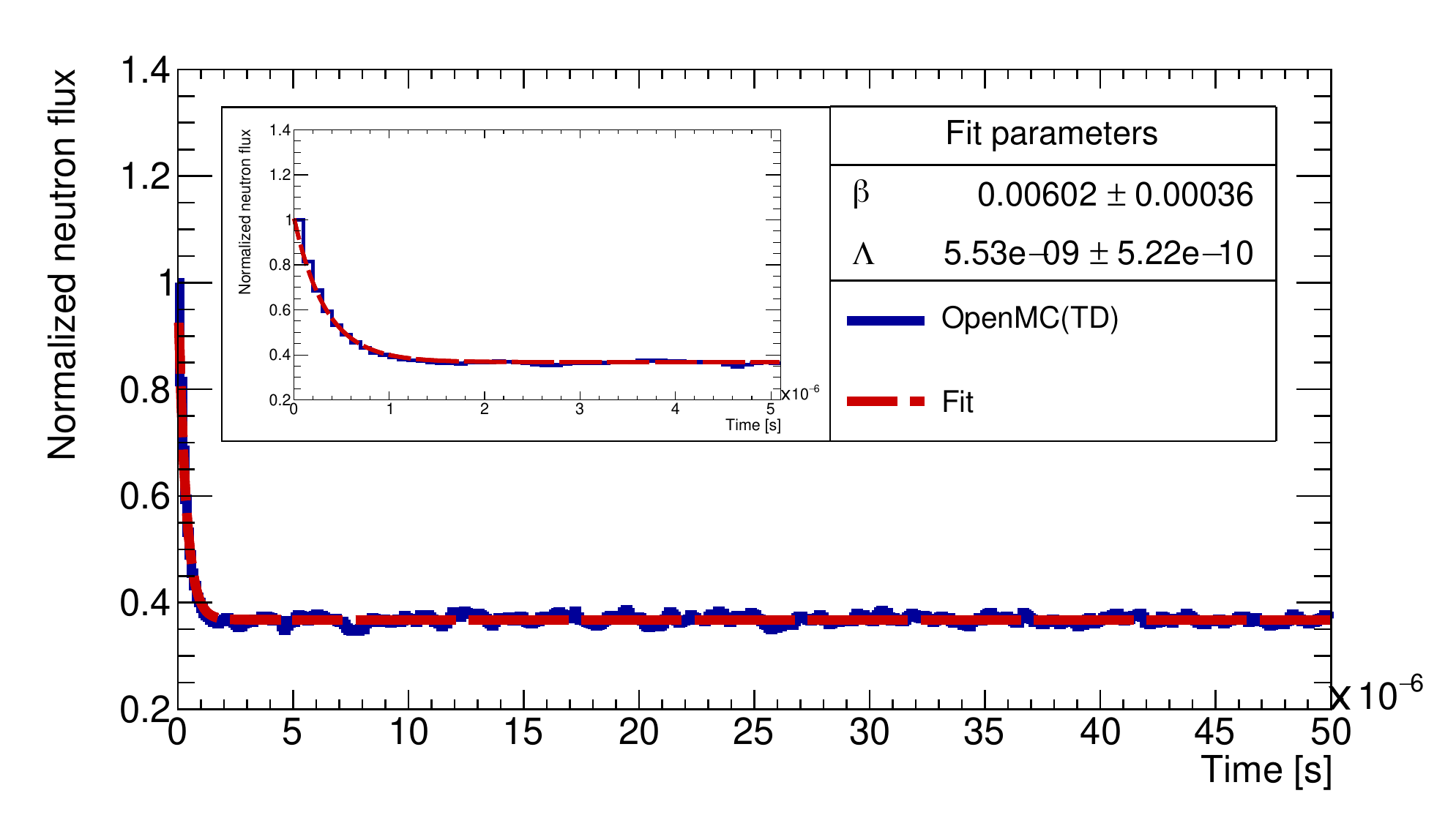}
	\caption{Study iii). Time evolution of the neutron flux in the studied subcritical configuration. The initial transient source is prepared in a critical state and at the beginning of the transient Monte Carlo simulation using OpenMC(TD), the system is made subcritical by decreasing $\delta_{U235}$. The time evolution of the neutron flux is shown in blue, while the fit obtained is shown in red. Between $0 < t < 5 \; \mu\text{s}$ the prompt drop can be observed, and then the decay of the neutron population slows. Inset figure shows the prompt drop zoomed for the first $5$ $\mu$s.}
	\label{fig:subcritical_iii}
\end{figure}

\begin{sloppypar}
Results obtained from the transient Monte Carlo simulation using OpenMC(TD) are shown in blue in Fig.~\ref{fig:subcritical_iii}, while the fit obtained by adjusting the results to Eq~\eqref{eq:solution_neutron_1precursor} are shown in red. In Fig.~\ref{fig:subcritical_iii} the prompt drop can be seen for the first $5 \; \mu\text{s}$, and then for $t>5 \; \mu\text{s}$ the decay of the neutron flux stabilizes.
\end{sloppypar}

Quantitatively, the fitted effective neutron generation time obtained was $\Lambda_{\mathit{fitted}} \!=\! 5.53(52)$~ns. By comparison, MCNP obtained value was $\Lambda_{\mathit{MCNP}}\!=\!5.74(1)$~ns, giving a $\sim 3.7 \%$ difference between both quantities.
The fitted effective delayed neutron fraction obtained was $\beta_{\mathit{eff}}^{(\mathit{fitted})}\!=\!0.00602(36)$. By comparison, MCNP obtained value was $\beta_{\mathit{eff}}^{(\mathit{MCNP})}\!=\!0.00644(6)$. Table~\ref{table:subcritical_iii} shows a summary of results obtained in this section.

\begin{table}[h]
	\centering
	\begin{tabular}{@{}ccccc@{}}
		\toprule
		\textbf{Parameter}  & \textbf{Calculated}   & \textbf{Fitted}        &   $\Delta$         \\ 
		\textbf{Unit}  &     \textbf{MCNP}      &   \textbf{OpenMC}       &              \\ \midrule
		$\Lambda$ [ns]  & $5.74(1)$ & $5.53(52)$ &   $3.7$$\%$ \\ 
		$\beta_{\mathit{eff}}$ [pcm]  & $644(6)$ & $602(36)$ &   $6.5$$\%$ \\  \bottomrule
	\end{tabular}
	\caption{Study iii). Values of the parameters obtained from running an OpenMC(TD) transient simulation in a subcritical configuration. The calculated values for the parameters were calculated using MCNP, while the OpenMC(TD) parameters were obtained by fitting Eq.~\eqref{eq:solution_neutron_1precursor} to the time evolution of the neutron flux.}
	\label{table:subcritical_iii}
\end{table}

\subsubsection{iv) $8$-group with energy distribution from JEFF-3.1.1}
\label{subsub:subcritical_8_group_energy_average}
An $8$-group precursor structure was simulated. Delayed neutrons energies were randomly sampled from one of the energy distributions from JEFF-$3$.$1$.$1$, shown in Appendix~\ref{app:group_spectra}.

This simulation was run using $3$ batches and the total simulation time was $0.05$~ms divided in $500$ time intervals of $100$~ns each. Population control (see Sec.~\ref{subsection:pop_control}) was applied at the end of each interval. 

\begin{figure}[ht!]
	\centering
	\includegraphics[width=\textwidth]{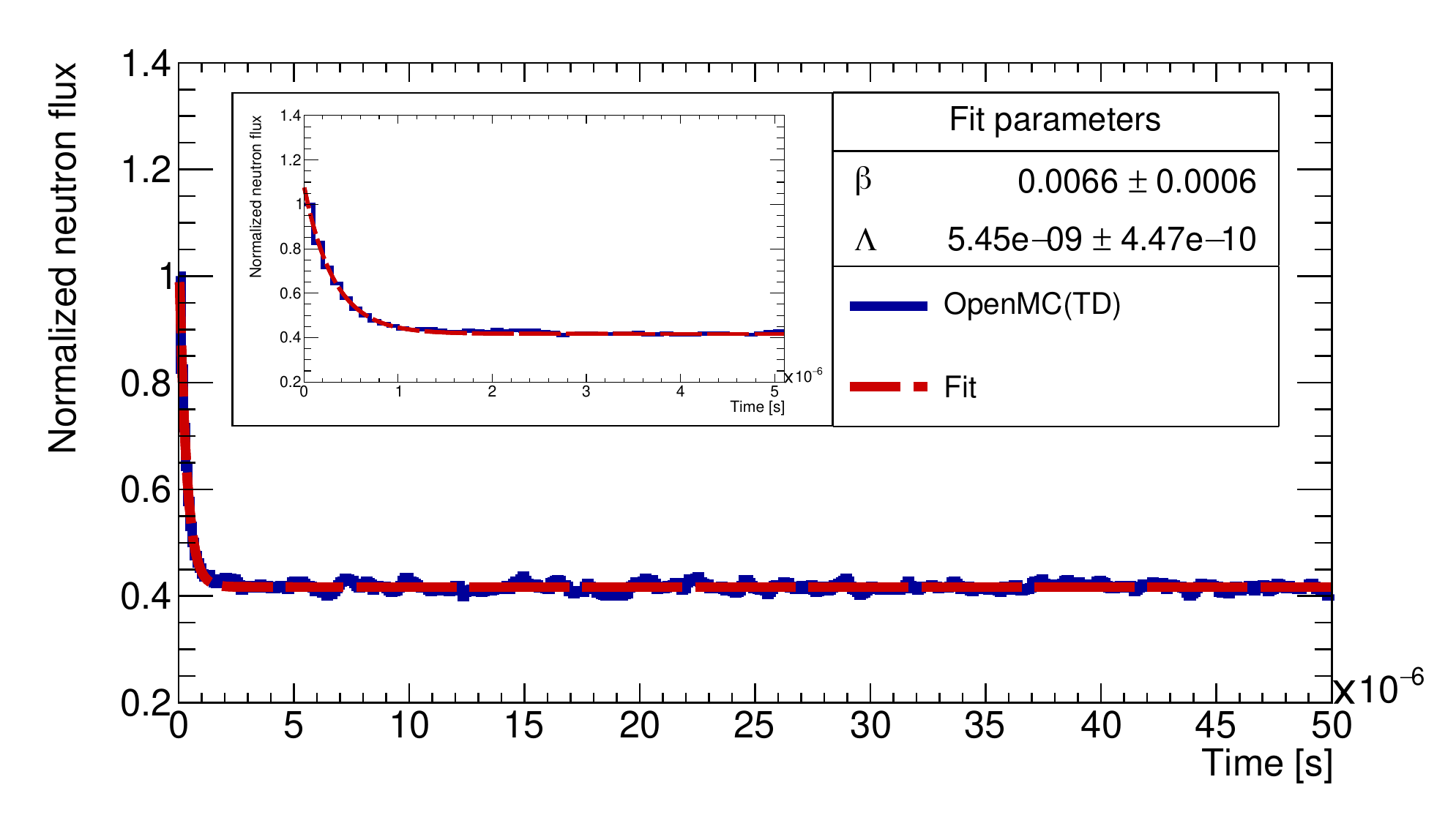}
	\caption{Study iv). Time evolution of the neutron flux in the studied subcritical configuration. The initial transient source is prepared in a critical state and at the beginning of the transient Monte Carlo simulation using OpenMC(TD), the system is made subcritical by decreasing $\delta_{U235}$. The time evolution of the neutron flux is shown in blue, while the fit obtained is shown in red. Between $0 < t < 5 \; \mu\text{s}$ the prompt drop can be observed, and then the decay of the neutron population slows. Inset figure shows the prompt drop zoomed for the first $5$ $\mu$s.}
	\label{fig:subcritical_iv}
\end{figure}

\begin{sloppypar}
Results obtained from the transient Monte Carlo simulation using OpenMC(TD) are shown in blue in Fig.~\ref{fig:subcritical_iv}, while the fit obtained by adjusting the results to Eq~\eqref{eq:solution_neutron_1precursor} are shown in red. In Fig.~\ref{fig:subcritical_iv} the prompt drop can be seen for the first $5 \; \mu\text{s}$, and then for $t>5 \; \mu\text{s}$ the decay of the neutron flux stabilizes.
\end{sloppypar}

Quantitatively, the fitted effective neutron generation time obtained was $\Lambda_{\mathit{fitted}} \!=\! 5.45(45)$~ns. By comparison, MCNP obtained value was $\Lambda_{\mathit{MCNP}}\!=\!5.74(1)$~ns, giving a $\sim 5.1 \%$ difference between both quantities.
The fitted effective delayed neutron fraction obtained was $\beta_{\mathit{eff}}^{(\mathit{fitted})}\!=\!0.00660(60)$. By comparison, MCNP obtained value was $\beta_{\mathit{eff}}^{(\mathit{MCNP})}\!=\!0.00644(6)$. Table~\ref{table:subcritical_iv} shows a summary of results obtained in this section.

\begin{table}[h!]
	\centering
	\begin{tabular}{@{}ccccc@{}}
		\toprule
		\textbf{Parameter}  & \textbf{Calculated}   & \textbf{Fitted}        &   $\Delta$         \\ 
		\textbf{Unit}  &     \textbf{MCNP}      &   \textbf{OpenMC}       &              \\ \midrule
		$\Lambda$ [ns]  & $5.74(1)$ & $5.53(52)$ &   $5.1$$\%$ \\ 
		$\beta_{\mathit{eff}}$ [pcm]  & $644(6)$ & $660(60)$ &   $2.5$$\%$ \\  \bottomrule
	\end{tabular}
	\caption{Study iv). Values of the parameters obtained from running an OpenMC(TD) transient simulation in a subcritical configuration. The calculated values for the parameters were calculated using MCNP, while the OpenMC(TD) parameters were obtained by fitting Eq.~\eqref{eq:solution_neutron_1precursor} to the time evolution of the neutron flux.}
	\label{table:subcritical_iv}
\end{table}

\subsubsection{v) $6$-group with average energy from ENDF-B/VIII.0}
\label{subsub:subcritical_6_group_energy_average}
A $6$-group structure was simulated. Delayed neutrons energies were randomly sampled according to $\beta_i / \beta$, from the listed average energies of the six precursor groups (see Table~\ref{table:6groups_precursors}).

This simulation was run using $3$ batches and the total simulation time was $0.05$~ms divided in $500$ time intervals of $100$~ns each. Population control (see Sec.~\ref{subsection:pop_control}) was applied at the end of each interval. 

\begin{figure}[ht!]
	\centering
	\includegraphics[width=\textwidth]{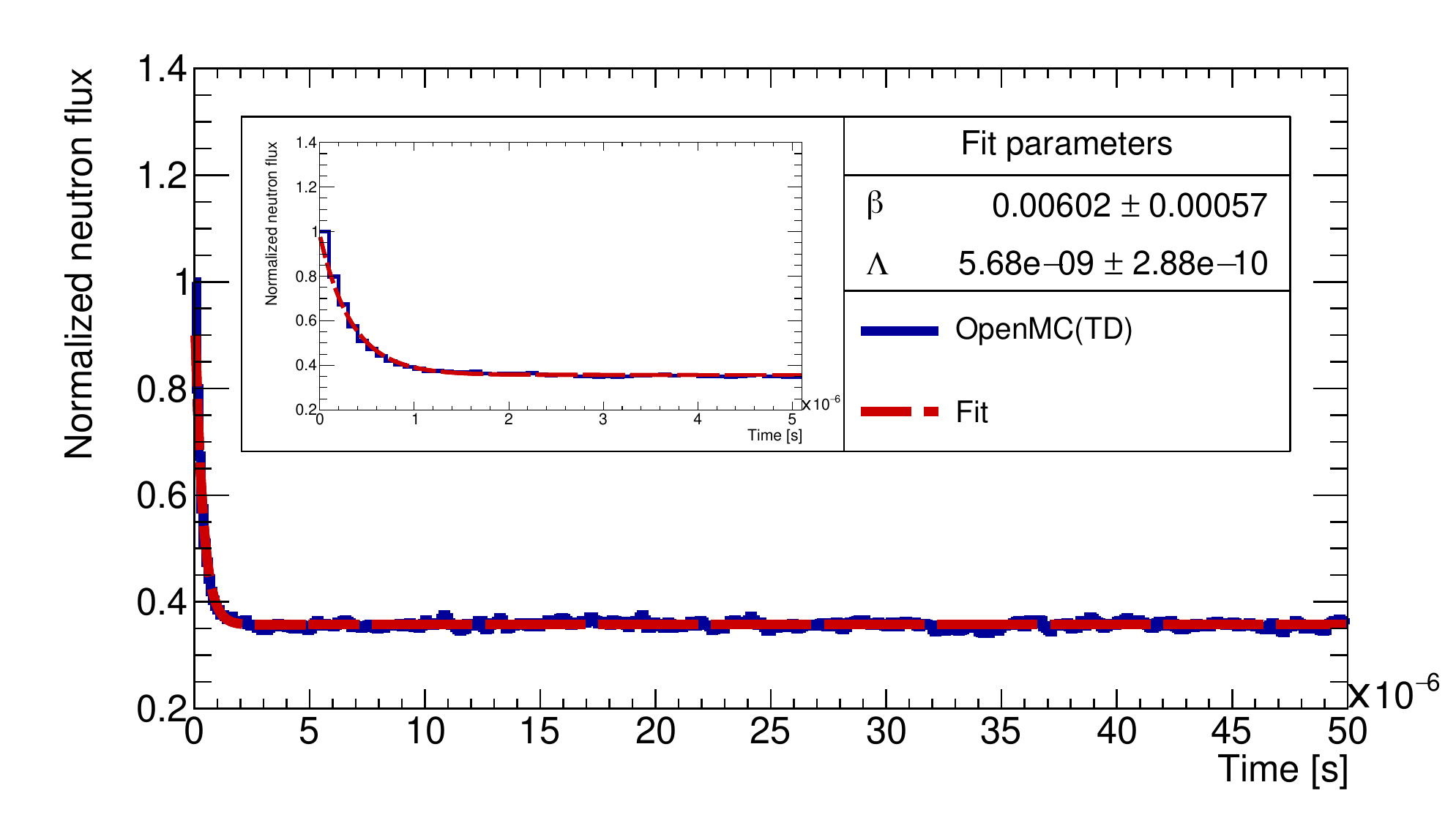}
	\caption{Study v). Time evolution of the neutron flux in the studied subcritical configuration. The initial transient source is prepared in a critical state and at the beginning of the transient Monte Carlo simulation using OpenMC(TD), the system is made subcritical by decreasing $\delta_{U235}$. The time evolution of the neutron flux is shown in blue, while the fit obtained is shown in red. Between $0 < t < 5 \; \mu\text{s}$ the prompt drop can be observed, and then the decay of the neutron population slows. Inset figure shows the prompt drop zoomed for the first $5$ $\mu$s.}
	\label{fig:subcritical_v}
\end{figure}

\begin{sloppypar}
Results obtained from the transient Monte Carlo simulation using OpenMC(TD) are shown in blue in Fig.~\ref{fig:subcritical_v}, while the fit obtained by adjusting the results to Eq~\eqref{eq:solution_neutron_1precursor} are shown in red. In Fig.~\ref{fig:subcritical_v} the prompt drop can be seen for the first $5 \; \mu\text{s}$, and then for $t>5 \; \mu\text{s}$ the decay of the neutron flux stabilizes.
\end{sloppypar}

Quantitatively, the fitted effective neutron generation time obtained was $\Lambda_{\mathit{fitted}} \!=\! 5.68(29)$~ns. By comparison, MCNP obtained value was $\Lambda_{\mathit{MCNP}}\!=\!5.74(1)$~ns, giving a $\sim 1 \%$ difference between both quantities.
The fitted effective delayed neutron fraction obtained was $\beta_{\mathit{eff}}^{(\mathit{fitted})}\!=\!0.00602(57)$. By comparison, MCNP obtained value was $\beta_{\mathit{eff}}^{(\mathit{MCNP})}\!=\!0.00644(6)$. Table~\ref{table:subcritical_v} shows a summary of results obtained in this section.

\begin{table}[h!]
	\centering
	\begin{tabular}{@{}ccccc@{}}
		\toprule
		\textbf{Parameter}  & \textbf{Calculated}   & \textbf{Fitted}        &   $\Delta$         \\ 
		\textbf{Unit}  &     \textbf{MCNP}      &   \textbf{OpenMC}       &              \\ \midrule
		$\Lambda$ [ns]  & $5.74(1)$ & $5.68(29)$ &   $1$$\%$ \\ 
		$\beta_{\mathit{eff}}$ [pcm]  & $644(6)$ & $602(57)$ &   $6.5$$\%$ \\  \bottomrule
	\end{tabular}
	\caption{Study v). Values of the parameters obtained from running an OpenMC(TD) transient simulation in a subcritical configuration. The calculated values for the parameters were calculated using MCNP, while the OpenMC(TD) parameters were obtained by fitting Eq.~\eqref{eq:solution_neutron_1precursor} to the time evolution of the neutron flux.}
	\label{table:subcritical_v}
\end{table}

\subsubsection{vi) $50$ individual precursors with average energies from ENDF-B/VIII.0}
\label{subsub:subcritical_50_individual_energy_average}

A $50$ individual precursor structure was simulated. Delayed neutrons were randomly sampled according to its importances $I_i\!\!=\!\!{(CY_i \; P_{n,i})}/{\nu_d}$, from the calculated average energies of the $50$ individual precursors used in this work (see Table~\ref{table:50-individual_precursor_structure}).

This simulation was run using $3$ batches and the total simulation time was $0.05$~ms divided in $500$ time intervals of $100$~ns each. Population control (see Sec.~\ref{subsection:pop_control}) was applied at the end of each interval. 

\begin{figure}[ht!]
	\centering
	\includegraphics[width=\textwidth]{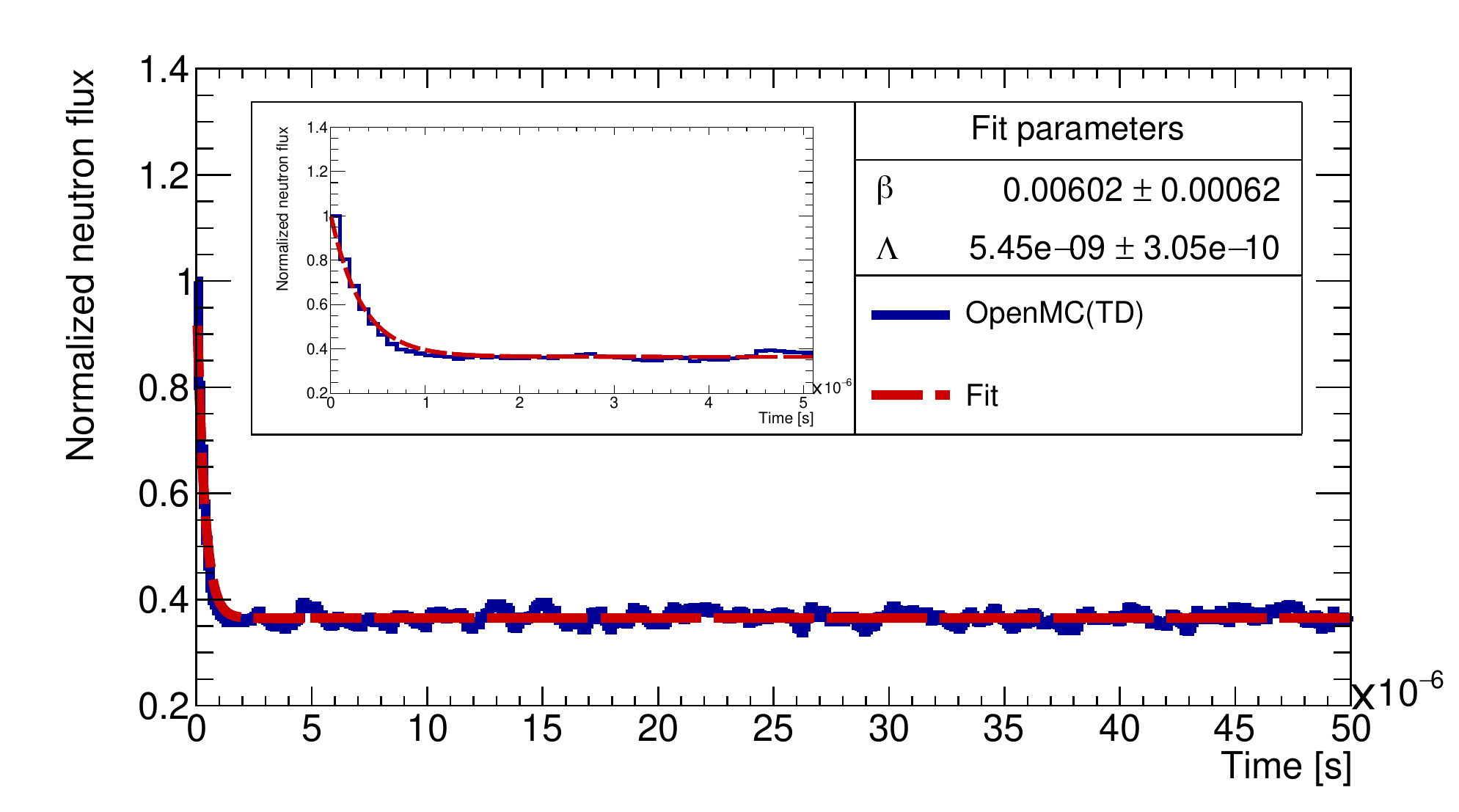}
	\caption{Study vi). Time evolution of the neutron flux in the studied subcritical configuration. The initial transient source is prepared in a critical state and at the beginning of the transient Monte Carlo simulation using OpenMC(TD), the system is made subcritical by decreasing $\delta_{U235}$. The time evolution of the neutron flux is shown in blue, while the fit obtained is shown in red. Between $0 < t < 5 \; \mu\text{s}$ the prompt drop can be observed, and then the decay of the neutron population slows. Inset figure shows the prompt drop zoomed for the first $5$ $\mu$s.}
	\label{fig:subcritical_vi}
\end{figure}

\begin{sloppypar}
Results obtained from the transient Monte Carlo simulation using OpenMC(TD) are shown in blue in Fig.~\ref{fig:subcritical_vi}, while the fit obtained by adjusting the results to Eq~\eqref{eq:solution_neutron_1precursor} are shown in red. In Fig.~\ref{fig:subcritical_vi} the prompt drop can be seen for the first $5 \; \mu\text{s}$, and then for $t>5 \; \mu\text{s}$ the decay of the neutron flux stabilizes.
\end{sloppypar}

\newpage
Quantitatively, the fitted effective neutron generation time obtained was $\Lambda_{\mathit{fitted}} \!=\! 5.45(31)$~ns. By comparison, MCNP obtained value was $\Lambda_{\mathit{MCNP}}\!=\!5.74(1)$~ns, giving a $\sim 5.3 \%$ difference between both quantities.
The fitted effective delayed neutron fraction obtained was $\beta_{\mathit{eff}}^{(\mathit{fitted})}\!=\!0.00602(62)$. By comparison, MCNP obtained value was $\beta_{\mathit{eff}}^{(\mathit{MCNP})}\!=\!0.00644(6)$. Table~\ref{table:subcritical_vi} shows a summary of results obtained in this section.

\begin{table}[h!]
	\centering
	\begin{tabular}{@{}ccccc@{}}
		\toprule
		\textbf{Parameter}  & \textbf{Calculated}   & \textbf{Fitted}        &   $\Delta$         \\ 
		\textbf{Unit}  &     \textbf{MCNP}      &   \textbf{OpenMC}       &              \\ \midrule
		$\Lambda$ [ns]  & $5.74(1)$ & $5.45(31)$ &   $5.3$$\%$ \\ 
		$\beta_{\mathit{eff}}$ [pcm]  & $644(6)$ & $602(57)$ &   $6.5$$\%$ \\  \bottomrule
	\end{tabular}
	\caption{Study vi). Values of the parameters obtained from running an OpenMC(TD) transient simulation in a subcritical configuration. The calculated values for the parameters were calculated using MCNP, while the OpenMC(TD) parameters were obtained by fitting Eq.~\eqref{eq:solution_neutron_1precursor} to the time evolution of the neutron flux.}
	\label{table:subcritical_vi}
\end{table}

It is important to notice that in this fast, unmoderated system, the prompt drop can be studied in detail given that short time step of $100$~ns are used in the simulation. This new Monte Carlo capability was shown to work in a monoenergetic system and here it is shown that also works in a continuous energy dependent system. Lastly, OpenMC(TD) code also correctly predicts the behavior of the system after the prompt drop, when the neutron flux changes slowly.  

\subsection{Supercritical configuration}
\label{subsection:energy_u235_vacuum_super}
Now a supercritical configuration was studied. The system was made supercritical by increasing the ${}^{235}$U density from $\delta_{U235}\!=\!4.496 \times 10^{-2}$~(atoms/b cm) to $\delta_{U235}\!=\!4.511 \times 10^{-2}$~(atoms/b cm), while mantaining the dimensions of the box constant, making the effective multiplication factor of the system $k_{\mathit{eff}}\!=\!1.00271 \pm 0.00003$. 

\subsubsection{i) First group with energy distribution from JEFF-3.1.1}
\label{subsub:supercritical_first_group_energy_real}

The first precursor group, characterized by a half-life $T_{1/2}\!=\!55.6$~s, was simulated. The delayed neutron energy was sampled from its neutron energy distribution, reported from JEFF-$3$.$1$.$1$. Group $1$ $\beta$-delayed neutron energy spectrum from JEFF-$3$.$1$.$1$. is shown in Fig.~\ref{fig:jeff311_group1_spectrum_1}, and the remaining delayed neutron group spectra can be found in Appendix~\ref{app:group_spectra}.

The simulation was run using $10$ batches and the total simulation time was $0.1$~ms divided in $1000$ time intervals of $100$~ns each. Population control (see Sec.~\ref{subsection:pop_control}) was applied at the end of each interval. 

\begin{figure}[h!]
	\centering
	\includegraphics[width=\textwidth]{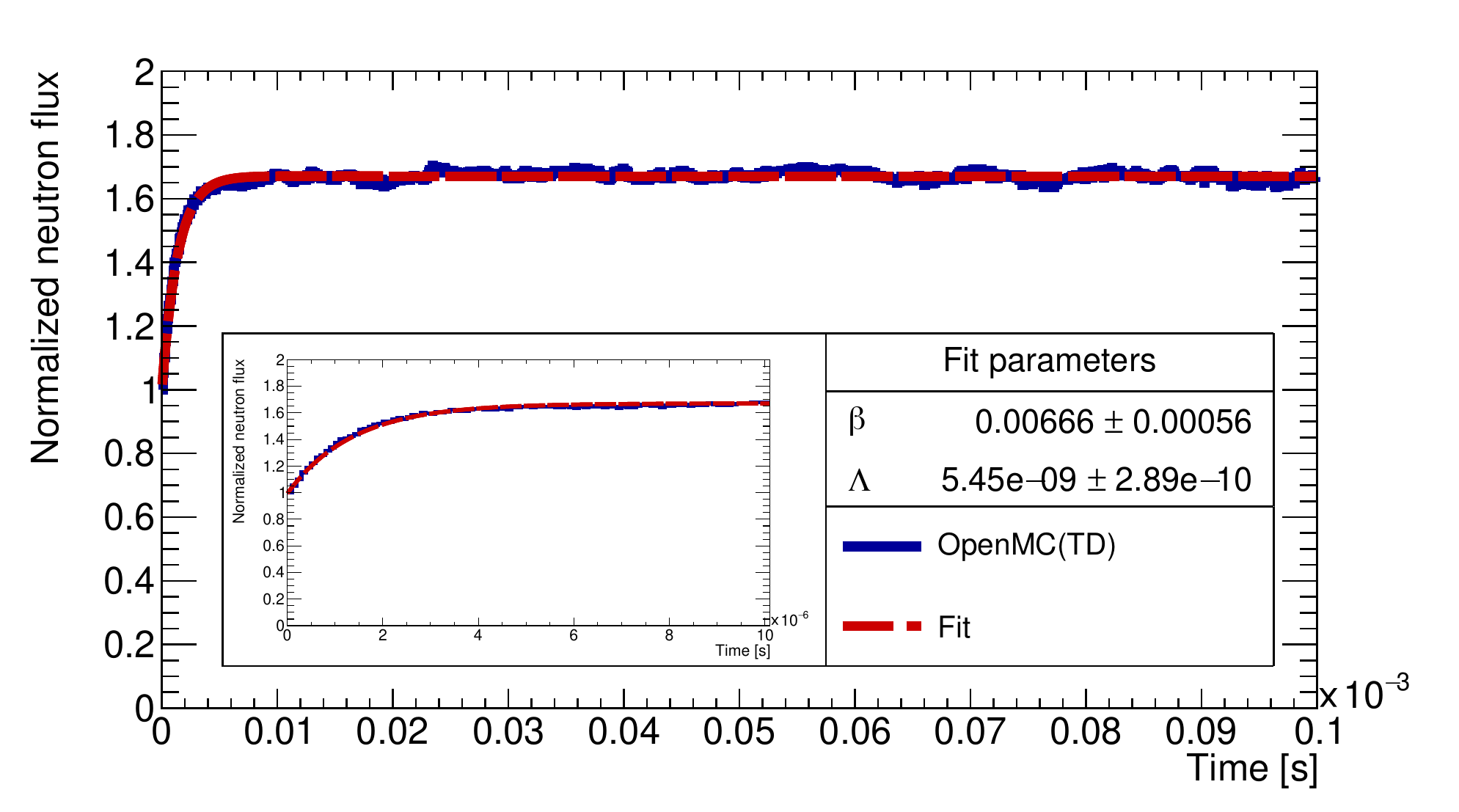}
	\caption{Study i). Time evolution of the neutron flux in the studied supercritical configuration. The initial transient source is prepared in a critical state and at the beginning of the transient Monte Carlo simulation using OpenMC(TD) the configuration is made supercritical by increasing $\delta_{U235}$. The time evolution of the neutron flux is shown in blue, while the fit obtained is shown in red. Between $0 < t < 1 \; \mu\text{s}$ the prompt jump can be observed, an then the growth of the neutron population slows. Inset figure shows the prompt jump zoomed for the first $10$ $\mu$s.}
	\label{fig:supercritical_vacuum_i}
\end{figure}

\begin{sloppypar}
Results obtained from the transient Monte Carlo simulation using OpenMC(TD) are shown in blue in Fig.~\ref{fig:supercritical_vacuum_i}, while the fit obtained by adjusting the results to Eq~\eqref{eq:solution_neutron_1precursor} are shown in red. In Fig.~\ref{fig:supercritical_vacuum_i} the prompt jump can be seen for the first $1 \; \mu\text{s}$, and then for $t>1 \; \mu\text{s}$ the growth of the neutron flux stabilizes.
\end{sloppypar}

Quantitatively, the fitted effective neutron generation time obtained was $\Lambda_{\mathit{fitted}}\!=\!5.45(57)$~ns. By comparison, MCNP obtained value was $\Lambda_{\mathit{MCNP}}\!=\!6.00(1)$~ns, giving $\sim\!9.2 \%$ difference between both quantities.
The fitted effective delayed neutron fraction obtained was $\beta_{\mathit{eff}}^{(\mathit{fitted})}\!=\!0.00666(56)$. MCNP obtained value was $\beta_{\mathit{eff}}^{(\mathit{MCNP})}\!=\!0.00651(6)$. Table~\ref{table:supercritical_i} shows a summary of results obtained in this section.

 
\begin{table}[h!]
	\centering
	\begin{tabular}{@{}ccccc@{}}
	\toprule
	\textbf{Parameter}  & \textbf{Calculated}   & \textbf{Fitted}        &   $\Delta$         \\ 
	\textbf{Unit}  &     \textbf{MCNP}      &   \textbf{OpenMC}       &              \\ \midrule
	$\Lambda$ [ns]  & $6.00(1)$ & $5.45(57)$ &   $9.2$$\%$ \\ 
	$\beta_{\mathit{eff}}$ [pcm]  & $651(6)$ & $666(56)$ &   $<1$$\%$ \\  \bottomrule
	\end{tabular}
\caption{Study i). Values of the parameters obtained from running an OpenMC(TD) transient simulation in a supercritical configuration. The calculated values for the parameters were calculated using MCNP, while the OpenMC(TD) parameters were obtained by fitting point kinetics solution to the time evolution of the neutron flux.}
\label{table:supercritical_i}
\end{table}

\subsubsection{ii) First group with average energy from JEFF-3.1.1}
\label{subsub:supercritical_first_group_energy_average}
The first precursor group was simulated, but in this case the delayed neutron was emitted with the average energy of the first group energy distribution reported from JEFF-$3$.$1$.$1$. This energy is $\bar{E}_{1g} \!=\! 212.31$~keV and it was calculated as the weighted average per eV from the distribution shown in Fig.~\ref{fig:subcritical_ii}.

\begin{figure}[h!]
	\centering
	\includegraphics[width=\textwidth]{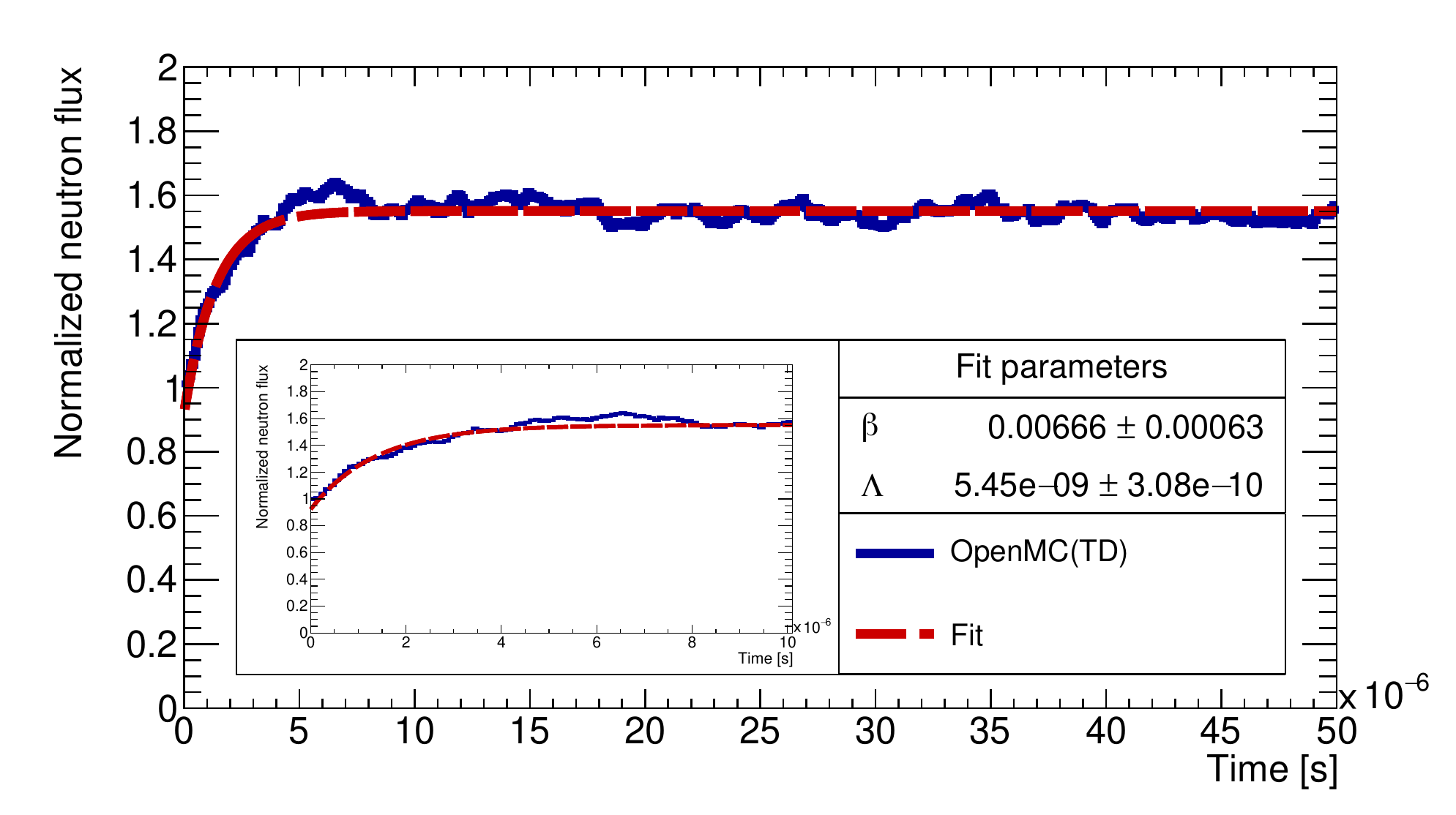}
	\caption{Study ii). Time evolution of the neutron flux in the studied supercritical configuration. The initial transient source is prepared in a critical state and at the beginning of the transient Monte Carlo simulation using OpenMC(TD) the configuration is made supercritical by increasing $\delta_{U235}$. The time evolution of the neutron flux is shown in blue, while the fit obtained is shown in red. Between $0 < t < 1 \; \mu\text{s}$ the prompt jump can be observed, an then the growth of the neutron population slows. Inset figure shows the prompt jump zoomed for the first $10$ $\mu$s.}
	\label{fig:supercritical_ii}
\end{figure}

This simulation was run using $3$ batches and the total simulation time was $0.05$~ms divided in $500$ time intervals of $100$~ns each. Population control (see Sec.~\ref{subsection:pop_control}) was applied at the end of each interval. 
\begin{sloppypar}
Results obtained from the transient Monte Carlo simulation using OpenMC(TD) are shown in blue in Fig.~\ref{fig:supercritical_ii}, while the fit obtained by adjusting the results to Eq~\eqref{eq:solution_neutron_1precursor} are shown in red. In Fig.~\ref{fig:supercritical_ii} the prompt jump can be seen for the first $1 \; \mu\text{s}$, and then for $t>1 \; \mu\text{s}$ the growth of the neutron flux stabilizes.
\end{sloppypar}

\newpage
Quantitatively, the fitted effective neutron generation time obtained was $\Lambda_{\mathit{fitted}} \!=\! 5.45(31)$~ns. By comparison, MCNP obtained value was $\Lambda_{\mathit{MCNP}}\!=\!6.00(1)$~ns, giving $\sim 9.2 \%$ difference between both quantities. The fitted effective delayed neutron fraction obtained was $\beta_{\mathit{eff}}^{(\mathit{fitted})}\!=\!0.00666(63)$. By comparison, MCNP obtained value was $\beta_{\mathit{eff}}^{(\mathit{MCNP})}\!=\!0.00651(6)$. Table~\ref{table:supercritical_ii} shows a summary of results obtained in this section.

\begin{table}[htbp]
	\centering
	\begin{tabular}{@{}ccccc@{}}
		\toprule
		\textbf{Parameter}  & \textbf{Calculated}   & \textbf{Fitted}        &   $\Delta$         \\ 
		\textbf{Unit}  &     \textbf{MCNP}      &   \textbf{OpenMC}       &              \\ \midrule
		$\Lambda$ [ns]  & $6.00(1)$ & $5.45(31)$ &   $9.2$$\%$ \\ 
		$\beta_{\mathit{eff}}$ [pcm]  & $651(6)$ & $666(63)$ &   $2.3$$\%$ \\  \bottomrule
	\end{tabular}
	\caption{Study ii). Values of the parameters obtained from running an OpenMC(TD) transient simulation in a supercritical configuration. The calculated values for the parameters were calculated using MCNP, while the OpenMC(TD) parameters were obtained by fitting point kinetics solution to the time evolution of the neutron flux.}
	\label{table:supercritical_ii}
\end{table}

\subsubsection{iii) $1$-group with average energy from ENDF-B/VIII.0}
\label{subsub:supercritical_one_group_energy_average}

A $1$-group precursor structure was simulated. Delayed neutrons were emitted with the average energy of the $6$-group precursor structure from ENDF-B/VIII.$0$. This energy is $\bar{E}_{6g} \!=\! 501.31$~keV and it was calculated as the weighted average of the reported average energies per group, according to
\begin{equation}
\bar{E}_{6g} = \sum_{i=1}^{6} \frac{\beta_i}{\beta} \bar{E}_i,
\label{eq:E6g2}
\end{equation}
where $E_i$ is the average energy for $i$-th group, see Table~\ref{table:6groups_precursors}.

This simulation was run using $3$ batches and the total simulation time was $0.05$~ms divided in $500$ time intervals of $100$~ns each. Population control (see Sec.~\ref{subsection:pop_control}) was applied at the end of each interval. 

\begin{figure}[h!]
	\centering
	\includegraphics[width=\textwidth]{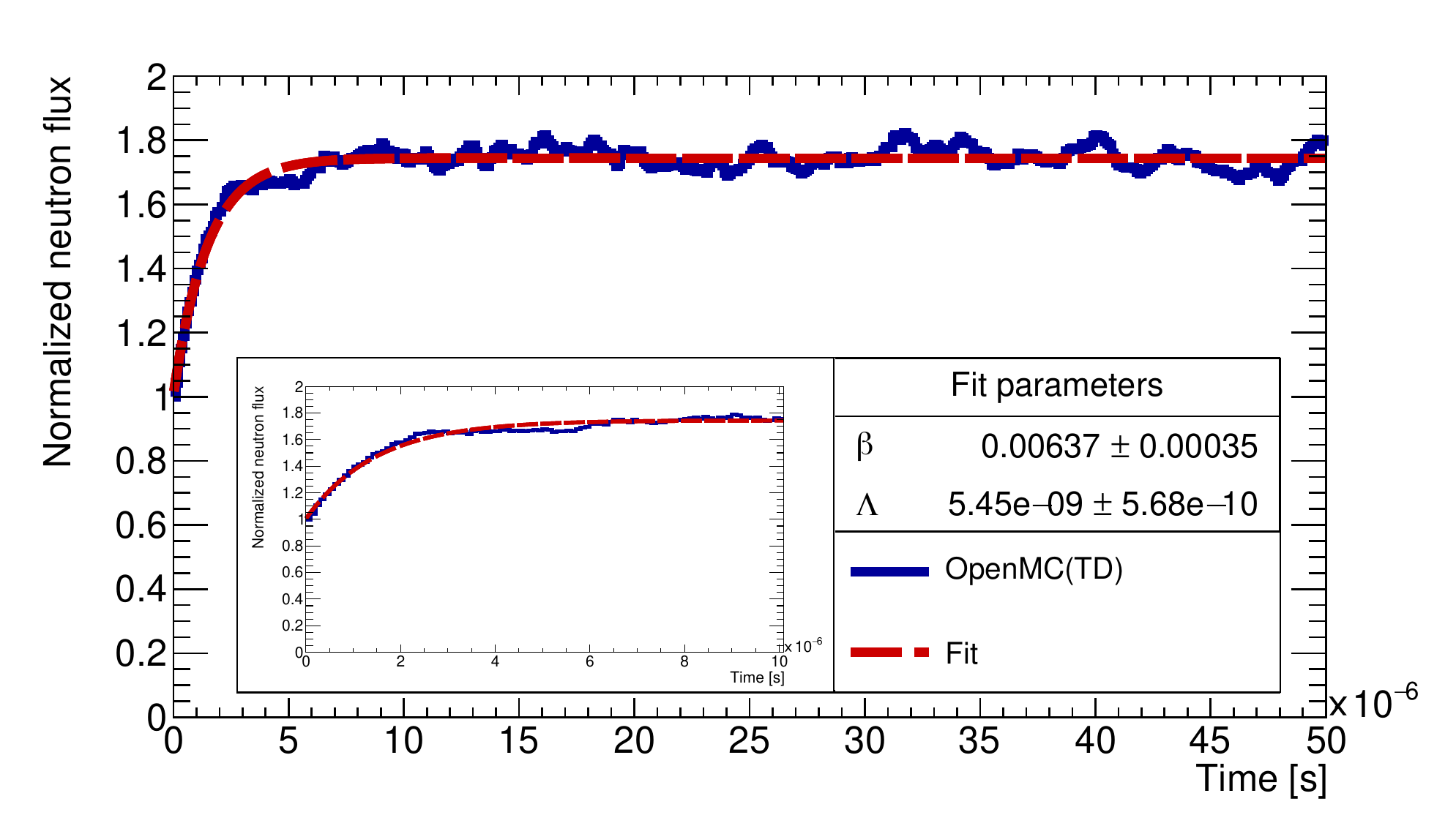}
	\caption{Study iii). Time evolution of the neutron flux in the studied supercritical configuration. The initial transient source is prepared in a critical state and at the beginning of the transient Monte Carlo simulation using OpenMC(TD) the configuration is made supercritical by increasing $\delta_{U235}$. The time evolution of the neutron flux is shown in blue, while the fit obtained is shown in red. Between $0 < t < 1 \; \mu\text{s}$ the prompt jump can be observed, an then the growth of the neutron population slows. Inset figure shows the prompt jump zoomed for the first $10$ $\mu$s.}
	\label{fig:supercritical_iii}
\end{figure}

\begin{sloppypar}
Results obtained from the transient Monte Carlo simulation using OpenMC(TD) are shown in blue in Fig.~\ref{fig:supercritical_iii}, while the fit obtained by adjusting the results to Eq~\eqref{eq:solution_neutron_1precursor} are shown in red. In Fig.~\ref{fig:supercritical_iii} the prompt jump can be seen for the first $1 \; \mu\text{s}$, and then for $t>1 \; \mu\text{s}$ the growth of the neutron flux stabilizes.
\end{sloppypar}

Quantitatively, the fitted effective neutron generation time obtained was $\Lambda_{\mathit{fitted}} \!=\! 5.45(57)$~ns. By comparison, MCNP obtained value was $\Lambda_{\mathit{MCNP}}\!=\!6.00(1)$~ns, giving $\sim 9.2 \%$ difference between both quantities.
The fitted effective delayed neutron fraction obtained was $\beta_{\mathit{eff}}^{(\mathit{fitted})}\!=\!0.00637(35)$. By comparison, MCNP obtained value was $\beta_{\mathit{eff}}^{(\mathit{MCNP})}\!=\!0.00651(6)$. Table~\ref{table:supercritical_iii} shows a summary of results obtained in this section.

\begin{table}[h!]
	\centering
	\begin{tabular}{@{}ccccc@{}}
		\toprule
		\textbf{Parameter}  & \textbf{Calculated}   & \textbf{Fitted}        &   $\Delta$         \\ 
		\textbf{Unit}  &     \textbf{MCNP}      &   \textbf{OpenMC}       &              \\ \midrule
		$\Lambda$ [ns]  & $6.00(1)$ & $5.45(57)$ &   $9.2$$\%$ \\ 
		$\beta_{\mathit{eff}}$ [pcm]  & $651(6)$ & $637(35)$ &   $2.2$$\%$ \\  \bottomrule
	\end{tabular}
	\caption{Study iii). Values of the parameters obtained from running an OpenMC(TD) transient simulation in a supercritical configuration. The calculated values for the parameters were calculated using MCNP, while the OpenMC(TD) parameters were obtained by fitting point kinetics solution to the time evolution of the neutron flux.}
	\label{table:supercritical_iii}
\end{table}
\newpage
\subsubsection{iv) $8$-group with energy distribution from JEFF-3.1.1}
\label{subsub:supercritical_8_group_energy_average}
An $8$-group precursor structure was simulated. Delayed neutrons energies were randomly sampled from one of the energy distributions from JEFF-$3$.$1$.$1$, shown in Appendix~\ref{app:group_spectra}.

\begin{figure}[ht]
	\centering
	\includegraphics[width=\textwidth]{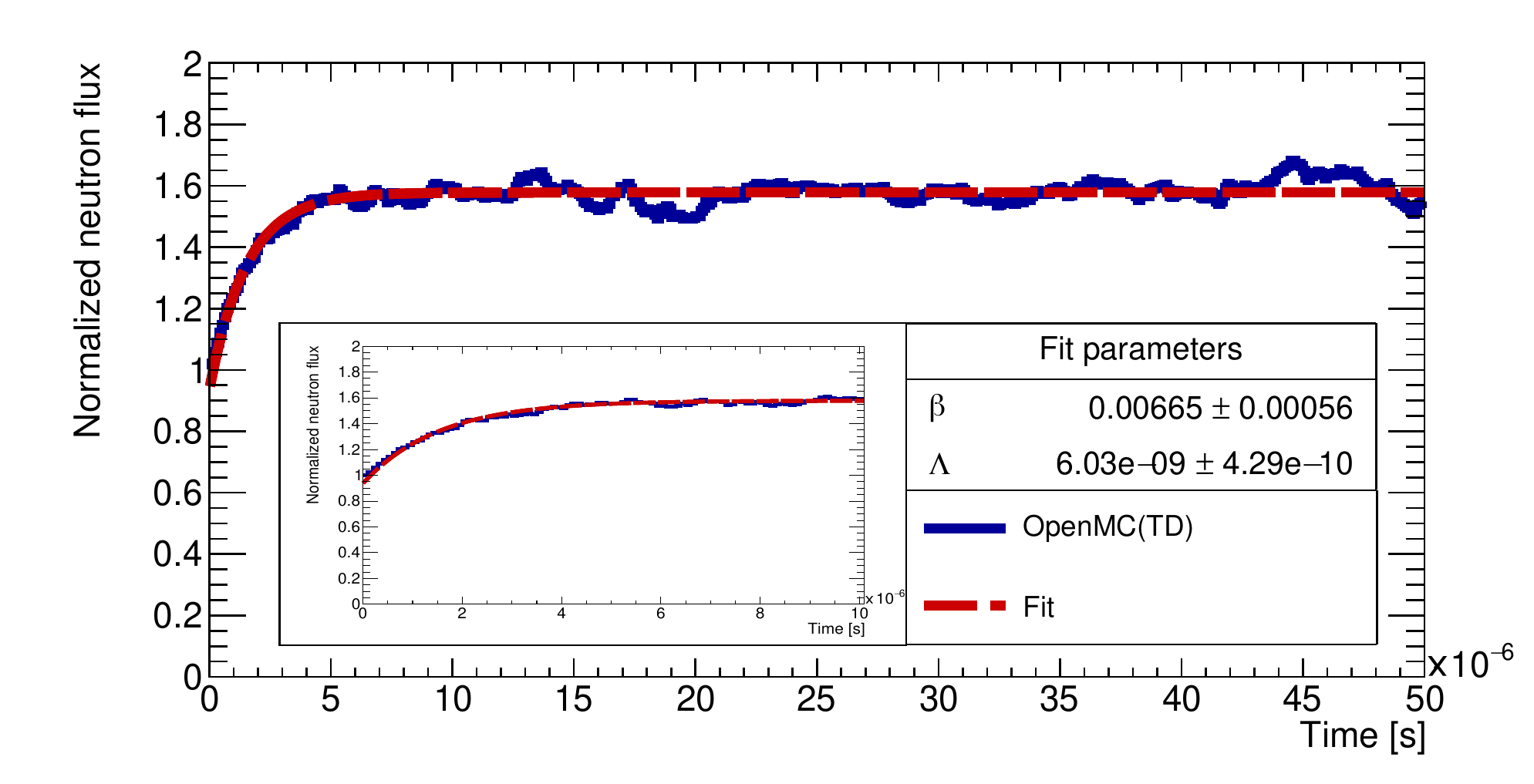}
	\caption{Study iv). Time evolution of the neutron flux in the studied supercritical configuration. The initial transient source is prepared in a critical state and at the beginning of the transient Monte Carlo simulation using OpenMC(TD) the configuration is made supercritical by increasing $\delta_{U235}$. The time evolution of the neutron flux is shown in blue, while the fit obtained is shown in red. Between $0 < t < 1 \; \mu\text{s}$ the prompt jump can be observed, an then the growth of the neutron population slows. Inset figure shows the prompt jump zoomed for the first $10$ $\mu$s.}
	\label{fig:supercritical_iv}
\end{figure}

This simulation was run using $3$ batches and the total simulation time was $0.05$~ms divided in $500$ time intervals of $100$~ns each. Population control (see Sec.~\ref{subsection:pop_control}) was applied at the end of each interval. 
\begin{sloppypar}
Results obtained from the transient Monte Carlo simulation using OpenMC(TD) are shown in blue in Fig.~\ref{fig:supercritical_iv}, while the fit obtained by adjusting the results to Eq~\eqref{eq:solution_neutron_1precursor} are shown in red. In Fig.~\ref{fig:supercritical_iv} the prompt jump can be seen for the first $1 \; \mu\text{s}$, and then for $t>1 \; \mu\text{s}$ the growth of the neutron flux stabilizes.
\end{sloppypar}

\newpage
Quantitatively, the fitted effective neutron generation time obtained was $\Lambda_{\mathit{fitted}} \!=\! 6.00(43)$~ns. By comparison, MCNP obtained value was $\Lambda_{\mathit{MCNP}}\!=\!6.00(1)$~ns, giving $\sim <1 \%$ difference between both quantities.
The fitted effective delayed neutron fraction obtained was $\beta_{\mathit{eff}}^{(\mathit{fitted})}\!=\!0.00665(56)$. By comparison, MCNP obtained value was $\beta_{\mathit{eff}}^{(\mathit{MCNP})}\!=\!0.00651(6)$. Table~\ref{table:supercritical_iv} shows a summary of results obtained in this section.

\begin{table}[h!]
	\centering
	\begin{tabular}{@{}ccccc@{}}
		\toprule
		\textbf{Parameter}  & \textbf{Calculated}   & \textbf{Fitted}        &   $\Delta$         \\ 
		\textbf{Unit}  &     \textbf{MCNP}      &   \textbf{OpenMC}       &              \\ \midrule
		$\Lambda$ [ns]  & $6.00(1)$ & $6.00(43)$ &   $<1$$\%$ \\ 
		$\beta_{\mathit{eff}}$ [pcm]  & $651(6)$ & $665(35)$ &   $2.2$$\%$ \\  \bottomrule
	\end{tabular}
	\caption{Study iv). Values of the parameters obtained from running an OpenMC(TD) transient simulation in a supercritical configuration. The calculated values for the parameters were calculated using MCNP, while the OpenMC(TD) parameters were obtained by fitting point kinetics solution to the time evolution of the neutron flux.}
	\label{table:supercritical_iv}
\end{table}
\newpage
\subsubsection{v) $6$-group with average energy from ENDF-B/VIII.0}
\label{subsub:supercritical_6_group_energy_average}
\begin{figure}[ht!]
	\centering
	\includegraphics[width=\textwidth]{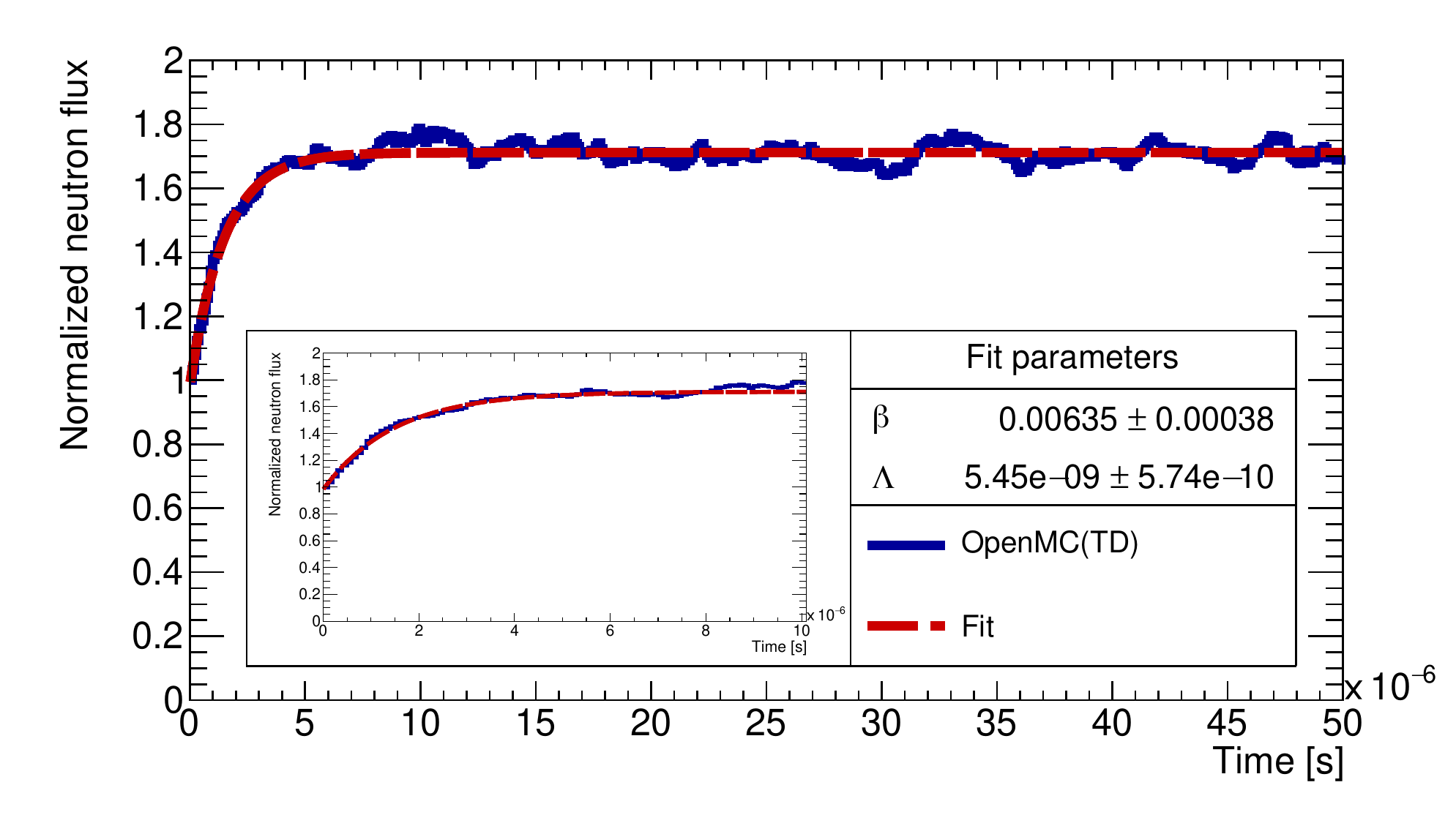}
	\caption{Study v). Time evolution of the neutron flux in the studied subcritical configuration. The initial transient source is prepared in a critical state and at the beginning of the transient Monte Carlo simulation using OpenMC(TD), the system is made subcritical by decreasing $\delta_{U235}$. The time evolution of the neutron flux is shown in blue, while the fit obtained is shown in red. Between $0 < t < 1 \; \mu\text{s}$ the prompt drop can be observed, and then the decay of the neutron population slows. Inset figure shows the prompt drop zoomed for the first $10$ $\mu$s.}
	\label{fig:supercritical_v}
\end{figure}

A $6$-group structure was simulated. Delayed neutrons energies were randomly sampled according to $\beta_i / \beta$, from the listed average energies of the six precursor groups (see Table~\ref{table:6groups_precursors}).
This simulation was run using $3$ batches and the total simulation time was $0.05$~ms divided in $500$ time intervals of $100$~ns each. Population control (see Sec.~\ref{subsection:pop_control}) was applied at the end of each interval. 
\begin{sloppypar}
Results obtained from the transient Monte Carlo simulation using OpenMC(TD) are shown in blue in Fig.~\ref{fig:supercritical_v}, while the fit obtained by adjusting the results to Eq~\eqref{eq:solution_neutron_1precursor} are shown in red. In Fig.~\ref{fig:supercritical_v} the prompt jump can be seen for the first $1 \; \mu\text{s}$, and then for $t>1 \; \mu\text{s}$ the growth of the neutron flux stabilizes.
\end{sloppypar}

\newpage
Quantitatively, the fitted effective neutron generation time obtained was $\Lambda_{\mathit{fitted}} \!=\! 5.45(57)$~ns. By comparison, MCNP obtained value was $\Lambda_{\mathit{MCNP}}\!=\!6.00(1)$~ns, giving $\sim 9.2 \%$ difference between both quantities.
The fitted effective delayed neutron fraction obtained was $\beta_{\mathit{eff}}^{(\mathit{fitted})}\!=\!0.00635(38)$. By comparison, MCNP obtained value was $\beta_{\mathit{eff}}^{(\mathit{MCNP})}\!=\!0.00651(6)$. Table~\ref{table:supercritical_v} shows a summary of results obtained in this section.
\newpage
\begin{table}[h!]
	\centering
	\begin{tabular}{@{}ccccc@{}}
		\toprule
		\textbf{Parameter}  & \textbf{Calculated}   & \textbf{Fitted}        &   $\Delta$         \\ 
		\textbf{Unit}  &     \textbf{MCNP}      &   \textbf{OpenMC}       &              \\ \midrule
		$\Lambda$ [ns]  & $6.00(1)$ & $5.45(57)$ &   $9.2$$\%$ \\ 
		$\beta_{\mathit{eff}}$ [pcm]  & $651(6)$ & $635(38)$ &   $2.5$$\%$ \\  \bottomrule
	\end{tabular}
	\caption{Study v). Values of the parameters obtained from running an OpenMC(TD) transient simulation in a supercritical configuration. The calculated values for the parameters were calculated using MCNP, while the OpenMC(TD) parameters were obtained by fitting point kinetics solution to the time evolution of the neutron flux.}
	\label{table:supercritical_v}
\end{table}

\newpage
\subsubsection{vi) $50$ individual precursors with average energies from ENDF-B/VIII.0}
\label{subsub:supercritical_50_individual_energy_average}

A $50$ individual precursor structure was simulated. Delayed neutrons were randomly sampled according to its importances $I_i\!=\!{(CY_i \; P_{n,i})}/{\nu_d}$, from the calculated average energies of the $50$ individual precursors used in this work (see Table~\ref{table:50-individual_precursor_structure}).

This simulation was run using $3$ batches and the total simulation time was $0.05$~ms divided in $500$ time intervals of $100$~ns each. Population control (see Sec.~\ref{subsection:pop_control}) was applied at the end of each interval. 

\begin{figure}[ht!]
	\centering
	\includegraphics[width=\textwidth]{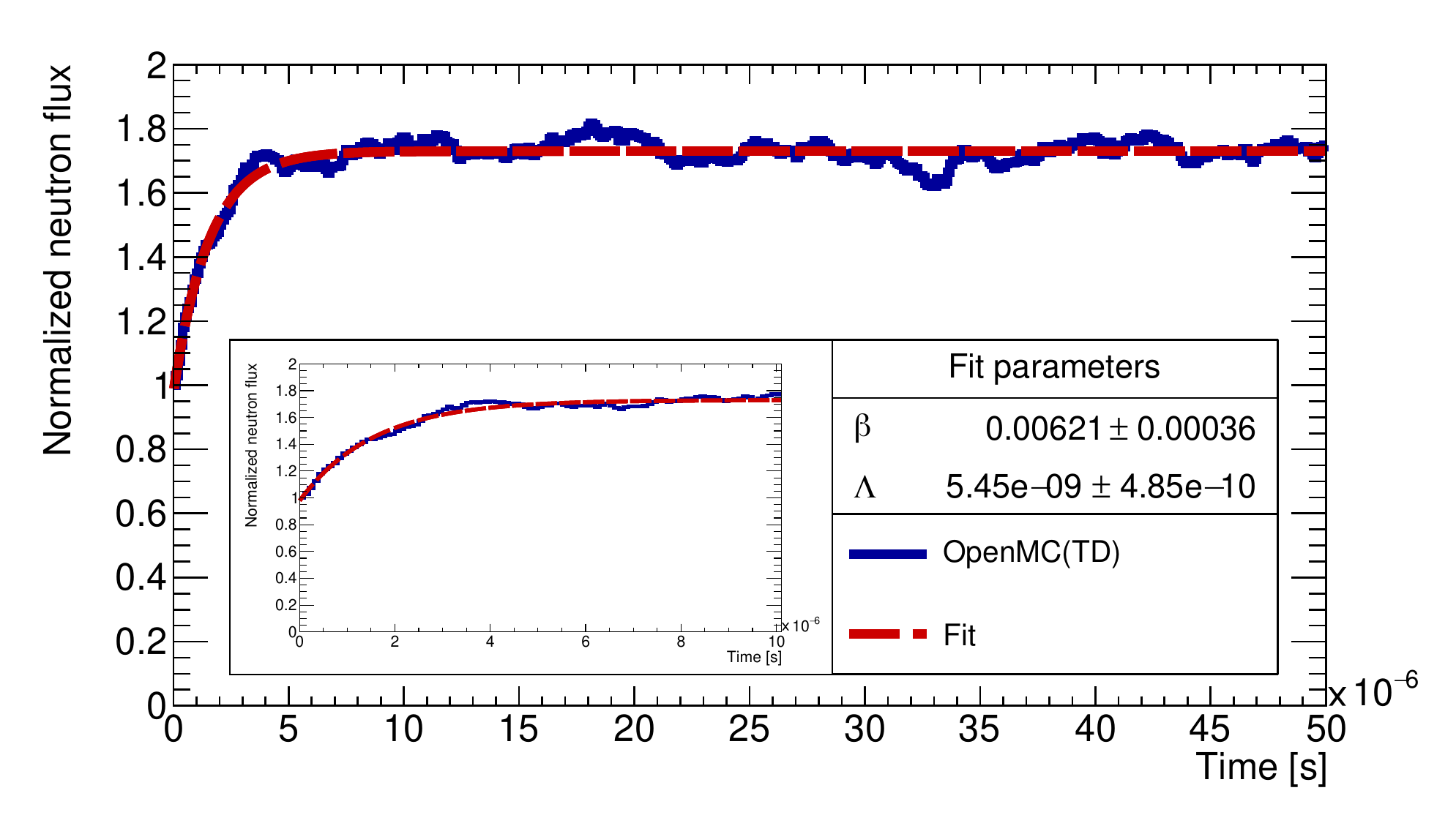}
	\caption{Study vi). Time evolution of the neutron flux in the studied subcritical configuration. The initial transient source is prepared in a critical state and at the beginning of the transient Monte Carlo simulation using OpenMC(TD), the system is made subcritical by decreasing $\delta_{U235}$. The time evolution of the neutron flux is shown in blue, while the fit obtained is shown in red. Between $0 < t < 1 \; \mu\text{s}$ the prompt drop can be observed, after which the decay of the neutron flux slows. Inset figure shows the prompt drop zoomed for the first $10$ $\mu$s.}
	\label{fig:supercritical_vi}
\end{figure}

\begin{sloppypar}
Results obtained from the transient Monte Carlo simulation using OpenMC(TD) are shown in blue in Fig.~\ref{fig:supercritical_v}, while the fit obtained by adjusting the results to Eq~\eqref{eq:solution_neutron_1precursor} are shown in red. In Fig.~\ref{fig:supercritical_vi} the prompt jump can be seen for the first $1 \; \mu\text{s}$, and then for $t>1 \; \mu\text{s}$ the growth of the neutron flux stabilizes.
\end{sloppypar}

Quantitatively, the fitted effective neutron generation time obtained was $\Lambda_{\mathit{fitted}} \!=\! 5.45(49)$~ns. By comparison, MCNP obtained value was $\Lambda_{\mathit{MCNP}}\!=\!6.00(1)$~ns, giving $\sim 9.2 \%$ difference between both quantities.
The fitted effective delayed neutron fraction was $\beta_{\mathit{eff}}^{(\mathit{fitted})}\!=\!0.00621(36)$. By comparison, the MCNP value was $\beta_{\mathit{eff}}^{(\mathit{MCNP})}\!=\!0.00651(6)$. Table~\ref{table:supercritical_vi} shows a summary of results obtained in this section.

\begin{table}[h!]
	\centering
	\begin{tabular}{@{}ccccc@{}}
		\toprule
		\textbf{Parameter}  & \textbf{Calculated}   & \textbf{Fitted}        &   $\Delta$         \\ 
		\textbf{Unit}  &     \textbf{MCNP}      &   \textbf{OpenMC}       &              \\ \midrule
		$\Lambda$ [ns]  & $6.00(1)$ & $5.45(49)$ &   $9.2$$\%$ \\ 
		$\beta_{\mathit{eff}}$ [pcm]  & $651(6)$ & $621(36)$ &   $4.6$$\%$ \\  \bottomrule
	\end{tabular}
	\caption{Study vi). Values of the parameters obtained from running an OpenMC(TD) transient simulation in a supercritical configuration. The calculated values for the parameters were obtained using MCNP, while the OpenMC(TD) parameters were obtained by fitting point kinetics solution to the time evolution of the neutron flux.}
	\label{table:supercritical_vi}
\end{table}

\newpage
As for the subcritical case, it is important to notice that in this fast, unmoderated system, the prompt jump can be studied in detail given that short time step of $0.1$~$\mu$s are used in the simulation. This new Monte Carlo capability was shown to work in a monoenergetic system and here it is shown that also works in a varying energy system. Lastly, the OpenMC(TD) code also correctly describes the behaviour of the system after the prompt jump, when the neutron flux changes slowly. 

In this section, values for: i) nuclear cross sections, ii) mean energies, iii) cumulative yields, iv) probability of delayed neutron emission and, v) decay constants were taken from nuclear databases JEFF-$3$.$1$.$1$ and ENDF-B/VIII.$0$. In this regard, JEFF-$3$.$1$.$1$ has reported the neutron energy spectra for each of the eight groups, but it does not have the delayed neutron energy spectra for the $269$ individual precursors. As it was seen in Section~\ref{subsection:nuclear_libraries}, there are discrepancies between both databases. In this work, it was necessary to use the cumulative yields from JEFF-$3$.$1$.$1$, and the probability of delayed neutron emission and average delayed neutron energy from ENDF-B/VIII.$0$. These discrepancies are finally reflected in the value obtained for the effective delayed neutron fraction ($\bar{\beta}_{\mathit{eff}} \!=\! 658(26)$~pcm for JEFF-$3$.$1$.$1$ and $\bar{\beta}_{\mathit{eff}}\!=\!602(29)$~pcm for ENDF-B/VIII.$0$, for the subcritical configuration; $\bar{\beta}_{\mathit{eff}} = 666(34)$~pcm for JEFF-$3$.$1$.$1$ and $\bar{\beta}_{\mathit{eff}}\!\!=\!\!631(21)$~pcm for ENDF-B/VIII.$0$, for the supercritical configuration). This shows how important is that every database counts with good and better nuclear data for individual precursors, either average energies or energy spectra.     

\section{Light-water moderated energy dependent system with individual precursor structure}
\label{section:energy_individual_precursors}
In this section the fast system studied in Section~\ref{section:energy_dependent} was modified by including a neutron moderator surrounding the ${}^{235}$U. The $\beta$-delayed neutron emission now was produced by individual precursors and results obtained were compared when emission is from the $6$-group precursor structure. 

Comparisons were made between simulations using the $6$-group structure and $50$ individual precursors, such as i) effective multiplication factor for a critical system (see Sec.~\ref{subsection:criticality_50precursors}), and ii) time evolution of the neutron flux in a transient simulation (see Sec.~\ref{subsection:critical_state_50precursors}). 

As a final test the $10$ most important\footnote{Importance was defined in Eq.~\eqref{eq:precursor_importance}, see Sec.~\ref{subsection:individual_precursor}.} precursors were removed from the $50$ individual precursors, in order to account for its effect on the time evolution of the neutron flux in comparison with the $6$-group structure (see Sec.~\ref{subsection:only_10_precursors}).  

The configuration was surrounded with a $4.29$~cm-thickness water moderator and made critical by setting the ${}^{235}$U density to $\delta_{U235}\!=\!3.2671 \times 10^{-2}$~(atoms/b cm). The continuous energy cross sections used were from JEFF-$3$.$1$.$1$~\cite{jeff311}. The dimensions of the box remained the same. Prior to the transient simulations presented in this section, a non-transient standard criticality calculation was run with $10^6$ neutrons, $5000$ batches and $300$ skipped cycles using OpenMC. The effective multiplication factor obtained was $k_{\mathit{eff}}\!=\!1.00025 \pm 0.00001$.

In this section, the code input were the density of the fissile material, which will be denoted as $\delta_{U235}$, the delayed neutron energy (sampled or averaged from spectra) and the number of precursors used ($50$ or $40$). Reactivity was inserted using the same method described in Sec.~\ref{section:energy_dependent}. The output values (observables) were the effective multiplication factor $k_{\mathit{eff}}$ and the time evolution of the neutron flux $\phi(t)$, like in the previous section. Since this is no longer a $1$-group precursor problem, there are no analytical solutions to the point kinetics equations. Nevertheless, resorting to the point kinetics approximation, a good estimation to the asymptotic decay constant for the neutron flux~\cite{KEEPIN1957IN2} can be found using the equation
\begin{equation}
	\alpha_D = \frac{\bar \lambda \rho}{\beta_{\mathit{eff}} - \rho},
	\label{eq:alphaD_asymptotic}
\end{equation}
where $\alpha_D$ is the asymptotic decay constant, $\bar \lambda$ is the average $\beta$-weighted decay constant\footnote{The average $\beta$-weighted decay constant is given by $\bar \lambda \!=\! \sum \frac{\beta_i}{\beta} \lambda_i$}, $\beta_{\mathit{eff}}$ is the effective delayed neutron fraction and $\rho$ is the system reactivity. Regarding the choice of the effective delayed neutron fraction, the average delayed neutron yield obtained when using the data from JEFF-$3$.$1$.$1$ in Eq.~\eqref{eq:nud_calculation} is $\nu_d \!=\! 1.48 \times 10^{-2}$, while the value obtained when using ENDF/B-VIII.$0$ is $\nu_d \!=\! 1.90 \times 10^{-2}$. If the average neutron yield is taken to be $\nu \!=\! 2.4355$~\cite{international2007iaea}, then the delayed neutron fraction, $\beta \!=\! {\nu_d}/{\nu}$, ranges from $\beta \!=\! 607$~pcm to $\beta\!=\!780$~pcm. In view of this, the value for the effective delayed neutron fraction was chosen to be $\beta_{\mathit{eff}}\!=\!700$~pcm.  The decay constant was $\bar{\lambda}\!=\!0.0784$~s${}^{-1}$. The reactivity of the system was obtained as fitted parameter, and then compared to the reactivity obtained from the initial non-transient criticality calculation ($\rho\!=\!(k_{\mathit{eff}}-1)/k_{\mathit{eff}}$).

\subsection{Criticality calculation using individual precursors}
\label{subsection:criticality_50precursors}
As it was shown in Section~\ref{section:mono-energetic_system} and Section~\ref{section:energy_dependent}, prior to every transient simulation with Monte Carlo code OpenMC(TD), a non-transient, standard steady state criticality calculation with OpenMC must be done in order to create the initial transient source, and assess the reactivity of the system. Since during the writing of this thesis there are no codes able to perform a criticality calculation using individual precursors as the source of $\beta$-delayed neutrons, in this work the capability to run criticality calculations using individual precursors instead of the $N$-group structure was also added to the OpenMC(TD) code. Criticality was achieved by mantaining ${}^{235}$U density at $\delta_{U235}\!=\!3.2671 \times 10^{-2}$~(atoms/b cm) obtaining $k_{\mathit{eff}}\!=\!1.00025(3)$ for the $6$-group precursor structure and $k_{\mathit{eff}}\!=\!1.00032(3)$ for the $50$-individual precursor structure.

Results obtained using the $6$-group structure and $50$ individual precursors are shown in Table~\ref{table:keff_groups}.

\begin{table}[h!]
	\centering
	\begin{tabular}{@{}ccccc@{}}
		\toprule
		            &       \textbf{6-groups}   & \textbf{$50$ precursors} &  \textbf{Difference}\\ \midrule
		$\boldsymbol{k_{\mathit{eff}}}$  & $1.00025(3)$ & $1.00032(3)$ &   $7(4)$ \\ \bottomrule
	\end{tabular}
	\caption{Effective multiplication factors obtained for the ${}^{235}$U cube when is thermalized by surrounding it with a water moderator of a $4.29$~cm thickness. Results for $6$-group structure and $50$ individual precursors.}
	\label{table:keff_groups}
\end{table}

The effective multiplication factor obtained for this system shows that this configuration is slightly supercritical and both results are in good agreement with each other, with a difference of $7$~pcm among them. 

\subsection{Critical configuration with 50 individual precursor structure}
\label{subsection:critical_state_50precursors}
A transient simulation using OpenMC(TD) was ran for the previous system (see Sec.~\ref{subsection:criticality_50precursors}) in a critical configuration comparing the time evolution of the neutron flux obtained when $6$-group and $50$ individual precursor structures were used. Both simulations were run using $2$ batches. Total simulation time was $4$~s divided in $400$~time intervals of $10$~ms each. Population control was applied at the end of each interval. The wall-clock time for the $6$-group precursor simulation was about $260.05$~h, while for the $50$ individual precursor simulation was $410.76$~h.  

\begin{figure}[h!]
	\centering
	\includegraphics[width=\textwidth]{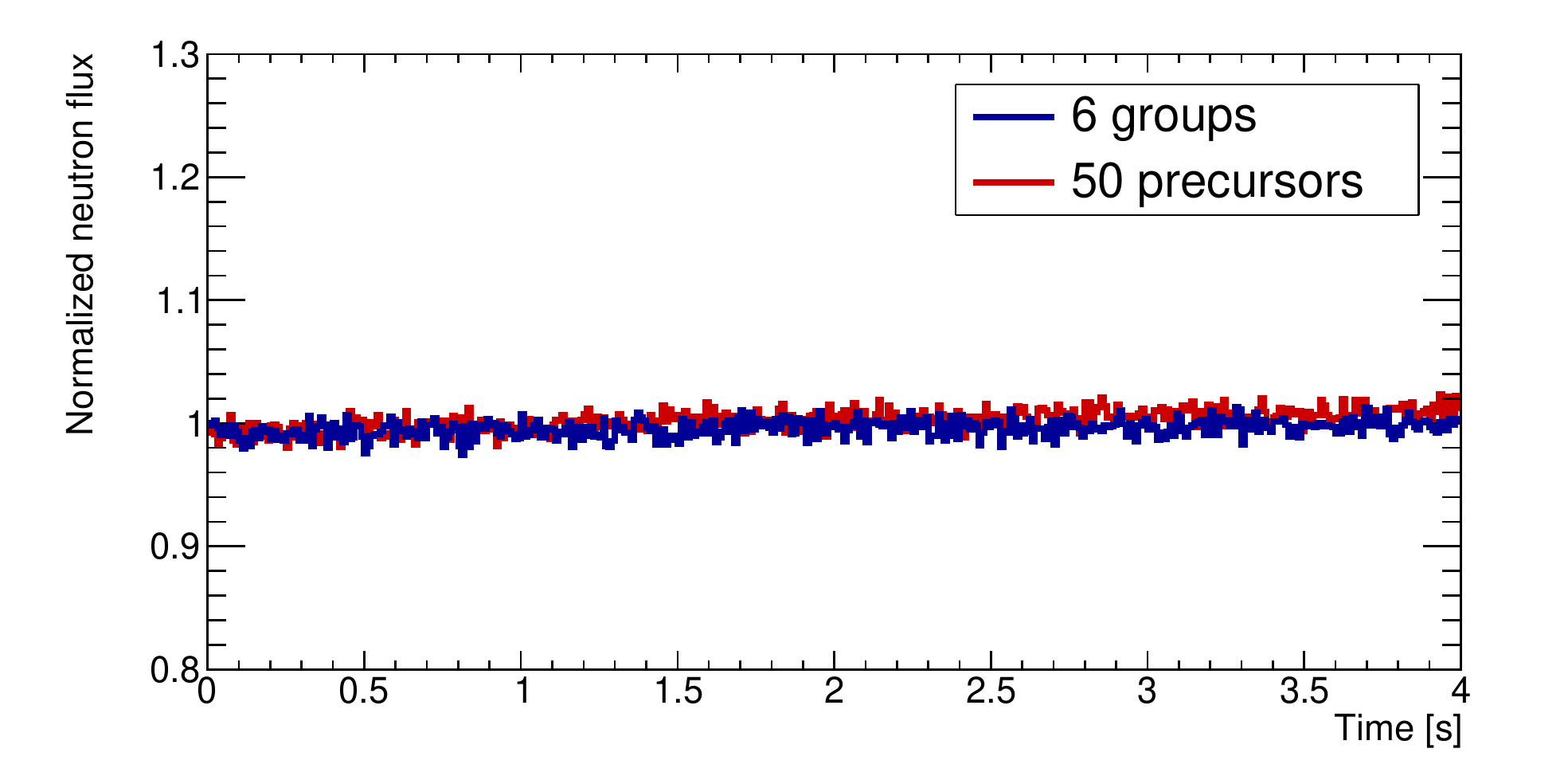}
	\caption{Time evolution of neutron flux in a water moderated box made of pure ${}^{235}$U, simulated with OpenMC. Results for $6$ groups are shown in blue, and results for $50$ individual precursors are shown in red. Both results show a slightly supercritical system, where neutron flux increases slowly in time, consistent with the $k_\mathit{eff}$ of a near critical configuration.}
	\label{fig:comparison_critical_6vs50_400steps_10E-2}
\end{figure}

Results obtained from transient Monte Carlo simulation using OpenMC(TD) for the time evolution of the neutron flux, for the $6$-group structure, are shown in blue in Fig.~\ref{fig:comparison_critical_6vs50_400steps_10E-2}, while results obtained for the $50$ individual precursor structure are shown in red. From Fig.~\ref{fig:comparison_critical_6vs50_400steps_10E-2} it can be seen qualitatively that both results show a slightly supercritical system, where the neutron flux increases slowly in time, which is consistent with the effective multiplication factor of a near critical configuration.
\newpage
Now, analyzing Fig.~\ref{fig:comparison_critical_6vs50_400steps_10E-2}, from a quantitative point of view, the reactivity value $\rho$ can be obtained as fitted parameter of $\phi(t) \sim e^{-\alpha_D t}$, for both $6$-group and $50$ individual precursor structure\footnote{The asymptotic decay constant $\alpha_D$ was defined in Eq.~\eqref{eq:alphaD_asymptotic}.}. These $\rho_{\mathit{fit}}^{(6)}\!=\!0.00017(368)$ and $\rho_{\mathit{fit}}^{(50)}\!=\!0.00036(347)$ fitted reactivity values were compared to the reactivity from the criticality calculation using OpenMC(TD), $\rho^{(6)}\!=\!(k_{\mathit{eff}}^{(6)}-1)/k_{\mathit{eff}}^{(6)}\!=\!0.00025(3)$ and $\rho^{(50)}\!=\!(k_{\mathit{eff}}^{(50)}-1)/k_{\mathit{eff}}^{(50)} \!=\!0.00032(3)$.
The reactivity value is usually obtained by running a criticality Monte Carlo calculation. In this work, using OpenMC(TD), this value can be obtained by fitting  $\phi(t) \sim e^{-\alpha_D t}$ to the time evolution of the neutron population.  A summary of the results obtained can be seen in Table~\ref{table:results_6vs50}.

\begin{table}[h!]
	\centering
	\begin{tabular}{@{}ccc@{}}
		\toprule
		 { }                 & \textbf{$6$-group}   & \textbf{$50$ individual}  \\ 
		 { }                 & \textbf{structure} & \textbf{structure}      \\ \midrule
		 $\boldsymbol{\rho}$ [pcm]  &    $25(3)$         &   $32(3)$               \\
		 $\boldsymbol{\rho_{\mathit{fit}}}$ [pcm] &  $17(368)$  &  $35(347)$  \\  \bottomrule
	\end{tabular}
	\caption{Results obtained for the reactivity of the water moderated energy dependent simulated system in a critical configuration using $6$-group and $50$ individual precursor structure.}
	\label{table:results_6vs50}
\end{table}

From examining the results obtained for the fitted parameters, it can be noticed that even when they are in good agreement with the calculated values, they possess quite large uncertainties. This is due to the fact that neutron population for $k_{\mathit{eff}}>1$ ($\rho>0$) shows an exponential growing behaviour. Simulations ran for $4$~s which was not enough time to reveal the exponential growing. To reduce the fitted uncertainties, the simulation time should increase to tens of seconds. This would increase the wall-clock time of the simulation. For instance, a simulation time of $50$~s, would take $216$ days ($7$ months and $6$ days), beyond reach for the purposes of this thesis.

Nevertheless, to further explore these uncertainties issues related to the simulation time, the fitted reactivities from the subcritical (where the neutron population was scored for $50$~s, see Sec.~\ref{subsection:mono_subcritical_state}) and critical (where neutron population was scored for $25$~s, see Sec.~\ref{subsection:mono_steady_state}) configurations of the monoenergetic fissile system from Sec.~\ref{section:mono-energetic_system}, were obtained by taking into account the time evolution of neutron population only for the first $4$~s. 

For the monoenergetic system with a subcritical configuration (see Sec.~\ref{subsection:mono_subcritical_state}), the previously fitted reactivity was $\rho_{\mathit{fit}}^{(50s)}\!=\!-0.01193(626)$, while the calculated value was $\rho\!=\!-0.01193(3)$. When taking into account the neutron population decay for only the first $4$~s, the obtained reactivity was $\rho_{\mathit{fit}}^{(4s)}\!=\!-0.01159(28906)$. Meanwhile $\rho_{\mathit{fit}}^{(50s)}$ and $\rho$ are in excellent agreement with each other, $\rho_{\mathit{fit}}^{(4s)}$ shows a difference of $2.85 \%$ with $\rho$. But the uncertainty of the calculated reactivity is $0.25 \%$, while the uncertainty of $\rho_{\mathit{fit}}^{(4s)}$ is almost $25$ times its value.  

For the monoenergetic system with critical configuration (see Sec.~\ref{subsection:mono_steady_state}), the previously fitted reactivity was $\rho_{\mathit{fit}}^{(25s)}\!=\!0.00013(70)$, while the calculated value was $\rho\!=\!0.00010(3)$. When taking into account the neutron population decay for only the first $4$~s, the obtained reactivity was $\rho_{\mathit{fit}}^{(4s)}\!=\!0.00009(597)$. In this case, the uncertainty of $\rho_{\mathit{fit}}^{(4s)}$ is almost $66$ times its value. 

It is important to remark that OpenMC(TD) is stable for both configurations and the rate of change for the neutron flux is consistent with the reactivities from the criticality calculations, considering that: i) continuous energy-dependent cross sections, ii) addition of a neutron moderator to the system and, iii) implementation of individual precursors.   

In summary, the time evolution of neutron flux in a water moderated box of ${}^{235}$U was obtained using $50$ individual precursors and the results were consistent with the initial calculated reactivities. The simulation was stable, and there was no divergence of the neutron fission chains, which means that population control worked as expected. 

\subsection{Critical configuration without the 10 most important precursors}
\label{subsection:only_10_precursors}
The final study was a critical configuration, but in this case the $10$ precursors with the largest importances $I_i$ (see Eq.~\eqref{eq:precursor_importance}) were removed from the previous individual precursor structure. This means that a $40$ individual precursor structure was used for this calculation. As in Sec.~\ref{subsection:critical_state_50precursors}, the simulation was run using OpenMC(TD) for $2$ batches and the total simulation time was $4$~s, divided in $400$ time intervals of $10$~ms each. Population control was applied at the end of each time step. The wall-clock time for this simulation was about $319.65$~h.

\begin{figure}[h!]
	\centering
	\includegraphics[width=\textwidth]{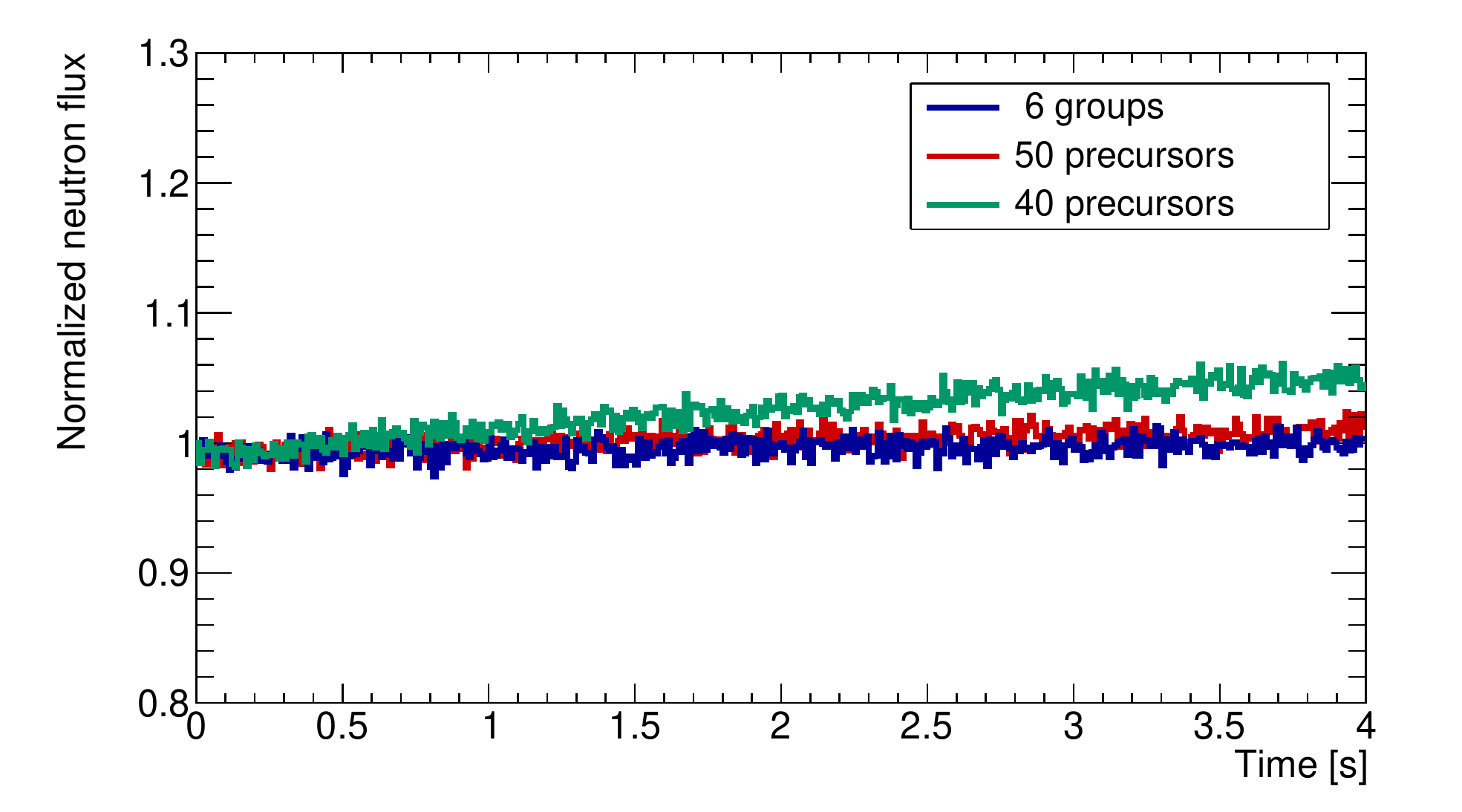}
	\caption{Time evolution for the neutron flux in a water moderated box made of pure ${}^{235}$U, simulated with OpenMC. When the $6$ groups are used results shown in blue. Results obtained when $50$ individual precursors are used are shown in red, and results obtained when using $40$ precursors are shown in green. In this case it can be seen that the time evolution of the neutron flux calculated using $40$ precursors clearly diverges from the previous results, showing that the neutron flux grows more rapidly in time.}
	\label{fig:comparison_critical_6vs50vs40_400steps_10E-2}
\end{figure}

Fig.~\ref{fig:comparison_critical_6vs50vs40_400steps_10E-2} shows results obtained from transient Monte Carlo simulation using OpenMC(TD) for the time evolution of the neutron flux. Results in blue are when $6$ individual precursor structure was used, in red, when $50$ individual precursor structure was used, while in green when $40$ individual precursor structure was used. In this case it can be seen that the time evolution of the neutron flux calculated using $40$ precursors clearly diverges from the previous results. The reason for this behaviour, is because by removing the $10$ most important precursors, the number of delayed neutrons emitted decreased, thus the period of the fissile system increased as explained in Sec.~\ref{subsection:importance_of_delayed_neutrons}.
   
This deviation from criticality for the case when $40$ precursors were used can be quantified by calculating the fitted reactivity. Indeed, the value of this parameter for the $40$ precursors calculation was $\rho_{40}\!=\!0.00111(270)$, showing that this system is no longer close to critical, but supercritical. A summary of the results obtained can be seen in Table~\ref{table:results_6vs50vs40}, where, for completeness, results for the $6$-precursor group and $50$ individual precursor structures are also shown.

\begin{table}[h!]
	\centering
	\begin{tabular}{cccc}
		\toprule
		{ }                 & \textbf{6-group}   & \textbf{50 individual}  & \textbf{40 individual} \\ 
		{ }                 & \textbf{structure} & \textbf{structure}   & \textbf{structure}    \\ \midrule
		$\boldsymbol{\rho_{\mathit{fit}}}$ [pcm] &  $17(368)$  &  $35(347)$ & $111(270)$ \\  \bottomrule
	\end{tabular}
	\caption{Results obtained for the reactivity of the water moderated energy dependent simulated system in a critical configuration using $6$-group, $50$ individual and $40$ individual precursor structures.}
	\label{table:results_6vs50vs40}
\end{table}

\pagestyle{empty}
\begin{landscape}
\begin{table}[htbp]
	\begin{center}
		\resizebox{\columnwidth}{!}{%
			\begin{tabular}{|c|c|c|l|r|r|l|c|l|l|l|}  
				\hline
				\textbf{System} & \textbf{Configuration} & \textbf{Precursor} & \textbf{Delayed} & \textbf{Simulation} & \textbf{Wall-clock} & \textbf{OpenMC(TD)} & \textbf{Compared} & \textbf{Calculated} & \textbf{Difference} & \textbf{Error} \\ 
				  &   & \textbf{structure} & \textbf{neutron energy} & \textbf{time} & \textbf{time} & \textbf{parameter result} & \textbf{with} & \textbf{result} & & \\ \hline
				\multirow{2}{*}{\begin{minipage}{4cm}{\textbf{Monoenergetic fissile}}\end{minipage}} 
				& Subcritical & $1$-group & Monoenergetic  &$50$~s  & $12.35$~h & $\rho\!=\!-1193(626)$~pcm & Point kinetics & $\rho\!=\!-1193(3)$~pcm & $|\Delta \rho|=0$~pcm & $\delta \rho=3$~pcm~(*)   \\\cline{2-11}
				& Critical & $1$-group  & Monoenergetic & $25$~s & $52.97$~h   & $\rho\!=\!13(70)$~pcm & Point kinetics & $\rho\!=\!10(3)$~pcm  &  $|\Delta \rho|\!=\!3$~pcm  & $\delta \rho\!=\!3$~pcm~(*) \\  
				
				\hline
				\multirow{24}{*}{\begin{minipage}{4cm}{\textbf{Energy dependent U235}}\end{minipage}} 
				& Subcritical & $1$-group & $\chi_1(E)$ from & $100$~$\mu$s & $44.32$~h & $\Lambda_{\mathit{eff}}\!=\!5.45(56)$~ns & Adjoint flux & $\Lambda_{\mathit{eff}}\!\!=\!\!5.74(1)$~ns & $|\Delta \Lambda_{\mathit{eff}}|\!\!=\!\!0.29$~ns & $\delta \Lambda_{\mathit{eff}}\!=\!0.56$~ns \\ 
				&                       &  & JEFF-$3$.$1$.$1$  &     & { } & $\beta_{\mathit{eff}}\!=\!648(38)$~pcm & MCNP & $\beta_{\mathit{eff}}\!=\!644(6)$~pcm & $|\Delta \beta_{\mathit{eff}}|\!=\!4$~pcm & $ \delta \beta_{\mathit{eff}}\!=\!38$~pcm \\\cline{3-11} 
				
				&             & $1$-group & $\bar{E}_{1g}\!=\!212.31$~keV  &$50$~$\mu$s & $3.51$~h & $\Lambda_{\mathit{eff}}\!=\!5.45(42)$~ns & Adjoint flux & $\Lambda_{\mathit{eff}}\!=\!5.74(1)$~ns & $|\Delta \Lambda_{\mathit{eff}}|\!=\!0.29$~ns  &  $ \delta \Lambda_{\mathit{eff}}\!=\!0.42$~ns  \\ 
				&             &           & from JEFF-$3$.$1$.$1$ &        &  & $\beta_{\mathit{eff}}\!=\!666(34)$~pcm & MCNP & $\beta_{\mathit{eff}}\!=\!644(6)$~pcm & $|\Delta \beta_{\mathit{eff}}|\!=\!22$~pcm & $ \delta \beta_{\mathit{eff}}\!=\!34$~pcm \\\cline{3-11} 
				
				&             & $1$-group & $\bar{E}_{6g}\!=\!501.31$~keV &   $50$~$\mu$s & $3.49$~h & $\Lambda_{\mathit{eff}}\!=\!5.53(52)$~ns & Adjoint flux & $\Lambda_{\mathit{eff}}\!=\!5.74(1)$~ns & $|\Delta \Lambda_{\mathit{eff}}|\!=\!0.21$~ns  &  $ \delta \Lambda_{\mathit{eff}}\!=\!0.52$~ns  \\ 
				&             &           &  from ENDF/B-VIII.$0$  &    & { } & $\beta_{\mathit{eff}}\!=\!602(36)$~pcm & MCNP & $\beta_{\mathit{eff}}\!=\!644(6)$~pcm & $|\Delta \beta_{\mathit{eff}}|\!=\!42$~pcm & $ \delta \beta_{\mathit{eff}}\!=\!36$~pcm \\\cline{3-11} 
				
	            &              & $8$-group & $\chi(E)$ from  & $50$~$\mu$s & $4.32$~h & $\Lambda_{\mathit{eff}}\!=\!5.45(45)$~ns & Adjoint flux & $\Lambda_{\mathit{eff}}\!=\!5.74(1)$~ns & $|\Delta \Lambda_{\mathit{eff}}|\!=\!0.29$~ns  &  $ \delta \Lambda_{\mathit{eff}}\!=\!0.45$~ns  \\ 
	            &             &           &  JEFF-$3$.$1$.$1$  &      & { } & $\beta_{\mathit{eff}}\!=\!660(60)$~pcm & MCNP & $\beta_{\mathit{eff}}\!=\!644(6)$~pcm & $|\Delta \beta_{\mathit{eff}}|\!=\!16$~pcm & $ \delta \beta_{\mathit{eff}}\!=\!60$~pcm \\\cline{3-11} 
	            
	            &             & $6$-group & $\bar{E}_i$ from & $50$~$\mu$s & $4.27$~h & $\Lambda_{\mathit{eff}}\!=\!5.68(29)$~ns & Adjoint flux & $\Lambda_{\mathit{eff}}\!=\!5.74(1)$~ns & $|\Delta \Lambda_{\mathit{eff}}|\!=\!0.06$~ns  &  $ \delta \Lambda_{\mathit{eff}}\!=\!0.29$~ns  \\ 
	            &             &           &  ENDF/B-VIII.$0$ &     & { } & $\beta_{\mathit{eff}}\!=\!602(57)$~pcm & MCNP & $\beta_{\mathit{eff}}\!=\!644(6)$~pcm & $|\Delta \beta_{\mathit{eff}}|\!=\!22$~pcm & $ \delta \beta_{\mathit{eff}}\!=\!57$~pcm \\\cline{3-11} 
	            
	            &             & $50$ individual &  $\bar{E}_i$ from & $50$~$\mu$s & $6.43$~h & $\Lambda_{\mathit{eff}}\!=\!5.45(31)$~ns & Adjoint flux & $\Lambda_{\mathit{eff}}\!=\!5.74(1)$~ns & $|\Delta \Lambda_{\mathit{eff}}|=0.29$~ns  &  $ \delta \Lambda_{\mathit{eff}}\!=\!0.31$~ns  \\ 
	            &             &           &  ENDF/B-VIII.$0$  &    & { } & $\beta_{\mathit{eff}}\!=\!602(57)$~pcm & MCNP & $\beta_{\mathit{eff}}\!=\!644(6)$~pcm & $|\Delta \beta_{\mathit{eff}}|\!=\!42$~pcm & $ \delta \beta_{\mathit{eff}}\!=\!57$~pcm \\\cline{2-11} 
	            
				& Supercritical & $1$-group & $\chi_1(E)$ from  & $100$~$\mu$s & $52.45$~h & $\Lambda_{\mathit{eff}}\!=\!5.45(29)$~ns & Adjoint flux & $\Lambda_{\mathit{eff}}\!=\!6.00(1)$~ns & $|\Delta \Lambda_{\mathit{eff}}|\!=\!0.55$~ns & $ \delta \Lambda_{\mathit{eff}}\!=\!0.29$~ns\\  
				&             &           &  JEFF-$3$.$1$.$1$  &      & {  } & $\beta_{\mathit{eff}}\!=\!666(56)$~pcm & MCNP & $\beta_{\mathit{eff}}\!=\!651(6)$~pcm & $|\Delta \beta_{\mathit{eff}}|\!=\!15$~pcm & $ \delta \beta_{\mathit{eff}}\!=\!56$~pcm \\ \cline{3-11} 
				
					&          & $1$-group &  $\bar{E}_{1g}\!=\!212.31$~keV & $50$~$\mu$s & $7.84$~h & $\Lambda_{\mathit{eff}}\!=\!5.45(31)$~ns & Adjoint flux & $\Lambda_{\mathit{eff}}\!=\!6.00(1)$~ns & $|\Delta \Lambda_{\mathit{eff}}|\!=\!0.55$~ns & $ \delta \Lambda_{\mathit{eff}}\!=\!0.57$~ns\\  
				&             &           &  from JEFF-$3$.$1$.$1$ &       & {  } & $\beta_{\mathit{eff}}\!=\!666(63)$~pcm & MCNP & $\beta_{\mathit{eff}}\!=\!651(6)$~pcm & $|\Delta \beta_{\mathit{eff}}|\!=\!15$~pcm & $ \delta \beta_{\mathit{eff}}\!=\!63$~pcm \\ \cline{3-11} 
				
					&         & $1$-group & $\bar{E}_{6g}\!=\!501.31$~keV  &$50$~$\mu$s & $7.81$~h & $\Lambda_{\mathit{eff}}\!=\!5.45(57)$~ns & Adjoint flux & $\Lambda_{\mathit{eff}}\!=\!6.00(1)$~ns & $|\Delta \Lambda_{\mathit{eff}}|\!=\!0.55$~ns & $ \delta \Lambda_{\mathit{eff}}\!=\!0.57$~ns\\  
				&             &           &   from ENDF/B-VIII.$0$  &     & {  } & $\beta_{\mathit{eff}}\!=\!637(35)$~pcm & MCNP & $\beta_{\mathit{eff}}\!=\!651(6)$~pcm & $|\Delta \beta_{\mathit{eff}}|\!=\!14$~pcm & $ \delta \beta_{\mathit{eff}}\!=\!35$~pcm \\ \cline{3-11} 
				
					&           & $8$-group & $\chi(E)$ from  &$50$~$\mu$s & $11.19$~h & $\Lambda_{\mathit{eff}}\!=\!6.03(43)$~ns & Adjoint flux & $\Lambda_{\mathit{eff}}\!=\!6.00(1)$~ns & $|\Delta \Lambda_{\mathit{eff}}|\!=\!0.03$~ns & $ \delta \Lambda_{\mathit{eff}}\!=\!0.43$~ns\\  
				&             &           & JEFF-$3$.$1$.$1$  &       & {  } & $\beta_{\mathit{eff}}\!=\!665(56)$~pcm & MCNP & $\beta_{\mathit{eff}}\!=\!651(6)$~pcm & $|\Delta \beta_{\mathit{eff}}|\!=\!14$~pcm & $ \delta \beta_{\mathit{eff}}\!=\!56$~pcm \\ \cline{3-11} 
				
					&          & $6$-group & $\bar{E}_i$ from &$50$~$\mu$s & $11.03$~h & $\Lambda_{\mathit{eff}}\!=\!5.45(57)$~ns & Adjoint flux & $\Lambda_{\mathit{eff}}\!=\!6.00(1)$~ns & $|\Delta \Lambda_{\mathit{eff}}|\!=\!0.55$~ns & $ \delta \Lambda_{\mathit{eff}}\!=\!0.57$~ns\\  
				&             &           &  ENDF/B-VIII.$0$  &      & {  } & $\beta_{\mathit{eff}}\!=\!635(38)$~pcm & MCNP & $\beta_{\mathit{eff}}\!=\!651(6)$~pcm & $|\Delta \beta_{\mathit{eff}}|\!=\!16$~pcm & $ \delta \beta_{\mathit{eff}}\!=\!38$~pcm \\ \cline{3-11} 
				
					&         & $50$ individual & $\bar{E}_i$ from &$50$~$\mu$s & $17.24$~h & $\Lambda_{\mathit{eff}}\!=\!5.45(49)$~ns & Adjoint flux & $\Lambda_{\mathit{eff}}\!=\!6.00(1)$~ns & $|\Delta \Lambda_{\mathit{eff}}|\!=\!0.55$~ns & $ \delta \Lambda_{\mathit{eff}}\!=\!0.49$~ns\\  
				&             &           &  ENDF/B-VIII.$0$ &        & {  } & $\beta_{\mathit{eff}}\!=\!621(36)$~pcm & MCNP & $\beta_{\mathit{eff}}\!=\!651(6)$~pcm & $|\Delta \beta_{\mathit{eff}}|\!=\!30$~pcm & $ \delta \beta_{\mathit{eff}}\!=\!36$~pcm \\ 
				\hline 
				\multirow{3}{*}{\begin{minipage}{4cm}{\textbf{Light-water  moderated energy dependent U235}}\end{minipage}} 
				& Critical & $6$-group      &   $\bar{E}_i$ from & $4$~s & $260.05$~h & $\rho\!=\!17(368)$~pcm & OpenMC & $\rho\!=\!25(3)$~pcm & $|\Delta \rho|\!=\!8$~pcm & $\delta \rho\!=\!3$~pcm~(*) \\ 
				&          & $50$ individual&   ENDF/B-VIII.$0$ &       & $410.76$~h & $\rho\!=\!35(347)$~pcm & Criticality &   & $|\Delta \rho|\!=\!10$~pcm  &  \\  
				&          & $40$ individual&    &  & $319.65$~h & $\rho\!=\!111(270)$~pcm &  & {  } & $|\Delta \rho|\!=\!86$~pcm &  \\ 
				\hline 
			\end{tabular}%
		} 
	\end{center}
	\caption{Summary of all results obtained with OpenMC(TD).(*) Error was obtained only from the calculated result using point kinetics equations or criticality calculation using OpenMC, given that OpenMC(TD) error could be improved as explained in Sec.~\ref{subsection:critical_state_50precursors}.}
	\label{table:final_results_summary}
\end{table}
\end{landscape}
\pagestyle{plain}

With these results, OpenMC(TD) code shows its potential as a Monte Carlo tool with the capability to explore how precursor data from nuclear databases impacts on results obtained in fissile systems. For instance, since the code can use, in principle, an arbitrary precursor structure, it could be studied how the kinetic parameters of a given system responds to changes in the cumulative yield, probability of neutron emission, delayed neutron yield or average delayed neutron energy emitted. In that sense OpenMC(TD) could become a reliable tool to prompt new experimental data on individual $\beta$-delayed neutron emitters.

Finally, as it was discussed in Sec.~\ref{subsection:critical_state_50precursors}, in order to reduce the associated uncertainties from results obtained, increased simulation times would be required. Regarding this, two possible solutions are proposed: i) use of high computing power to run the simulations: since the code is already parallelized, it would benefit by having a greater number of cores available. Of course, this would require access to infrastucture, such as supercomputer clusters and ii) implement the variance reduction technique known as ``implicit fission'' in OpenMC(TD), here, the neutron either has a scattering interation or a fission interaction, and the weight of the neutron is multiplied by the mean number of fission neutrons produced in the event. By using this technique, there is no production of new neutrons during fission, thus reducing the calculation time; this would require modifications and testings of the code, but it would be feasible and it could positively impact the current calculation times using the same infrastucture used in this thesis.  

\chapter{Summary and conclusions}
\label{ch:summary_conclusions}
 
The objective of this work was to explicitly include, in a Monte Carlo simulation, the time dependence related to the $\beta$-delayed neutron emission from individual precursors. In order to achieve this, a modified version of OpenMC Monte Carlo simulation code was developed to include transient capabilities in neutron transport and the option to use individual precursors as $\beta$-delayed neutron emitters. This code has been named OpenMC(TD) or Time-Dependent OpenMC.

OpenMC(TD), in addition to the original OpenMC, includes: i) neutron time labeling and tracking; ii) monitoring of time dependent parameters in the simulation such as neutron flux, reaction rates, neutron current, and total neutron population; iii) simulation time interval division depending on the detail required for the studied physics case; iv) a new particle called precursor, which is not transported and acts as a $\beta$-delayed neutron emitter; v) individual precursor properties from nuclear databases such as precursor cumulative neutron yield, delayed neutron emission probability, $\beta$-delayed decay constant and average number of delayed neutrons produced per fission $\nu_d$; vi) either precursor $N$-group grouping capabilities or individual precursor treatment; vii) forced decay of precursor within each time interval; viii) population control at the end of each time interval using the combing method; and ix) a transient source routine to initialize transient simulations.

To approach the time modelling of neutron transport and interactions in a experimental nuclear reactor, a fissile system was simulated. OpenMC(TD) was tested in successively complex systems. Different observables such as reactivity $\rho$, effective delayed neutron fraction $\beta_{\mathit{eff}}$ and effective prompt generation time $\Lambda_{\mathit{eff}}$, obtained with OpenMC(TD) were compared with calculated results, either with exact point kinetics solutions ($1$-group, $6$-group, $8$-group and $50$ individual precursor structure) or asymptotic decay constant $\alpha_D$ ($6$-group and $50$ and $40$ individual precursor structure). A summary of the OpenMC(TD) results obtained for the systems, configurations and precursor structures studied in this work is shown in Table~\ref{table:final_results_summary}. 

For the monoenergetic system, using the $1$-group precursor structure, differences between OpenMC(TD) and the compared results using point kinetics equations were within the error of the point kinetics result. Nevertheless, large uncertainties were obtained for the reactivity of the subcritical and critical configurations, using OpenMC(TD). 

For the light-water moderated energy dependent ${}^{235}$U system, using the $6$-group, $50$ and $40$ individual precursor structure, differences between OpenMC(TD) and the compared results using criticality calculations with the standard $6$-group precursor structure, were greater than the error of the criticality calculation. The simulation time of $4$~s was too short to describe the asymptotical critical behaviour of the system, when the time evolution of the neutron flux increases gradually. 

For the energy dependent ${}^{235}$U system, discrepancies were found in the value obtained for the effective delayed neutron fraction using JEFF-$3$.$1$.$1$ and ENDF-B/VIII.$0$ nuclear databases, showing the importance of appropriate nuclear data for individual precursors. 
In this case, the simulation time was $100$~$\mu$s, with a time interval of $100$~ns, describing neutron flux prompt changes (prompt drop or prompt jump for the subcritical or supercritical configuration, respectively) within the first $10$~$\mu$s. Both total simulation time and time interval chosen for these cases were adequate to properly describe the transient behaviour of the neutron flux in these systems. 

Results and its errors can be improved in accuracy by running the simulation with longer wall-clock times at CSICCIAN cluster; by applying to computing time outside the institution or by implementing an implicit fission scheme in OpenMC(TD). 

Resuming the discussion about the possibility of using OpenMC(TD) to simulate a full system, such as a nuclear reactor core, according to what has been learned and developed in this thesis, this would require i) a complete model of the core geometry materials, its densities, nuclear cross sections, and to replicate this process with another code, such as MCNP, for its subsequent comparison with respect to $k_{\mathit{eff}}$; ii) to read a geometry file at the beginning of each time interval, simulating in this way the insertion or extraction of the control rods\footnote{This geometry file will contain the control rods positions at different time.}; and iii) a comparison with experimental measurements of reactivity changes.

The OpenMC(TD) code, developed in this thesis, shows its potential as a Monte Carlo tool with the capability to explore how precursor data from nuclear databases impacts on results obtained in fissile systems. In that sense, OpenMC(TD) could become a reliable tool to prompt new experimental data on individual $\beta$-delayed neutron emitters.
\newpage
\noindent \textbf{Future work}

Results obtained with OpenMC(TD) could be compared not only to results obtained from other codes, but also with experimental results from transient measurements in nuclear reactors. In the current state of the code, results obtained with simulation times of the order of tens of seconds, including individual precursors, would require computational times of the order of months. To decrease this time, reduction variance methods additional to forced decay, combing, and implicit absorption need to be implemented. One of the possible reduction variance methods that could be implemented, is implicit fission~\cite{SJENITZER20112195}. The implementation of this method would allow to increase the simulation time, as well as decreasing the time intervals, reducing the associated uncertainties of the obtained results from the OpenMC(TD) simulation.
Some future problems to study could be reactivity insertion in: i) moderated energy dependent system with individual precursors, to quantitatively assess the relative importance of precursors and thus, prompt experimental measurements of $\beta$-delayed emitter nuclei, and then; ii) reactor fuel and core model, in order to obtain its kinetic parameters and compare with reactivity measurements in a region of the reactor core, using a reactivimeter.

There exists other types of time-dependent problems of special interest in reactor physics, such as burn-up fuel calculations.  Although this is a time-dependent calculation, it requires knowledge of the isotopic abundance of fissile material present in fuel elements during the fuel period of use, which for an experimental nuclear reactor, is of the order of a few years. Another problem to study could be the coupling of the time evolution of isotopic abundance obtained using reaction rates calculated with the Bateman equations, with the OpenMC(TD) code, validated with experimental measurements during the time when the fuel is used. 

Nonetheless, the study of the inclusion of time dependence in Monte Carlo methods, would allow to explore other problems where the fuel materials and precursors are not fixed in space, but in movement during the operation time of the nuclear reactor, like fuel in IV-th generation nuclear reactors, such as Molten Salt Fast Reactors, where the salt (fuel) moves through the circuit in about $4$~s~\cite{brovchenko} and transient calculations are needed to take into account this circulation.

\appendix
\chapter{Delayed neutron group spectra}
\label{app:group_spectra}

In this appendix the $8$-group $\beta$-delayed neutron spectra are shown. 

\begin{figure}[h!]
	\centering
	\includegraphics[width=0.8\textwidth]{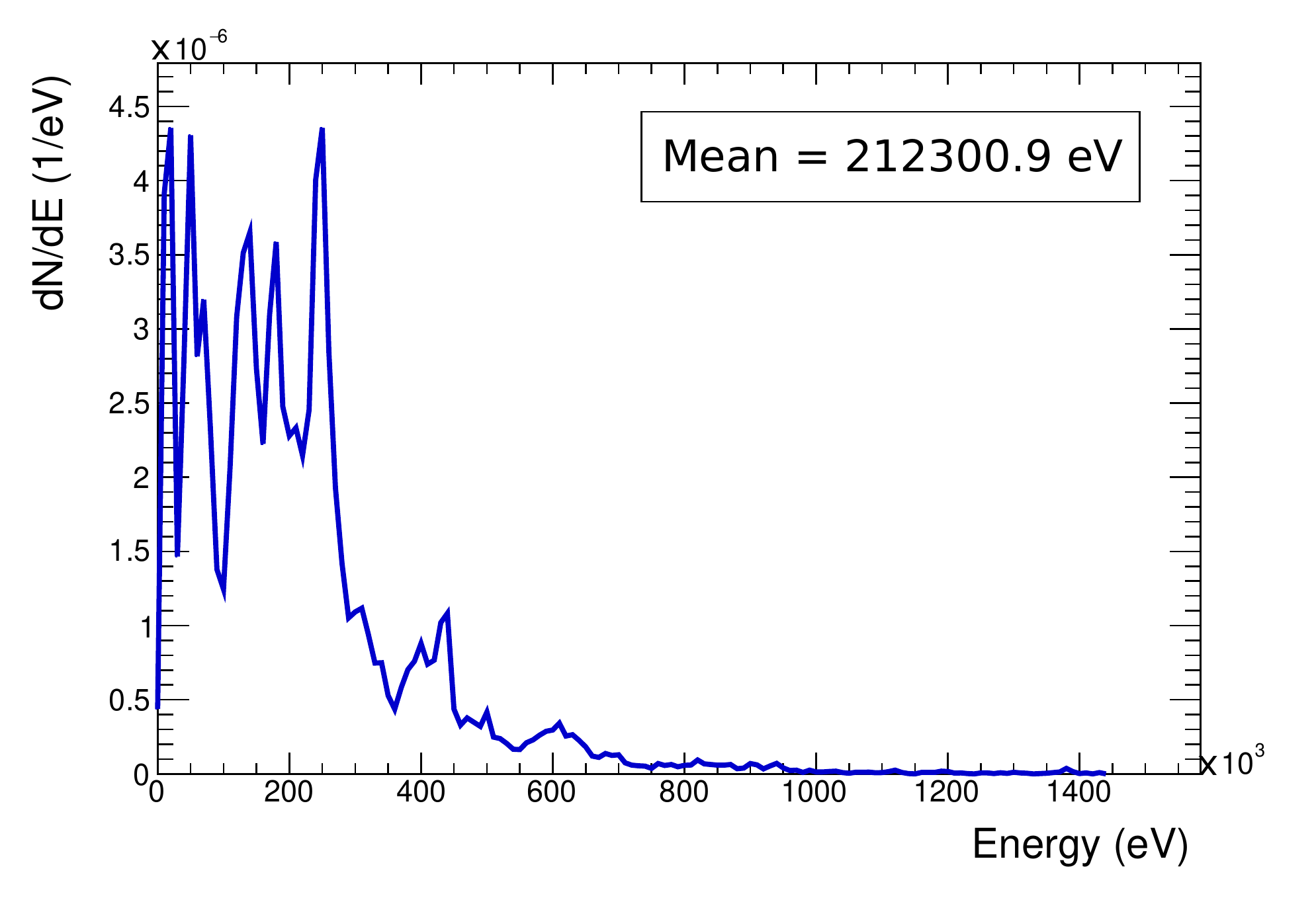}
	\caption{Group $1$, $\beta$-delayed neutron energy spectrum from JEFF-$3$.$1$.$1$.}
	\label{fig:jeff311_group1_spectrum}
\end{figure}

\begin{figure}[h!]
	\centering
	\includegraphics[width=0.8\textwidth]{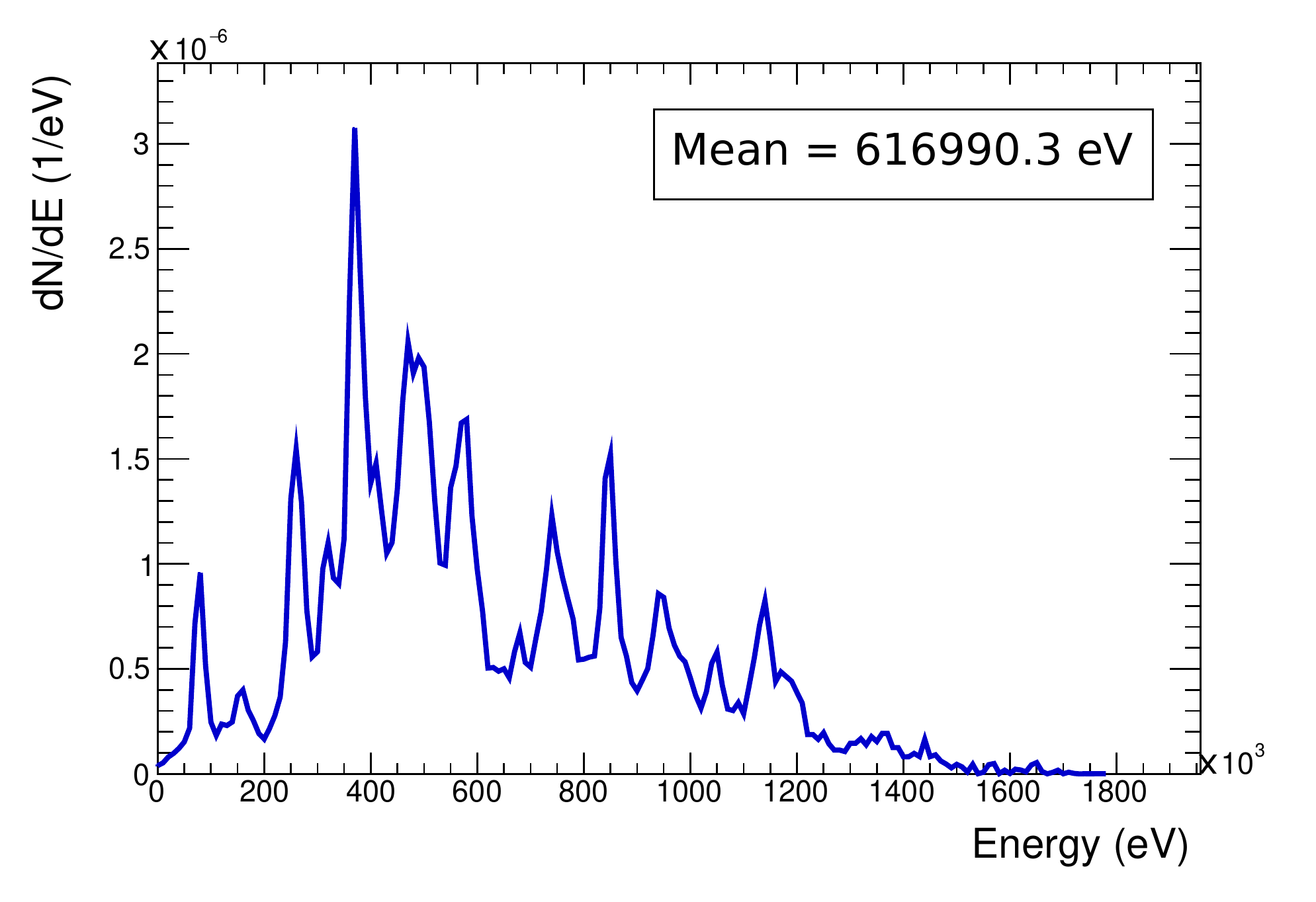}
	\caption{Group $2$, $\beta$-delayed neutron energy spectrum from JEFF-$3$.$1$.$1$.}
	\label{fig:jeff311_group2_spectrum}
\end{figure}

\begin{figure}[h!]
	\centering
	\includegraphics[width=0.8\textwidth]{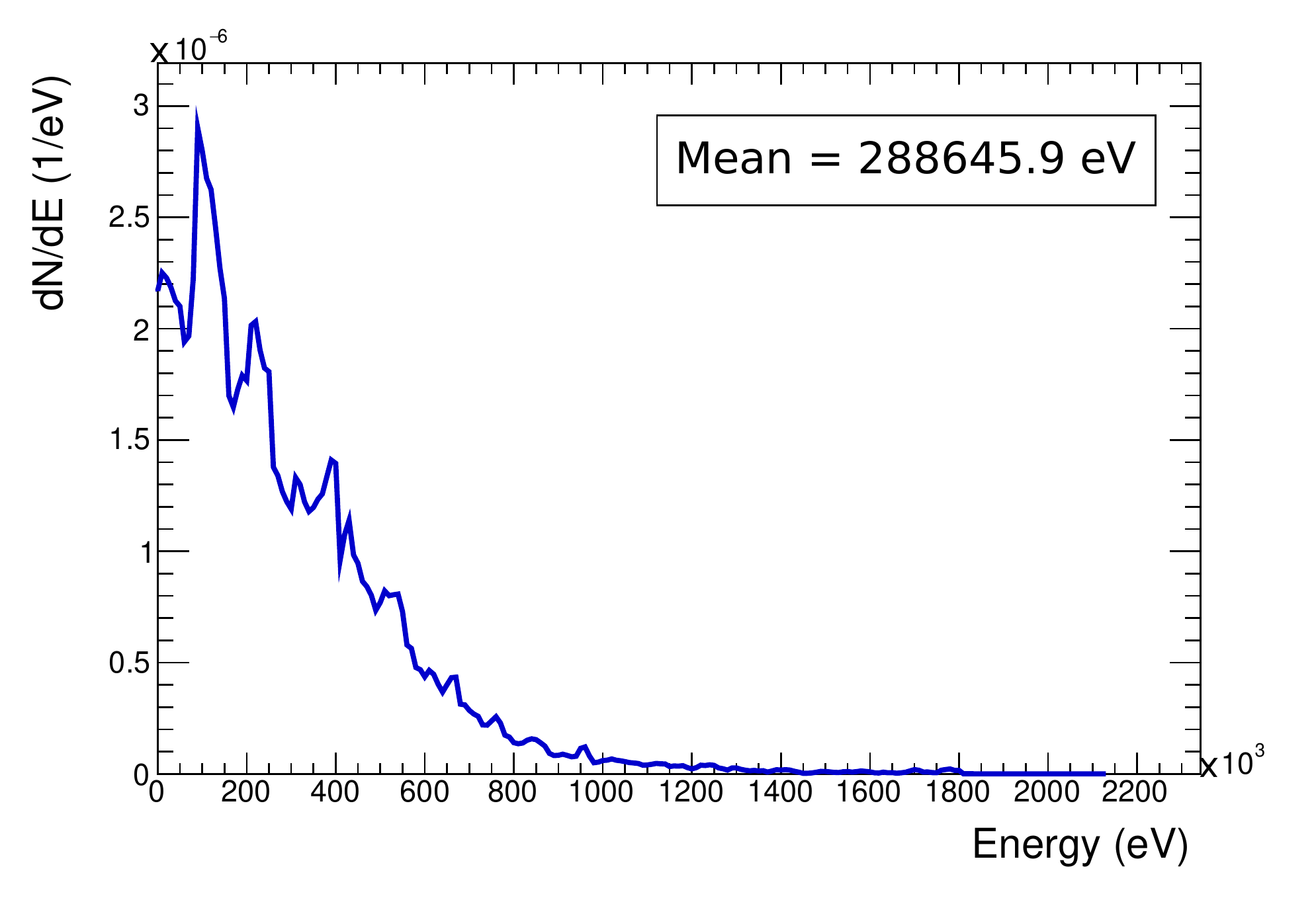}
	\caption{Group $3$, $\beta$-delayed neutron energy spectrum from JEFF-$3$.$1$.$1$.}
	\label{fig:jeff311_group3_spectrum}
\end{figure}

\begin{figure}[h!]
	\centering
	\includegraphics[width=0.8\textwidth]{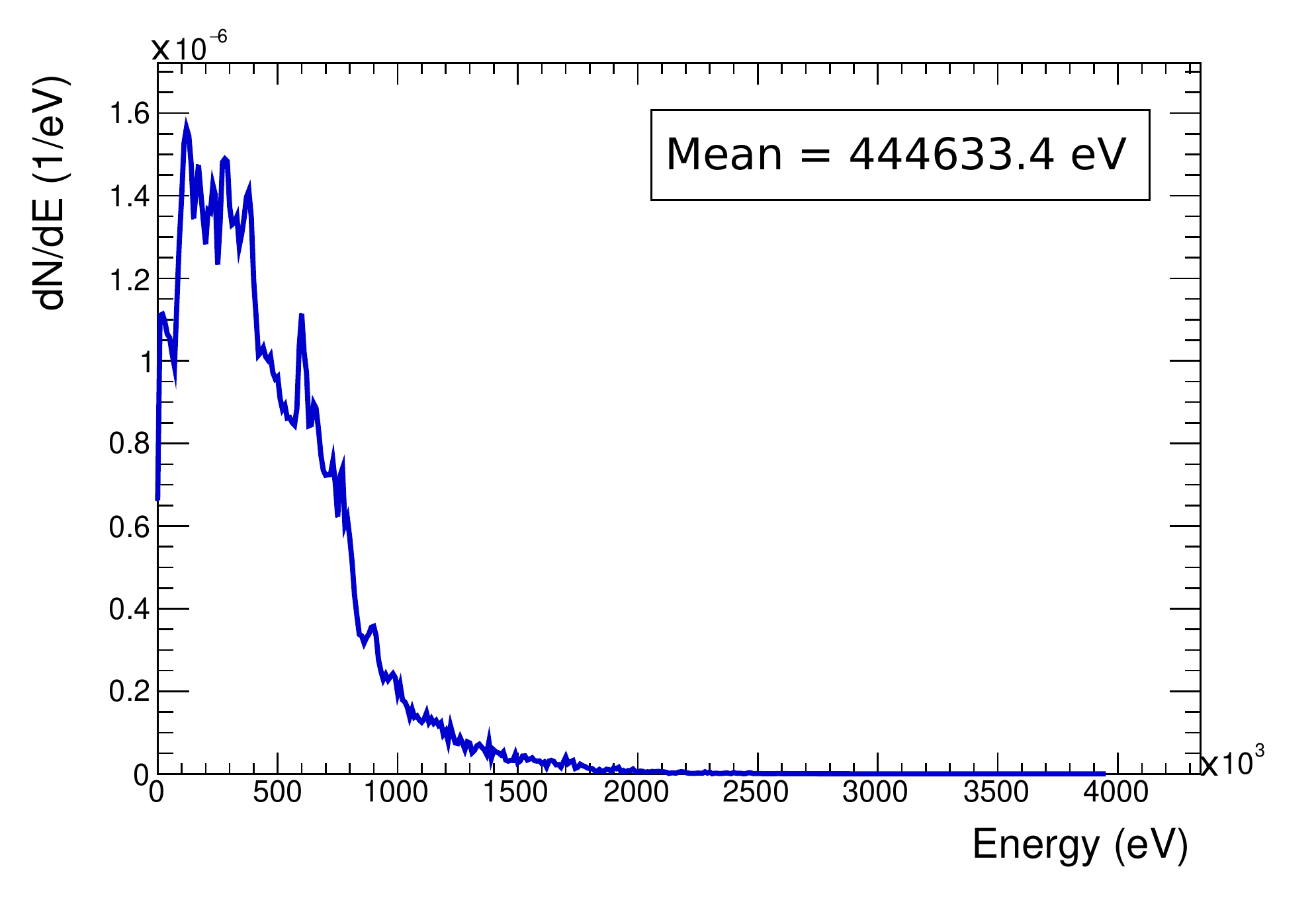}
	\caption{Group $4$, $\beta$-delayed neutron energy spectrum from JEFF-$3$.$1$.$1$.}
	\label{fig:jeff311_group4_spectrum}
\end{figure}

\begin{figure}[h!]
	\centering
	\includegraphics[width=0.8\textwidth]{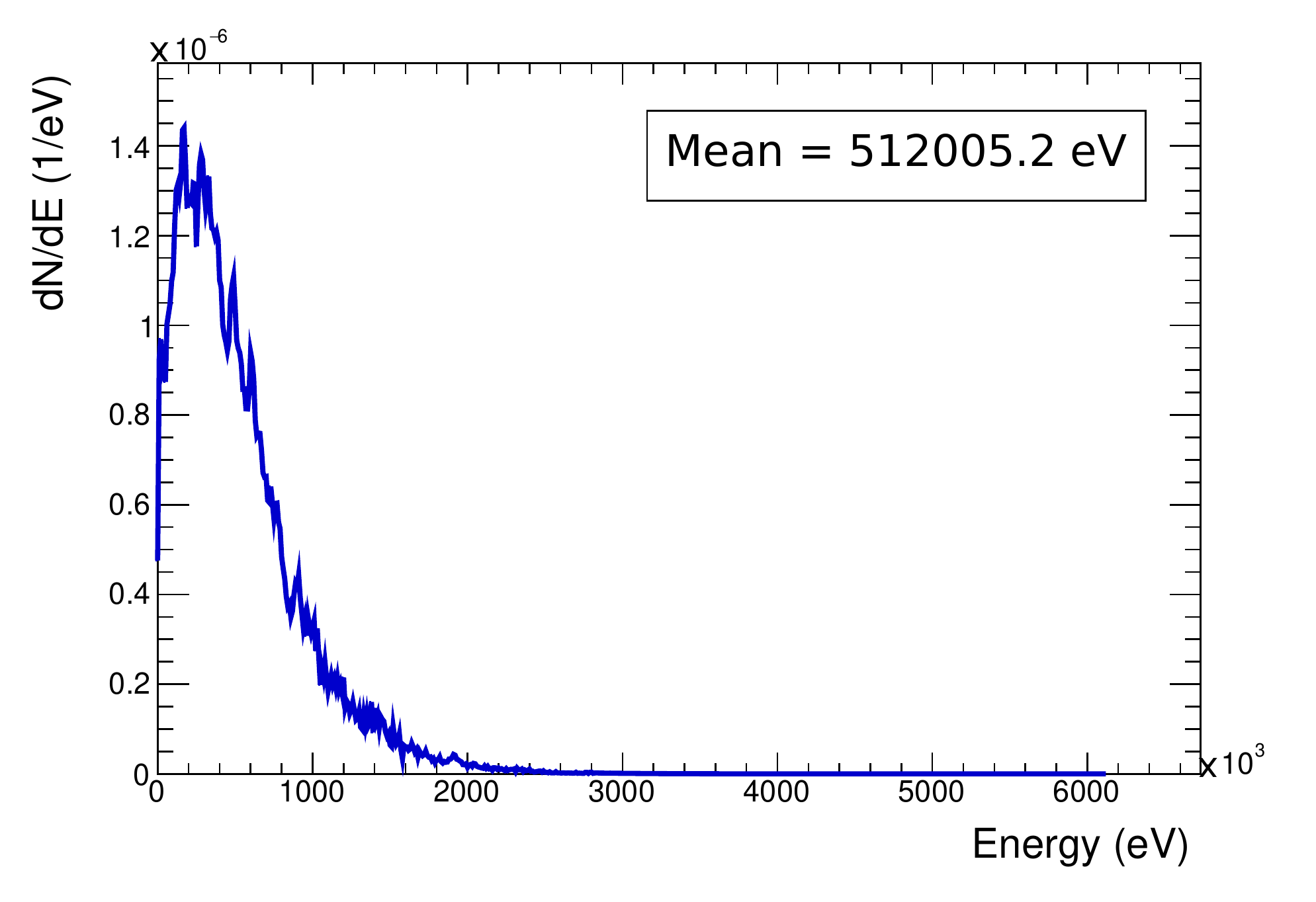}
	\caption{Group $5$, $\beta$-delayed neutron energy spectrum from JEFF-$3$.$1$.$1$.}
	\label{fig:jeff311_group5_spectrum}
\end{figure}

\begin{figure}[h!]
	\centering
	\includegraphics[width=0.8\textwidth]{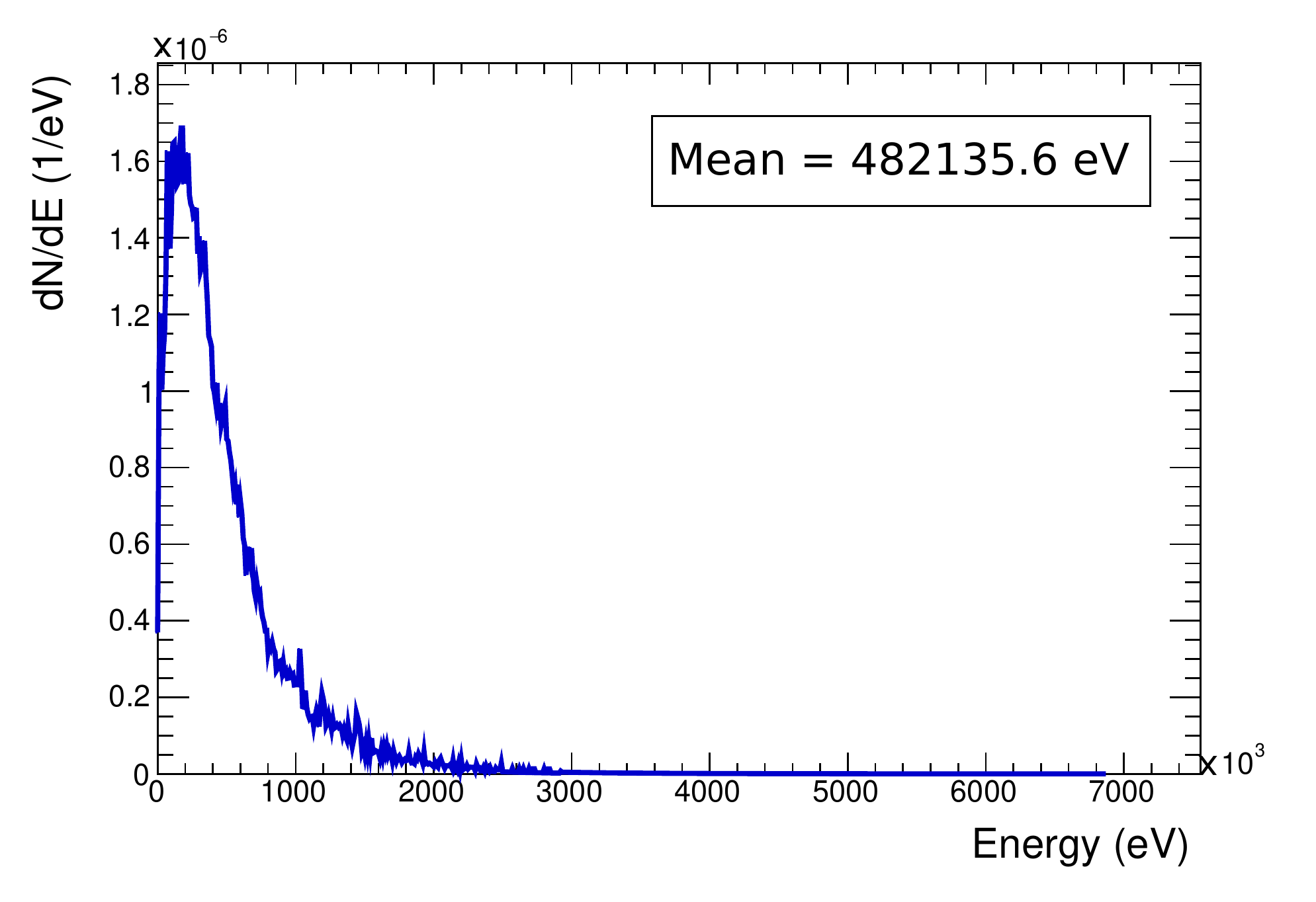}
	\caption{Group $6$, $\beta$-delayed neutron energy spectrum from JEFF-$3$.$1$.$1$.}
	\label{fig:jeff311_group6_spectrum}
\end{figure}

\begin{figure}[h!]
	\centering
	\includegraphics[width=0.8\textwidth]{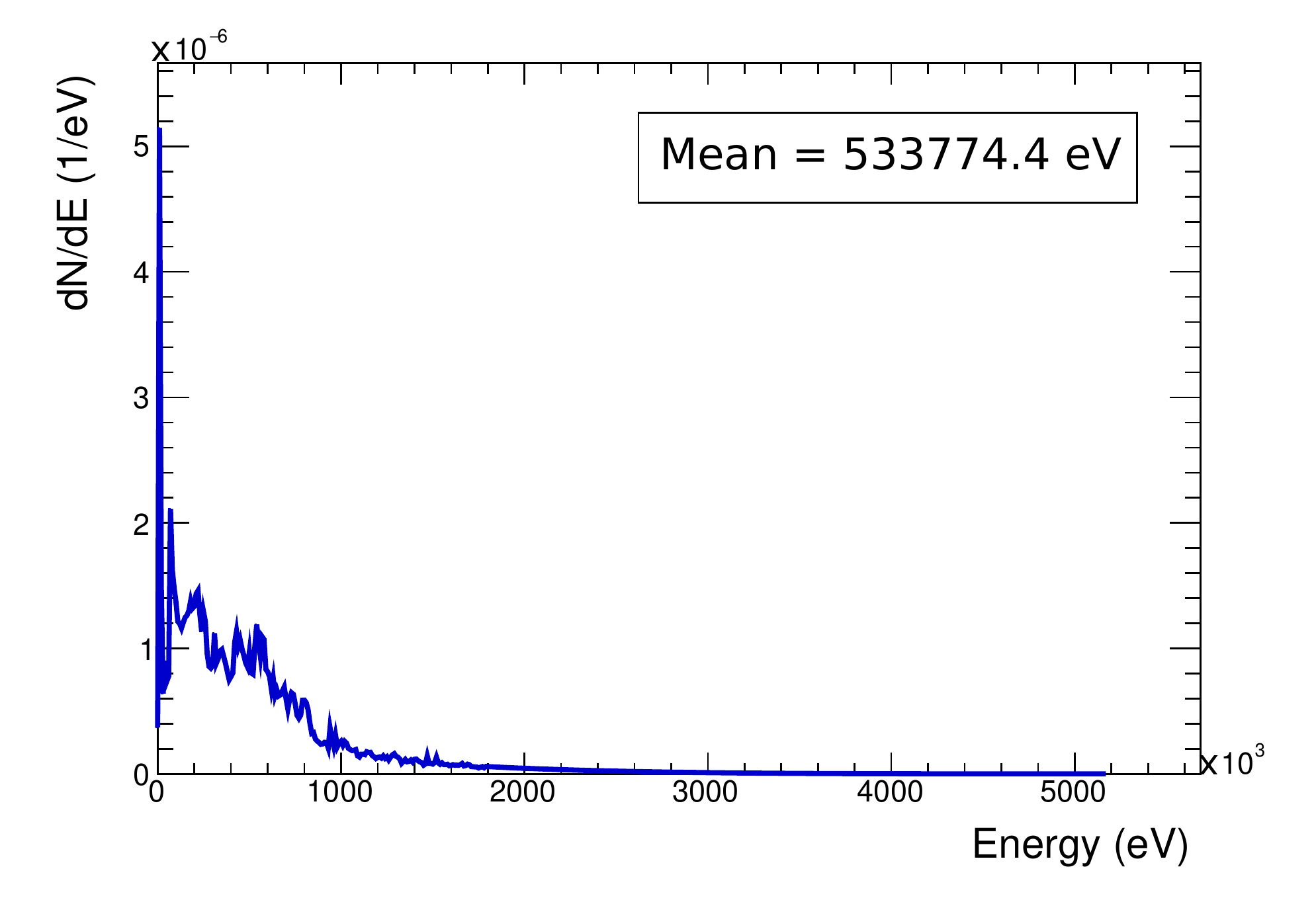}
	\caption{Group $7$ $\beta$-delayed neutron energy spectrum from JEFF-$3$.$1$.$1$.}
	\label{fig:jeff311_group7_spectrum}
\end{figure}

\begin{figure}[h!]
	\centering
	\includegraphics[width=0.8\textwidth]{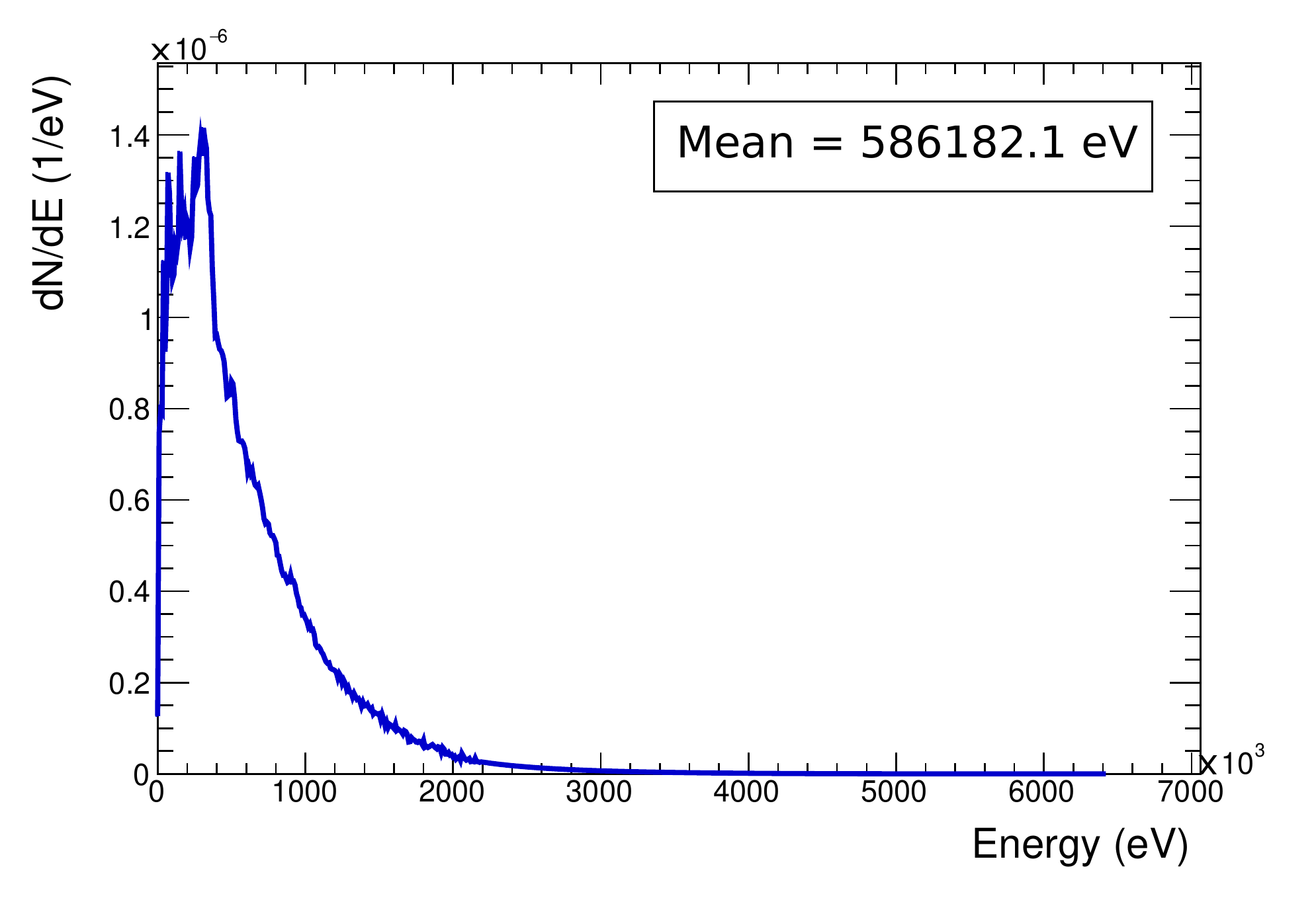}
	\caption{Group $8$ $\beta$-delayed neutron energy spectrum from JEFF-$3$.$1$.$1$.}
	\label{fig:jeff311_group8_spectrum}
\end{figure}
\chapter{Solutions of the Point Neutron Kinetics Equations for 1-group precursor approximation}
\label{app:monoenergetic_system}
If the $6$-group precursor groups are replaced by $1$-group precursor with an effective yield fraction $\beta$ and an effective decay constant given by
\begin{equation}
	\bar{\lambda} = \sum_i \frac{\beta_i \lambda_i}{\beta},
	\label{eq:lambda_effective}
\end{equation}
then the point kinetics equations become
\begin{equation}
	\frac{dn}{dt} = \frac{\rho - \beta}{\Lambda}n + \lambda C,
	\label{eq:neutron_population_1precursor}
\end{equation}
and
\begin{equation}
	\frac{dC}{dt}=\frac{\beta}{\Lambda} n - \lambda C.
	\label{eq:precursor_concetration_1precursor}
\end{equation}

\noindent The solutions to Eq.~\eqref{eq:neutron_population_1precursor} and Eq.~\eqref{eq:precursor_concetration_1precursor} are the time evolution of the neutron and precursor population, $n(t)$ and $C(t)$, given by
\begin{equation}
	n(t) = n_0 \left [ \frac{\rho}{\rho - \beta} \exp{(\alpha_P \, t)} - \frac{\beta}{\rho - \beta} \exp{( \alpha_D \, t)}\right ],
	\label{eq:solution_neutron_1precursor}
\end{equation}
and
\begin{equation}
C(t) = n_0 \left [ \frac{\rho \beta}{ (\rho - \beta)^2} \exp{(\alpha_P \, t)} - \frac{\beta}{\Lambda\lambda} \exp{( \alpha_D \, t)}\right ],
\label{eq:solution_precursor_1precursor}
\end{equation}
where the term $\alpha_P$ defined as
\begin{equation}
	\alpha_P = \frac{\rho - \beta}{\Lambda},
	\label{eq:alpha_P}
\end{equation}
is related to the fast readjustement of the prompt neutron population, which happens on the neutron generation timescale, given a change in the reactivity. 
On the other hand, the term $\alpha_D$, defined as
\begin{equation}
\alpha_D = \frac{\lambda \rho}{\rho - \beta},
\label{eq:alpha_D}
\end{equation}
corresponds in general to the slower change in the neutron population due to the delayed source of neutrons, characterized by the precursor decay constant $\lambda$.
\chapter{Monoenergetic fissile system with 1-group precursor structure}
\label{app:specifications_monoenergetic_system}
\begin{figure}[h]
	\centering
	\includegraphics[width=.4\textwidth]{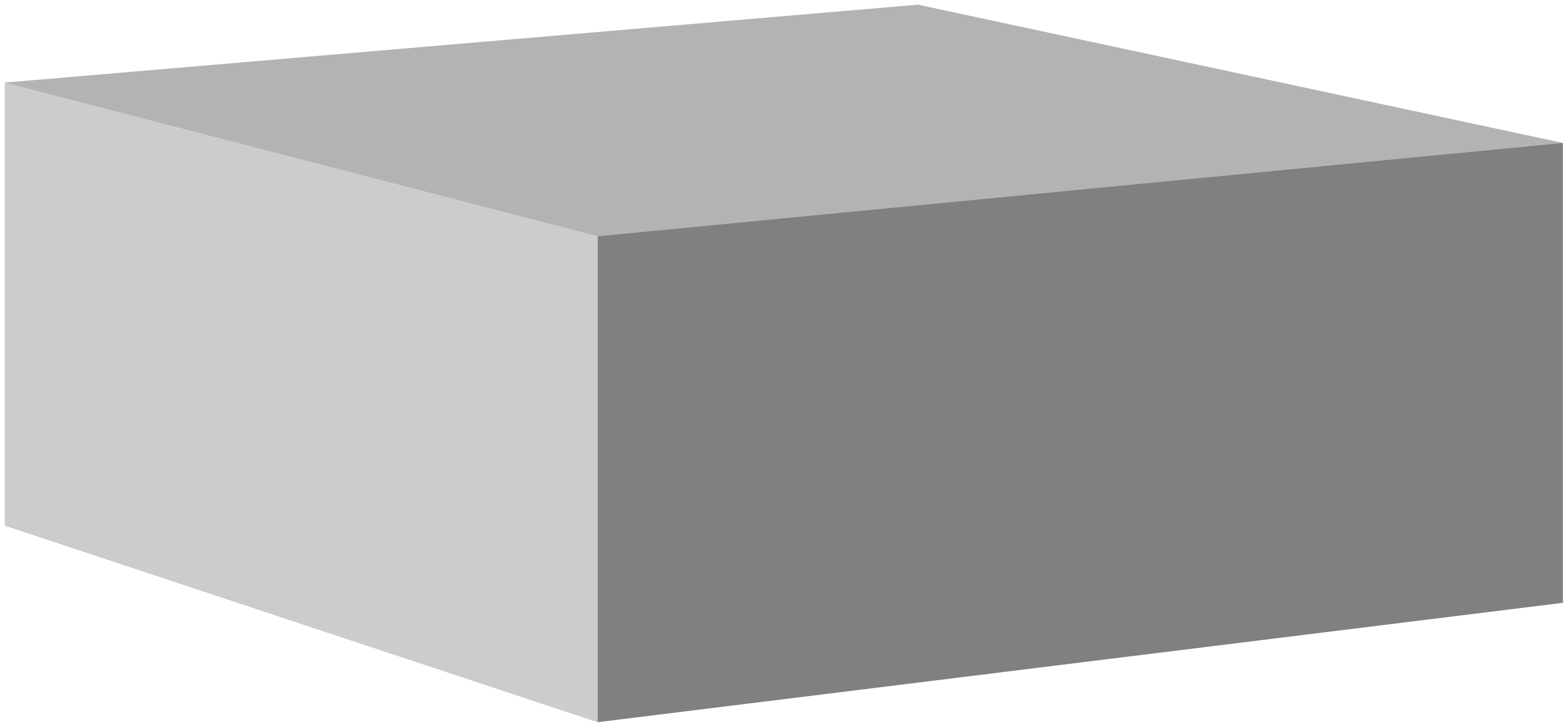}
	\caption{Box of $10 \text{ cm}\times 12 \text{ cm} \times 20 \text{ cm}$ simulated in Sec.~\ref{section:mono-energetic_system} and Sec.~\ref{section:energy_dependent}.}
	\label{fig:box}
\end{figure}

The system described in Sec.~\ref{section:preliminar_work} and in Sec.~\ref{section:mono-energetic_system} of this work consists of a rectangular box of $10 \text{ cm}\times 12 \text{ cm} \times 20 \text{ cm}$ (see Fig.~\ref{fig:box}) surrounded by vacuum. All neutrons have the same velocity, making this a mono-energetic problem. Total neutron yield is fixed, the material cross-sections are constant and there is one precursor group. The system parameters are shown in Table~\ref{tab:parameters_mono_system}. 
\vspace{2cm}

\begin{table}[h!]
	\centering
	\begin{tabular}{@{}cc@{}}
		\toprule
		\textbf{Parameter} & \textbf{Value} \\
		\midrule
		$\beta_{\mathit{eff}} $	& $0.00685$  \\ 
		$\lambda$~(s${}^{-1}$)	& $0.0784$  \\ 
		$\nu$	&  $2.5$\\ 
		$\Sigma_t$~(cm${}^{-1}$)	& $1.0$  \\ 
		$\Sigma_f$~(cm${}^{-1}$)	& $0.25$ \\ 
		$\Sigma_a$~(cm${}^{-1}$)	& $0.5882$ \\ 
		$\Sigma_s$~(cm${}^{-1}$)	& $0.4118$ \\ 
		$v$~(cm$/$s)	&  $2.2 \times 10^{4}$ \\ 
		\bottomrule
	\end{tabular}
	\caption{Material cross sections and parameters of the monoenergetic system.}
	\label{tab:parameters_mono_system}
\end{table}
\newpage
\noindent The mean neutron generation time $\Lambda$ is given by
\begin{equation}
\Lambda = (\Sigma_f \, v \, \nu)^{-1}.
\label{eq:Lambda_simple}
\end{equation} 
Here, $\Sigma_f$ is the macroscopic fission cross section, $v$ is neutron speed and $\nu$ is the average number of neutrons produced per fission. In general, this expression for the neutron generation depends on the energy, but for a monoenergetic system with constant cross sections, this value is exact. 
\chapter{Summary of the simulations performed in this work}
\label{app:summary_calculations}

Simulations presented in this work were run either in the CSICCIAN cluster from the Chilean Nuclear Energy Commission, which is comprised of $32$ cores of Intel(R) Xeon(R) CPU E$5$-$2640$ v$2$ @ $2.00$GHz processors, and $8$ Gb RAM, or in the LIN cluster from the Chilean Nuclear Energy Commission, which is comprised of $48$ cores of Intel(R) Xeon(R) CPU E$5$-$2670$ v$3$ @ $2.30$GHz processors, and $64$ Gb RAM. 

Tables presented in this Appendix summarize the details of each transient simulation presented in this work.
\newpage
\noindent \textbf{Section~\ref{section:mono-energetic_system}: Monoenergetic fissile system with $1$-group precursor structure}

\noindent All simulations presented in Table~\ref{table:app_mono_subcritical} were ran in the CSICCIAN cluster.
\begin{table}[h!]
\centering
\resizebox{\columnwidth}{!}{%
\begin{tabular}{@{}lccccccc@{}}
\toprule
\textbf{Configuration}& \textbf{Number} & \textbf{Number} & \textbf{Number} & \textbf{Number} & \textbf{Time interval} & \textbf{Simulation} & \textbf{Wall-clock} \\   
{ }  &\textbf{neutrons} & \textbf{precursors} & \textbf{batches} & \textbf{time intervals} & \textbf{length ($\boldsymbol{ms}$)} & \textbf{time ($\boldsymbol{s}$)} & \textbf{time ($\boldsymbol{h}$)} \\ \midrule
Subcritical          & $10^5$  & $9 \times 10^5$ & $60$ &   $500$ & $100$ & $50$ & $12.35$ \\ 
Critical             & $10^5$  & $9 \times 10^5$ & $4$ &    $250$ & $100$ & $25$ & $52.97$ \\  
Reactivity insertion & $10^5$  & $9 \times 10^5$ & $25$ &   $5000$ & $10$ & $50$ & $66.18$ \\  \bottomrule
\end{tabular}%
}
\caption{Summary of simulation parameters for monoenergetic fissile system in subcritical, critical, and reactivity insertion configurations, using $1$-group precursor structure.}
\label{table:app_mono_subcritical}
\end{table}

\noindent \textbf{Section~\ref{subsection:energy_u235_vacuum_sub}: Energy dependent system - Subcritical configuration}

\noindent All simulations in Table~\ref{table:app_energy_subcritical} were simulated in CSICCIAN cluster, except for study vi, which was simulated in LIN cluster.

\begin{table}[h!]
	\centering
	\resizebox{\columnwidth}{!}{%
		\begin{tabular}{@{}cccccccc@{}}
			\toprule
			\textbf{Study}& \textbf{Number} & \textbf{Number} & \textbf{Number} & \textbf{Number} & \textbf{Time interval} & \textbf{Simulation} & \textbf{Wall-clock} \\   
			{ }  &\textbf{neutrons} & \textbf{precursors} & \textbf{batches} & \textbf{time intervals} & \textbf{length (n$\boldsymbol{s}$)} & \textbf{time ($\boldsymbol{ms}$)} & \textbf{time ($\boldsymbol{h}$)} \\ \midrule
			i & $10^5$  & $9 \times 10^5$ & $22$ &   $1000$ & $100$ & $0.1$ & $44.32$ \\ 
			ii & $10^5$  & $9 \times 10^5$ & $3$ &   $500$ & $100$ & $0.05$ & $3.51$ \\  
			iii & $10^5$  & $9 \times 10^5$ & $3$ &   $500$ & $100$ & $0.05$ & $3.49$ \\  
			iv & $10^5$  & $9 \times 10^5$ & $3$ &   $500$ & $100$ & $0.05$ & $4.32$ \\  
			v & $10^5$  & $9 \times 10^5$ & $3$ &   $500$ & $100$ & $0.05$ & $4.27$ \\  
			vi & $10^5$  & $9 \times 10^5$ & $3$ &   $500$ & $100$ & $0.05$ & $6.43$ \\  \bottomrule
		\end{tabular}%
	}
	\caption{Summary of simulation parameters for energy dependent system in a subcritical configuration, using different precursor structures.}
	\label{table:app_energy_subcritical}
\end{table}
\newpage
\noindent \textbf{Section~\ref{subsection:energy_u235_vacuum_super}: Energy dependent system - Supercritical configuration}

\noindent All simulations in Table~\ref{table:app_energy_supercritical} were simulated in CSICCIAN cluster, except for the study vi, which was simulated in LIN cluster.
\begin{table}[h!]
	\centering
	\resizebox{\columnwidth}{!}{%
		\begin{tabular}{@{}cccccccc@{}}
			\toprule
			\textbf{Study}& \textbf{Number} & \textbf{Number} & \textbf{Number} & \textbf{Number} & \textbf{Time interval} & \textbf{Simulation} & \textbf{Wall-clock} \\   
			{ }  &\textbf{neutrons} & \textbf{precursors} & \textbf{batches} & \textbf{time intervals} & \textbf{length ($\boldsymbol{s}$)} & \textbf{time ($\boldsymbol{ms}$)} & \textbf{time ($\boldsymbol{h}$)} \\ \midrule
			i & $10^5$  & $9 \times 10^5$ & $10$ &   $1000$ & $100$ & $0.1$ & $52.45$ \\ 
		ii & $10^5$  & $9 \times 10^5$ & $3$ &   $500$ & $100$ & $0.05$ & $7.84$ \\  
		iii & $10^5$  & $9 \times 10^5$ & $3$ &   $500$ & $100$ & $0.05$ & $7.81$ \\  
		iv & $10^5$  & $9 \times 10^5$ & $3$ &   $500$ & $100$ & $0.05$ & $11.19$ \\  
		v & $10^5$  & $9 \times 10^5$ & $3$ &   $500$ & $100$ & $0.05$ & $11.03$ \\  
		vi & $10^5$  & $9 \times 10^5$ & $3$ &   $500$ & $100$ & $0.05$ & $17.24$ \\  \bottomrule
		\end{tabular}%
	}
	\caption{Summary of simulation parameters for energy dependent system in a supercritical configuration, using different precursor structures.}
	\label{table:app_energy_supercritical}
\end{table}

\noindent \textbf{Section~\ref{section:energy_individual_precursors}: Energy-dependent system with individual precursors and neutron moderator}

\noindent All simulations in Table~\ref{table:app_critical_moderated} were simulated in LIN cluster.
\begin{table}[h!]
	\centering
	\resizebox{\columnwidth}{!}{%
		\begin{tabular}{@{}lccccccc@{}}
			\toprule
			\textbf{Precursor}& \textbf{Number} & \textbf{Number} & \textbf{Number} & \textbf{Number} & \textbf{Time interval} & \textbf{Simulation} & \textbf{Wall-clock} \\   
			\textbf{structure}  &\textbf{neutrons} & \textbf{precursors} & \textbf{batches} & \textbf{time intervals} & \textbf{length ($\boldsymbol{ms}$)} & \textbf{time ($\boldsymbol{s}$)} & \textbf{time ($\boldsymbol{h}$)} \\ \midrule
			$6$-group       & $10^5$  & $9 \times 10^5$ & $2$ &   $400$ & $10$ & $400$ & $260.05$ \\ 
			$40$ individual & $10^5$  & $9 \times 10^5$ & $2$ &   $400$ & $10$ & $400$ & $410.75$ \\  
			$50$ individual & $10^5$  & $9 \times 10^5$ & $2$ &   $400$ & $10$ & $400$ & $319.65$ \\  \bottomrule
		\end{tabular}%
	}
	\caption{Summary of simulation parameters for light-water moderated, energy dependent system in a critical configuration, using $6$-group, $40$ individual, and $50$ individual precursor structures.}
	\label{table:app_critical_moderated}
\end{table}

\chapter{Individual precursor data}
\label{app:precursors_used_in_this_work}

In this appendix the different precursor structures used in this work are presented. 

\noindent \textbf{$6$-group precursor structure}

\begin{longtable}{|c|c|c|c|}
\hline
\textbf{Group} & \textbf{$\lambda$ ($s^{-1}$)} & \textbf{$\beta_i / \beta$} & \textbf{$\bar{E}$ (eV)} \\ \hline
$1$ & $0.0127$ &  $0.038$ & $400318$ \\ \hline
$2$ & $0.0317$ &  $0.213$ & $466542$ \\ \hline 
$3$ & $0.1156$ &  $0.188$ & $437634$ \\ \hline
$4$ & $0.311$  &  $0.407$ & $552428$ \\ \hline
$5$ & $1.397$  &  $0.128$ & $513201$ \\ \hline
$6$ & $3.872$  &  $0.026$ & $535234$ \\ \hline
\caption{$6$-group precursor structure used in this work.}
\label{table:6-group_precursor_structure}
\end{longtable}
\newpage
\noindent \textbf{$50$ individual precursor structure}

\begin{longtable}{|c|c|c|c|c|c|c|}
	\hline
	\textbf{Number} & \textbf{Z} & \textbf{Symbol} & \textbf{A} & \textbf{$\lambda$ ($s^{-1}$)} & \textbf{$I_i$} & \textbf{$\bar{E}$ (eV)} \\ 
	\hline 
 $1$&   $53$ &         I   &    $  137 $    &  $  0.0282917 $        &      $  0.1617$            &       $   624755  $   \\    
 $2$&	$35$ &        Br   &    $  89  $    &  $  0.1575335 $        &      $  0.1125$            &       $   512800  $   \\     
 $3$&	$37$ &        Rb   &    $  94  $    &  $  0.2565312 $        &      $  0.0915$            &       $   437334  $   \\     
 $4$&	$35$ &        Br   &    $  88  $    &  $  0.0425505 $        &      $  0.0740$            &       $   246533  $   \\      
 $5$&   $35$ &        Br   &    $  90  $    &  $  0.3610142 $        &      $  0.0733$            &       $   643126  $   \\     	
 $6$&	$33$ &        As   &    $  85  $    &  $  0.3429724 $        &      $  0.0478$            &       $   701816  $   \\     	
 $7$&	$53$ &         I   &    $  138 $    &  $  0.1112596 $        &      $  0.0471$            &       $   373547  $   \\     
 $8$&   $39$ &         Y   &    $  98 $m    &  $  0.3465736 $        &      $  0.0417$            &       $   214585  $   \\   
 $9$&   $53$ &         I   &    $  139 $    &  $  0.3040119 $        &      $  0.0401$            &       $   406239  $   \\     
 $10$&   $37$ &        Rb   &    $  95   $   &   $ 1.8351792  $       &       $ 0.0357$           &        $  524538   $  \\     
 $11$&   $37$ &        Rb   &    $  93  $    &  $  0.1186896 $        &      $  0.0317$            &       $   400904  $   \\         
 $12$&   $35$ &        Br   &    $  87  $    &  $  0.0124555 $        &      $  0.0314$            &       $   209628  $   \\    
 $13$&	$35$ &        Br   &    $  91   $   &   $ 1.2812332  $       &       $ 0.0279 $           &        $  886967   $  \\     
 $14$&   $39$ &         Y   &    $  99  $    &  $  0.4715287 $        &      $  0.0247$            &       $   437844  $   \\      
 $15$&   $51$ &        Sb   &    $  135 $    &  $  0.4128333 $        &      $  0.0244$            &       $   879204  $   \\  
 $16$&   $52$ &        Te   &    $  136 $    &  $  0.0396084 $        &      $  0.0157$            &       $   286456  $   \\        
 $17$&   $55$ &        Cs   &    $  143 $    &  $  0.3870169 $        &      $  0.0151$            &       $   256420  $   \\      
 $18$&	$53$ &         I   &    $  140  $   &   $ 0.8059851  $       &       $ 0.0102 $           &        $  414845   $  \\      
 $19$&   $52$ &        Te   &    $  137 $    &  $  0.2783724 $        &      $  0.0086$            &       $   373558  $   \\      
 $20$&   $37$ &        Rb   &    $  96   $   &   $ 3.4145181  $       &       $ 0.0084 $           &        $  415322   $  \\     
 $21$&   $55$ &        Cs   &    $  145  $   &   $ 1.1808299  $       &       $ 0.0073 $           &        $  335719   $  \\     
 $22$&   $33$ &        As   &    $  86   $   &   $ 0.7334891  $       &       $ 0.0071 $           &        $  553183   $  \\     
 $23$&   $55$ &        Cs   &    $  144 $    &  $  0.6973312 $        &      $  0.0061$            &       $   312643  $   \\     
 $24$&   $36$ &        Kr   &    $  93  $    &  $  0.5389947 $        &      $  0.0060$            &       $   448103  $   \\     
 $25$&   $35$ &        Br   &    $  92   $   &   $ 2.0208373  $       &       $ 0.0042 $           &        $ 1117783   $  \\     
 $26$&   $37$ &        Rb   &    $  97   $   &   $ 4.0990371  $       &       $ 0.0040 $           &        $  513170   $  \\     
 $27$&   $53$ &         I   &    $  141  $   &   $ 1.6119702  $       &       $ 0.0038 $           &        $  274291   $  \\     
 $28$&   $52$ &        Te   &    $  138 $    &  $  0.4951051 $        &      $  0.0037$            &       $   661109  $   \\     
 $29$&   $36$ &        Kr   &    $  94   $   &   $ 3.2695622  $       &       $ 0.0034 $           &        $  431444   $  \\     
 $30$&    $34$ &        Se   &    $  89   $   &   $ 1.6906029  $       &       $ 0.0033 $           &        $  588763   $  \\     
 $31$&   $51$ &        Sb   &    $  136  $   &   $ 0.7509720  $       &       $ 0.0030 $           &        $  948910   $  \\     
 $32$&   $39$ &         Y   &    $  101  $   &   $ 1.5403271  $       &       $ 0.0024 $           &        $  426877   $  \\     
 $33$&   $39$ &         Y   &    $  100  $   &   $ 0.9430574  $       &       $ 0.0022 $           &        $  423207   $  \\  
 $34$&   $39$ &         Y   &    $  98   $   &   $ 1.2648671  $       &       $ 0.0019 $           &        $  218506   $  \\                  
 $35$&    $55$ &        Cs   &    $  142 $    &  $  0.4116076 $        &      $  0.0016$            &       $   257130  $   \\            
 $36$&	$34$ &        Se   &    $  87  $    &  $  0.1260268 $        &      $  0.0016$            &       $   145104  $   \\
 $37$&	$38$ &        Sr   &    $  98   $   &   $ 1.0614811  $       &       $ 0.0015 $           &        $  245100   $  \\     
 $38$&   $41$ &        Nb   &    $  105 $    &  $  0.2349651 $        &      $  0.0014$            &       $   183361  $   \\     
 $39$&   $54$ &        Xe   &    $  142 $    &  $  0.5635343 $        &      $  0.0014$            &       $   220063  $   \\     
 $40$&   $50$ &        Sn   &    $  134 $    &  $  0.6601402 $        &      $  0.0014$            &       $   574126  $   \\     
 $41$&   $33$ &        As   &    $  88   $   &   $ 6.1888141  $       &       $ 0.0014 $           &        $  538995   $  \\     
 $42$&   $55$ &        Cs   &    $  141 $    &  $  0.0279045 $        &      $  0.0011$            &       $   214649  $   \\     
 $43$&   $32$ &        Ge   &    $  84  $    &  $  0.7265694 $        &      $  0.0011$            &       $   561448  $   \\     
 $44$&   $33$ &        As   &    $  87   $   &   $ 1.2377628  $       &       $ 0.0011 $           &        $  382320   $  \\     
 $45$&    $57$ &        La   &    $  149 $    &  $  0.6601402 $        &      $  0.0009$            &       $   458595  $   \\     
 $46$&    $35$ &        Br   &    $  93   $   &   $ 6.7955606  $       &       $ 0.0008 $           &        $  602909   $  \\
 $47$&    $55$ &        Cs   &    $  146  $   &   $ 2.1593370  $       &       $ 0.0007 $           &        $  426509   $  \\     
 $48$&    $39$ &         Y   &    $  97  $    &  $  0.1848392 $        &      $  0.0006$            &       $   182608  $   \\     
 $49$&	$31$ &        Ga   &    $  81  $    &  $  0.5695540 $        &      $  0.0006$            &       $   369630  $   \\     
 $50$&	$41$ &        Nb   &    $  106  $   &   $ 0.7453195  $       &       $ 0.0006 $           &        $  294738   $  \\     
	\hline
	\caption{$50$ individual precursor structure used in this work. Precursors are ordered by importance.}
	\label{table:50-individual_precursor_structure}
\end{longtable}

\noindent \textbf{$40$ individual precursor structure}

\begin{longtable}{|c|c|c|c|c|c|c|}
	\hline
	\textbf{Number} & \textbf{Z} & \textbf{Symbol} & \textbf{A} & \textbf{$\lambda$ ($s^{-1}$)} & \textbf{$I_i$} & \textbf{$\bar{E}$ (eV)} \\ 
	\hline
$2$&	$37$	& Rb 	&$93	$&$0.1186896	$&$0.1155	$&$400904$\\	
$1$&    $35$	& Br  	&$87	$&$0.0124555	$&$0.1144	$&$209628$   \\
$3$&	$35$	& Br	&$91	$&$1.2812332	$&$0.1016	$&$886967$\\
$4$&	$39$	& Y 	&$99	$&$0.4715287	$&$0.0898	$&$437844$\\
$5$&    $51$	& Sb 	&$135	$&$0.4128333	$&$0.0887	$&$879204$\\
$7$&	$52$	& Te 	&$136	$&$0.0396084	$&$0.0571	$&$286456$\\
$6$&	$55$	& Cs 	&$143	$&$0.3870169	$&$0.0550	$&$256420$\\
$8$&    $53$	& I 	&$140	$&$0.8059851	$&$0.0370	$&$414845$\\
$9$&    $52$	& Te 	&$137	$&$0.2783724	$&$0.0312	$&$373558$\\
$10$&   $37$	& Rb 	&$96	$&$3.4145181	$&$0.0304	$&$415322$\\
$11$&   $55$	& Cs 	&$145	$&$1.1808299	$&$0.0268	$&$335719$\\
$12$&   $33$	& As 	&$86	$&$0.7334891	$&$0.0260	$&$553183$\\
$13$&   $55$	& Cs 	&$144	$&$0.6973312	$&$0.0224	$&$312643$\\
$14$&   $36$	& Kr 	&$93	$&$0.5389947	$&$0.0220	$&$448103$\\
$15$&	$35$	& Br 	&$92	$&$2.0208373	$&$0.0154	$&$1117783$\\
$16$&	$37$	& Rb 	&$97	$&$4.0990371	$&$0.0147	$&$513170$\\
$17$&	$53$	& I 	&$141	$&$1.6119702	$&$0.0139	$&$274291$\\
$18$&	$52$	& Te   	&$138	$&$0.4951051	$&$0.0136	$&$661109$\\
$19$&	$36$	& Kr 	&$94	$&$3.2695622	$&$0.0125	$&$431444$\\
$20$&	$34$	& Se 	&$89	$&$1.6906029	$&$0.0121	$&$588763$     \\
$21$&	$51$	& Sb 	&$136	$&$0.7509720	$&$0.0110	$&$948910$\\
$22$&	$39$	& Y 	&$101	$&$1.5403271	$&$0.0088	$&$426877$\\
$23$&	$39$	& Y 	&$100	$&$0.9430574	$&$0.0081	$&$423207$\\
$24$&	$39$	& Y 	&$98	$&$1.2648671	$&$0.0070	$&$218506$\\
$26$&	$55$	& Cs 	&$142	$&$0.4116076	$&$0.0059	$&$257130$\\
$25$&	$34$	& Se 	&$87	$&$0.1260268	$&$0.0059	$&$145104$\\
$27$&	$38$	& Sr 	&$98	$&$1.0614811	$&$0.0055	$&$245100$\\
$28$&	$41$	& Nb 	&$105   $&$0.2349651	$&$0.0051    $&$183361$\\
$29$&	$54$	& Xe 	&$142	$&$0.5635343	$&$0.0051	$&$220063$\\
$30$&	$50$	& Sn 	&$134	$&$0.6601402	$&$0.0051	$&$574126$   \\
$31$&	$33$	& As 	&$88	$&$6.1888141   $&$0.0051	$&$538995$\\
$32$&	$55$	& Cs 	&$141	$&$0.0279045	$&$0.0040	$&$214649$\\
$33$&	$32$	& Ge 	&$84	$&$0.7265694	$&$0.0040	$&$561448$\\
$34$&	$33$	& As 	&$87	$&$1.2377628	$&$0.0040	$&$382320$\\
$35$&	$57$	& La 	&$149	$&$0.6601402	$&$0.0033	$&$458595$\\
$36$&	$35$	& Br 	&$93	$&$6.7955606	$&$0.0029	$&$602909$\\
$37$&	$55$	& Cs 	&$146	$&$2.1593370	$&$0.0026	$&$426509$\\
$38$&	$39$	& Y 	&$97	$&$0.1848392	$&$0.0022	$&$182608$\\
$39$&	$31$	& Ga 	&$81	$&$0.5695540	$&$0.0022	$&$369630$\\
$40$&	$41$	& Nb 	&$106	$&$0.7453195	$&$0.0022	$&$294738$\\
\hline
	\caption{$40$ individual precursor structure used in this work. Precursors are ordered by importance.}
	\label{table:40-individual_precursor_structure}
\end{longtable}

\bibliographystyle{unsrt}
\bibliography{Biblio/ms}

\end{document}